\documentclass[aps,prb,floatfix,epsfig,twocolumn,showpacs,preprintnumbers]{revtex4}
\usepackage{amsmath,amssymb,graphicx,bm,epsfig}
\usepackage{color}

\newcommand{\eps}{\epsilon}
\newcommand{\vR}{{\mathbf{R}}}
\renewcommand{\vr}{{\mathbf{r}}}
\newcommand{\hr}{{\hat{\textbf{r}}}}
\newcommand{\vk}{{\mathbf{k}}}
\newcommand{\vK}{{\mathbf{K}}}

\newcommand{\cG}{{\cal G}}

\newcommand{\cL}{{\cal L}}
\newcommand{\Tr}{\mathrm{Tr}}

\renewcommand{\Im}{\textrm{Im}}
\renewcommand{\Re}{\textrm{Re}}
\newcommand{\cA}{{\cal A}}

\begin{document}

\title{Dynamical Mean-Field Theory within the Full-Potential Methods:
  Electronic structure of Ce-115 materials}

\author{Kristjan Haule, Chuck-Hou Yee, Kyoo Kim}
\affiliation{Department of Physics, Rutgers University, Piscataway, NJ 08854, USA}
\date{\today}

\begin{abstract}
We implemented the charge self-consistent combination of Density
Functional Theory and Dynamical Mean Field Theory (DMFT) in two
full-potential methods, the Augmented Plane Wave and the Linear
Muffin-Tin Orbital methods.
We categorize the commonly used projection methods in terms of the
causality of the resulting DMFT equations and the amount of partial
spectral weight retained.
The detailed flow of the Dynamical Mean Field algorithm is described,
including the computation of response functions such as transport
coefficients.
We discuss the implementation of the impurity solvers based on
hybridization expansion and an analytic continuation method for
self-energy.  We also derive the formalism for the bold continuous
time quantum Monte Carlo method.
We test our method on a classic problem in strongly correlated
physics, the isostructural transition in Ce metal.
We apply our method to the class of heavy fermion materials
CeIrIn$_5$, CeCoIn$_5$ and CeRhIn$_5$ and show that the Ce $4f$
electrons are more localized in CeRhIn$_5$ than in the other two, a
result corroborated by experiment.  We show that CeIrIn$_5$ is the most 
itinerant and has a very anisotropic hybridization, pointing mostly
towards the out-of-plane In atoms.  In CeRhIn$_5$ we stabilized the
antiferromagnetic DMFT solution below $3\,$K, in close agreement with the
experimental N\'eel temperature.
\end{abstract}
\date{\today}
\maketitle

\section{Introduction}

One of the most active areas of condensed matter theory is the
development of new algorithms to simulate and predict the behavior of
materials exhibiting strong correlations.
Recent developments in the dynamical mean-field theory
(DMFT)\cite{Antoine}, a powerful many-body approach, hold great
promise for more accurate and realistic descriptions of physical
properties of this challenging class of materials.

The crucial step towards realistic description of strongly correlated
materials was the formulation of
DFT+DMFT\cite{Anisimov,Lichtenstein,PhysicsToday}, a method formed by
the combination of density functional theory (DFT) and DMFT (for a
review see Ref.~\onlinecite{our-rmp}).  To date, this method already
has substantially advanced our understanding of the physics of the
Mott transition in real materials and demonstrated its ability to
explain phenomena including the structural phase diagrams of
actinides~\cite{V6,V7,PuAm}, phonon response~\cite{V8}, optical
conductivity~\cite{V9,FeAs}, valence and x-ray
absorption~\cite{V10,PuCris,Xray} and transport~\cite{V11} of
archetypal strongly correlated materials.

At present, much effort is devoted to the development of a robust and
precise implementation of DFT+DMFT using state of the art DFT
electronic structure
codes\cite{Savrasov04,Lichtenstein_new,KKR,Wannier,Wannier2} and
advanced impurity solvers\cite{Werner,Rubtsov,CTQMC,Werner2}.  This article
describes in detail the implementation of this method within
full-potential codes. There are three major issues that arise in
DFT+DMFT implementations: i) quality of the basis set, ii) quality of
the impurity solvers, and iii) choice of correlated orbitals onto
which the full Green's function is projected.  Modern DFT
implementations largely resolve the first issue, recent development of
new impurity
solvers~\cite{Rubtsov,Werner,CTQMC,SUNCA,XDai,SergejIPT,LichSolver}
have focused attention on the second, while the third is rarely
discussed in the literature.  Many DFT+DMFT proposals in the
literature are based on downfolding to low energy model
Hamiltonians~\cite{Anisimov,Anisimov2,Wannier,Wannier2}, which requires an atomic set of
orbitals and treats the kinetic operator on the level of an effective
tight binding model. In contrast, we avoid the ambiguities of
downfolding and instead keep the kinetic part of the Hamiltonian and
electronic charge expressed in a highly accurate full potential basis
set. The advantage of our method is its ability to perform fully
self-consistent electronic charge calculations.  We concentrate here
on the Linear Augmented Plane Wave basis (LAPW)~\cite{LAPW-book} as
implemented in the Wien2K code \cite{Wien2K} and the LMTO basis as
implemented in LmtArt~\cite{Savrasov96}, in combination with the
impurity solvers based on the hybridization expansion
\cite{SUNCA,Werner,CTQMC,Werner2}.

The first half of the article introduces the basic steps of
implementing the DFT+DMFT algorithm and provides a pedagogical
introduction to the method. Section \ref{basis_set} is devoted to a
crucial element of the DFT+DMFT formalism, namely the projection of the full
electronic Green's function to the correlated subset. We show that the
projection used in the LDA+U method leads to non-causal DFT+DMFT
equations, while the projection on to the solution of the Schr\"odinger
equation within the Muffin-Tin (MT) spheres misses electronic spectral
weight. We propose a new projection that leads to causal DMFT
equations and captures all electronic spectral weight. Section \ref{Sfunctional} derives the
DFT+DMFT equations from a Baym-Kadanoff-like functional formalism.
Section \ref{algorithm} provides a detailed flowchart of all the steps of the
algorithm. In section
\ref{complex-tetrahedra} we discuss the necessary changes to the
tetrahedron method when used in the context of DMFT. Section
\ref{transport} described the algorithm to compute transport properties
within DFT+DMFT. Section \ref{Impurity-solvers} describes the impurity
solvers based on the hybridization expansion, the One Crossing
Approximation (OCA), and the \textit{Bold} continuous time quantum Monte
Carlo algorithm ($b$-CTQMC). Finally, section \ref{ancont} discusses a new
algorithm for analytic continuation of the self-energy from the imaginary
to real axis.

In the second half of the article, we describe the results obtained by
applying our new implementation of DFT+DMFT to several correlated
materials.  As a first test of the algorithm, in section \ref{Ce-test}
we present its application to elemental cerium.  Section \ref{115} is
devoted to a class of heavy fermion materials, CeRhIn$_5$, CeCoIn$_5$
and CeIrIn$_5$, dubbed Ce-155 materials. We 
show the difference in the electronic structure among these three
materials and demonstrate that the Ce $4f$ electrons are most localized in
CeRhIn$_5$ and order antiferromagnetically below $T_N\approx 3\,$K, in
agreement with experiment, while the Ce $4f$ electrons are most itinerant
in CeIrIn5. We explain the origin of the subtle difference between the
three Ce-115 compounds from the electronic structure point of view.

\section{Projection on to correlated orbitals within full-potential methods}
\label{basis_set}

\begin{figure}[hbt]
\centering{
  \includegraphics[width=0.99\linewidth]{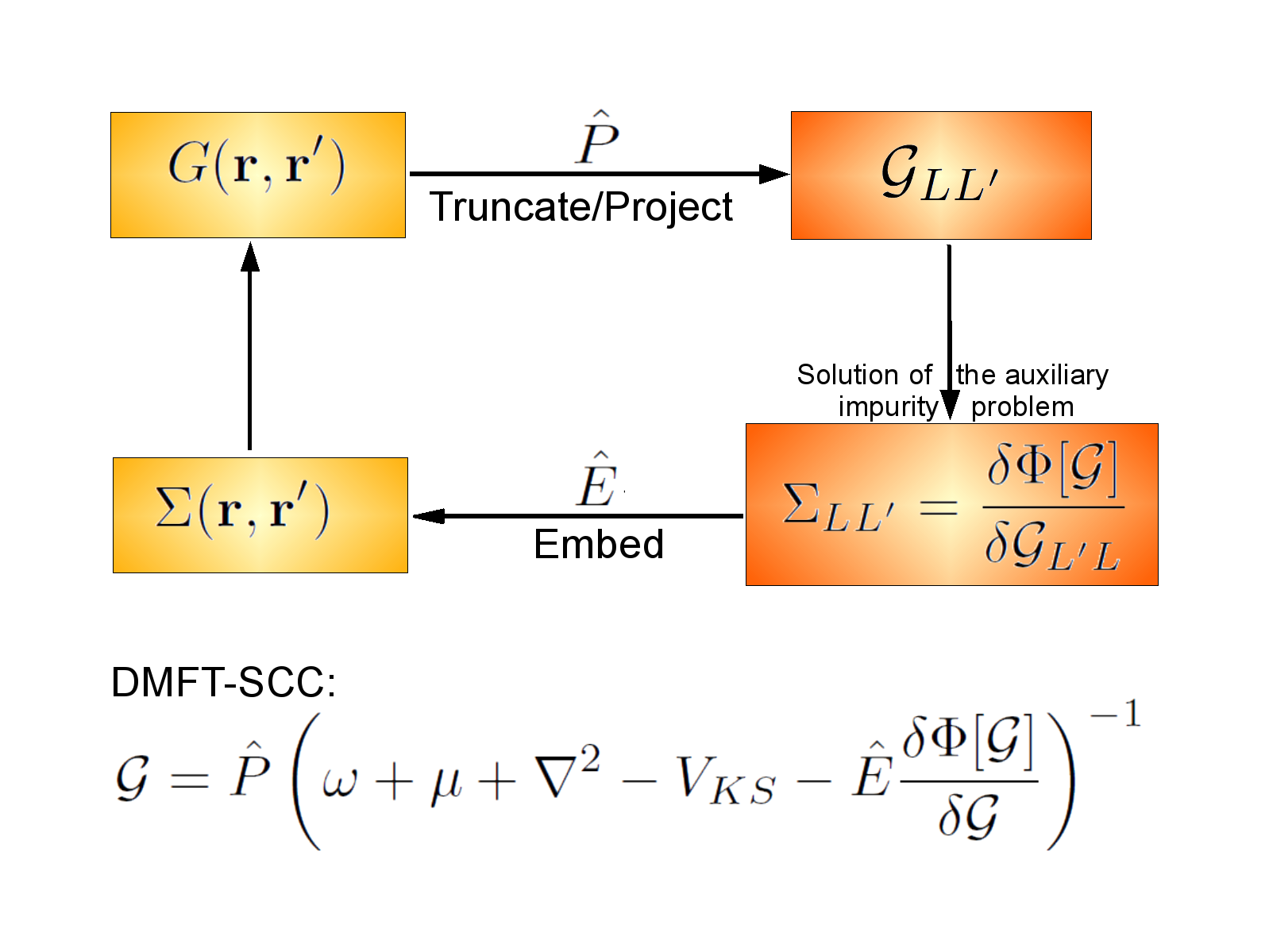}
  }
\caption{ Schematic diagram of the projection-embedding step in the
  DFT+DMFT algorithm. The full Green's function of the solid
  $G(\vr,\vr')$ is truncated to its local counterpart $\hat{P}G =
  \cG_{LL'}$. The impurity solution delivers an effective local
  potential, which is embedded ($\hat{E}$) into the Dyson equation of
  the solid. The DMFT self-consistency condition (DMFT-SCC) connects
  the two.  }
\label{sketch0}
\end{figure}

DFT+DMFT contains some aspects of band theory, adding a
``frequency-dependent local potential'' to the Kohn-Sham Hamiltonian. It
also contains some aspects of quantum chemistry, carrying out an exact
local configuration interaction procedure by summing all local
diagrams, which requires the definition of an ``atomic-like'' or ``local''
Green's function.  The operation of extracting the local Green's
function $\cG(\vr,\vr')$ from the full Green's function $G(\vr,\vr')$
is called projection (or truncation).  The reverse operation of
expressing the local time-dependent potential $\Sigma(\omega)$, derived
from the solution of the atomic problem in the presence of a
mean-field environment, is called embedding.  The various DFT+DMFT
implementations differ not only in the choice of basis set, but
also in the choice of the projection-embedding step. These ingredients are
sketched schematically in Fig.~\ref{sketch0}. The projection-embedding step
connects the atomic and solid state physics, and its proper
definition is a conceptual issue of DFT+DMFT method.
%

In the current formulation of DFT+DMFT\cite{our-rmp,Held,Lich}, one
must define the correlated orbitals to which the Coulomb correlation is applied, i.e.,
$\Sigma(\vr,\vr')=\sum_{\xi\xi'}\chi_\xi(\vr)\Sigma_{\xi\xi'}\chi_{\xi'}^*(\vr')$,
where $\chi_\xi(\vr)$ is a localized orbital.  Usually, this is
achieved by transforming the DFT Hamiltonian to a set of localized 
Wannier orbitals.  These Wannier orbitals are then identified as the
local correlated orbitals of DMFT.
Various choices of these orbitals were proposed in the literature, including
tight-binding LMTO's \cite{Anisimov,Lichtenstein}, non-orthogonal
LMTO's \cite{Savrasov04}, Nth-order Muffin-Tin orbitals
\cite{Pavarini}, numerically-orthogonalized LMTO's \cite{Purovskii},
and maximally-localized Wannier orbitals \cite{Wannier2,Korotin}.
The basis functions must fully respect the symmetries of the problem
and be atom-centered, rather than bond-centered.  Hence
maximally-localized Wannier functions~\cite{MLWO} are not a good
starting point for DMFT. 

Localized basis sets are a better starting point for our purposes, but
the non-orthogonality of these sets pose a serious
challenge. Straighforward orthogonalization mixes the character of the
orbitals, resulting in mixed the partial occupancies and partial density of states,
leading to incorrect partial electron counts. For example, within
modern DFT implementations, cerium metal has approximately one
$4f$ electron. Na\"ive orthogonalization results in a considerably higher
$4f$ electron count, leading to an unphysical DMFT solution.

Even more challenging is the formulation of the good localized orbitals
in full-potential basis sets. Here, multiple basis functions are
used to obtain more variational freedom.  To implement DMFT in such basis
sets, the group of orbitals representing the correlated electrons in
the solid must be contracted to form a single set of atomic-like heavy
orbitals, i.e., one $4f$ orbital per Ce atom, one $3d$ orbital
per Fe atom, etc.

A straighforward projection on to the orbital angular momentum
eigenfunctions $Y_{lm}\equiv Y_L$ leads to non-causal DMFT equations,
which result in an unphysical auxiliary impurity problem. The second
often-employed choice is the projection on to the solution of the
Schr\"odinger (Dirac) equation inside the MT sphere
$u_l(E_\nu,r)Y_L(\hr)$. While this choice is certainly superior to the
straighforward projection, it does not take into account the
contributions due to the energy derivative of the radial wave function
$\dot{u}_l(E_\nu,r)Y_L(\hr)$ and the localized orbitals (LO) at other
energies $u_l(E'_\nu,r)Y_L(\hr)$, and hence misses some electronic
spectral weight of the correlated orbital. Alternative choices are
possible which simultaneously capture all spectral weight and obey
causality.  We implemented one of them and we believe it is superior
to other choices in the literature. 

The central objects of DMFT are the local Green's function and the
local self-energy of the orbitals within the correlated subset.  We
specify the projection scheme by the projection operator
$P(\vr\vr',\tau LL')$, which defines the mapping between real-space
objects and their orbital counterparts $(\vr,\vr') \rightarrow (L,L')$
(see Fig.~\ref{sketch0}).  The operator $\hat{P}$ acts on the full
Green's function $G(\vr,\vr')$ and gives the correlated Green's
function $\cG^{\tau}_{LL'}$~\cite{Savrasov04,our-rmp}
\begin{eqnarray}
\cG_{LL'}^{\tau} = \int d\vr d\vr' P(\vr\vr',\tau LL')G(\vr\vr').
\label{GLL1}
\end{eqnarray}
The integrals over $\vr$ and $\vr'$ are performed inside the sphere of
size $S$ around the correlated atom at position $\tau$. The subscript $L$ can
index spherical harmonics $lm$, cubic harmonics, or relativistic
harmonics $jm_j$, depending on the system symmetry. We always
choose the basis which minimizes the off-diagonal elements of the correlated
Green's function in order to reduce the minus-sign problem in
Monte-Carlo impurity solvers.  In general, $\hat{P}$ is a
multidimensional tensor with one pair of indices in the space of
local correlated orbitals $(\tau L L')$ and the other pair in the space of
the full basis set, which can be expressed in a real space $(\vr\vr')$ or
Kohn-Sham $(\vk,ij)$ basis, where $i$ and $j$ are band indices.

The inverse process of embedding $\hat{E}$, i.e. the mapping
between the correlated orbitals and real-space $(L,L') \rightarrow
(\vr,\vr')$, is defined by the same four-index tensor.  However,
instead of integrals over real-space, its application is through a
discrete sum over the local degrees of freedom,
\begin{eqnarray}
\Sigma(\vr,\vr') =
\sum_{\tau LL'\in H}P(\vr'\vr,\tau L'L)\Sigma^\tau_{LL'}
\label{SRR1}
\end{eqnarray}
Here $LL'\in H$ means to only sum over correlated orbitals. In
actinides, the sum would run over $5f$ orbitals, in lanthanides over
$4f$ and in transition metals over $3d$ orbitals. $\tau$ runs over all
atoms in the solid and $\vr$ over the full space.
Note that within the correlated Hilbert subspace, the embedding and
projection should give unity $\hat{P}\hat{E}=I$, i.e.,
\begin{eqnarray}
\int d\vr d\vr' P(\vr\vr',\tau L_1 L_2)P(\vr'\vr,\tau' L_3 L_4) =
\delta_{L_1 L_4}\delta_{L_2 L_3}\delta_{\tau\tau'},
\label{PPunity}
\end{eqnarray}
while the projection from the full Hilbert space to the correlated
set, followed by embedding, gives the correlated local Green's function
in real space $\hat{E}\hat{P} G(\vr\vr')=\cG(\vr,\vr')$
\begin{eqnarray}
&& \cG(\vr,\vr')=
\label{corrG}\\ 
&& \sum_{\tau LL'\in H}P(\vr'\vr,\tau L' L)\int d\vr_1 d\vr_2 P(\vr_1\vr_2,\tau L L')G(\vr_1\vr_2)\nonumber
\end{eqnarray}
which is the central object of the functional definition of the DMFT
described below.  In general, the two operators $\hat{P}$ and
$\hat{E}$ could be different, but they must satisfy the condition
Eq.~(\ref{PPunity}).

The two simplest projections, namely, the projection on to the orbital
angular momentum functions $Y_L$, and the projection on to the solution
of the Schr\"odinger equation, can be explicitly written as
\begin{eqnarray}
P^0(\vr\vr',\tau LL') &=& Y_{L}(\hr_\tau)\delta(r-r')Y_{L'}^*(\hr_\tau')
\label{PP00}\\
P^1(\vr\vr',\tau LL') &=& Y_{L}(\hr_\tau) u_l^0(r_\tau) u_{l'}^0(r'_{\tau})Y^*_{L'}(\hr'_\tau)
\label{PP01}
\end{eqnarray}
where $\vr_\tau=\vr-R_\tau$ is the vector defined with the origin
placed at the atomic position $R_\tau$, and $u^0_l(r)$ is the solution of 
the radial Schr\"odinger equation for angular momentum $l$ at a fixed
energy $E_\nu$.

In the following, we will show that the projection $P^0$, used in some
implementations of DMFT \cite{Lichtenstein_new}, captures the
full spectral weight of the correlated character $L$, but leads to
non-causal DMFT equations. On the other hand $P^1$ gives causal DMFT
equations, but misses some spectral weight.

In our view, a good DFT+DMFT implementation should satisfy the
following conditions
\begin{itemize}

\item[(1)] \textit{Correct correlated spectral weight}:
The projected density of states, computed from the projected Green's function,
\begin{equation}
  \rho_L(\omega) = \frac{1}{2\pi i}[{\cG^\dagger}_{LL}(\omega)-\cG_{LL}(\omega)],
\end{equation}
should capture the partial electronic weight inside a given MT sphere
at all frequencies, i.e., $\rho_L(\omega) \stackrel{!}{=}
\rho^\text{LDA}_L(\omega)$.  In particular, $\cG_{LL'}$ must include
the electronic weight contained in $\dot{u}_l$ and local
orbitals. Projection should not include any weigh of other
character, nor miss correlated weight.

\item[(2)] \textit{DMFT equations are causal}:
For any causal self-energy $\Sigma$, the DMFT self-consistency
condition 
\begin{multline}
  \frac{1}{\omega-E_{imp}-\Sigma-\Delta}= \\ \sum_\vk P_{\vk}[(\omega+\mu-H_\vk^{DFT}-E_\vk\Sigma)^{-1}]
\label{causal1}
\end{multline}
should give a causal hybridization function $\Delta(\omega)$.  Here we
used projections $P_\vk$ in momentum space as opposed to their
real-space definitions in Eqs.~(\ref{GLL1}) and (\ref{PP00}),
(\ref{PP01}). 

\item[(3)] \textit{Sufficient accuracy of the hybridization function}:
  The hybridization function is usually very sensitive to the choice of
  the projector.  Therefore, we require that in the relevant low energy
  region, the hybridization function is similar to its DFT
  counterpart.  Explicitly, $\Delta(\omega) = \omega-E_{imp}-
  (P G_0)^{-1}$ must be sufficiently close to its DFT estimate, $\Delta(\omega)
  = \omega-E_{imp} - (P^0 G_0)^{-1}$.  Here $G_0(\vr,\vr')$ stands for the
  full Green's function $G(\vr,\vr')$ when $\Sigma=0$.
  The choice of $\Sigma=0$ is dictated by the fact that the hybridization
  $\Delta$, computed by $P^0$ is not well behaved for $\Sigma\ne 0$,
  as we will show below.
The motivation for using P0 in the above equation is that we want to
project the full Hilbert space to a correlated subset with pure
angular momentum, either \textit{f} or \textit{d}, but not to a mixure of
characters.
  
\item[(4)] \textit{Good representation of kinetic energy and
  electronic density}: Finally, it is crucial to faithfully represent
  the kinetic energy operator $\nabla^2$ and electronic density in
  real space, a feat most modern DFT implementations achieve.
  The DFT+DMFT implementation should not reduce the precision already
  achieved in DFT underlying code. 
  
\end{itemize}

Downfolding to only a few low energy bands clearly violates the
condition number (3), since the hybridization outside the downfolded
window vanishes.  A more severe problem is that downfolding
approximates the kinetic energy operator by expressing it in 
a small atomic-like basis set, hence condition (4) is violated.
Therefore, we will focus our discussion on DFT+DMFT implemented within
full-potential basis sets where all bands are kept at each stage of
the calculation.  Downfolding to a sufficiently large energy
window may sometimes be helpful due to its conceptual simplicity, but
this approach can not compute the electronic charge self-consistently,
as is possible in our implementation.  Moreover, the localized
orbitals chosen in the downfolding procedure combined with the limited
number of hoppings retained often cannot faithfully represent the
original Kohn-Sham bands.


To be more concrete, we will give the proofs of the
\textit{``weight loss problem''} and \textit{``causality problem''} within the
full-potential LAPW basis.  The equivalent derivation is possible for
the full-potential LMTO basis.  Inside the MT spheres, the
full-potential LAPW  basis functions can be written~\cite{LAPW-book}
\begin{eqnarray}
\chi_{\vk+\vK}(\vr) = \sum_{L\tau\kappa}A^{\tau\kappa}_{\vk+\vK,L}u_l^{\tau\kappa}(r_\tau)Y_L(\hr_\tau)
\end{eqnarray}
where $\kappa=0$ corresponds to the solution of the Schr\"odinger
equation $u_l(E_\nu,r_\tau)$ at a fixed energy $E_\nu$, $\kappa=1$ to
the energy derivative of the same solution $\dot{u}_l(E_\nu,r_\tau)$,
and $\kappa=2, 3, \ldots$ to a localized orbitals at additional linearization energies
$E_\nu', E_\nu'', \ldots$. Here $\tau$ runs over the atoms in the
unit cell.

The Kohn-Sham states $\psi_{i\vk}(\vr)$ are superpositions of the
basis functions
\begin{eqnarray}
\psi_{i\vk}(\vr) = \sum_\vK C_{i\vK}^\vk \;\chi_{\vk+\vK}(\vr)
\end{eqnarray}
and take the following form inside the MT spheres:
\begin{eqnarray}
\psi_{i\vk}(\vr) = \sum_{\tau L \kappa} \cA_{iL}^{\tau\kappa}(\vk) u_l^{\tau\kappa}(r_\tau)Y_L(\hr_\tau)
\end{eqnarray}
where
$\cA_{iL}^{\tau\kappa}(\vk) = \sum_\vK A_{\vk+\vK,L}^{\tau\kappa}C_{i\vK}^\vk$,
or equivalently,
$\int d\hr_\tau Y_L^*(\hr_\tau)\psi_{i\vk}(\vr)=\sum_{\kappa}\cA^{\tau\kappa}_{i L}(\vk)u_{l}^{\tau\kappa}(r_\tau)$.

The projectors (\ref{PP00}) and (\ref{PP01}) can be expressed in the
Kohn-Sham basis:
\begin{eqnarray}
P_\vk(ij,\tau LL')=\int d\vr d\vr' \psi^*_{i\vk}(\vr) P(\vr\vr',\tau LL')\psi_{j\vk}(\vr').
\end{eqnarray}
Hence, projector $P^0$ takes the form
\begin{eqnarray}
&& P_\vk^0(ij,\tau LL') \nonumber \\
&&= \int d\vr d\vr' \psi^*_{i\vk}(\vr)
Y_{L}(\hr_\tau)\delta(r-r')Y_{L'}^*(\hr'_\tau)\psi_{j\vk}(\vr')
\nonumber \\
&&=
\sum_{\kappa\kappa'}\cA_{i L}^{\tau\kappa *}(\vk)\cA_{j L'}^{\tau \kappa'}(\vk)
\langle u_l^{\tau\kappa}|u_{l'}^{\tau\kappa'}\rangle
\end{eqnarray}
Using projector $P^0$, we get the following expression for the partial density of states
\begin{equation}
D_{\tau L}(\omega) = 
\sum_{\kappa\kappa'\vk i}\cA_{i L}^{\tau\kappa *}(\vk)\cA_{i L}^{\tau \kappa'}(\vk)
\langle u_l^{\tau\kappa}|u_{l}^{\tau\kappa'}\rangle\delta(\omega+\mu-\varepsilon_{\vk i})
\label{DOS0}
\end{equation}
which exactly coincides with the DFT partial DOS.   Hence $P^0$
satisfies the condition number (1).  However, it does not lead to
causal DMFT equations.

To show that, consider the limit of a diverging self-energy,
$\Sigma\rightarrow-i\infty$, as is relevant for the Mott insulators.
Despite the diverging $\Sigma$, the projection must still produce a
finite hybridization.  In the case when all the bands at the energy of
the pole are correlated, the hybridization should vanish.  In this
limit, the DMFT self-consistency condition (\ref{causal1}) takes the
form
\begin{eqnarray}
&& ({\Sigma^{\tau}}+\Delta)^{-1}_{LL'}\nonumber\\
&&  =\sum_{\vk ij} P_\vk(ji,\tau LL')\left[\sum_{L_2 L_3\tau'}P_\vk(::,\tau'L_2L_3)\Sigma^{\tau'}_{L_3 L_2}\right]^{-1}_{ij}
\label{Causal1}
\end{eqnarray}
where $::$ stands for the two band indices constituting a matrix in
$ij$ to be inverted. Since $\Delta$ is finite while $\Sigma$ 
diverges, we neglect $\Delta$ to obtain the condition for
causal projection,
\begin{eqnarray}
\delta_{LL''}=\sum_{\vk ij,\tau L'} P_\vk(ji,\tau LL')\Sigma^\tau_{L'L''}\times \nonumber\\
\times \left[\sum_{L_2 L_3\tau'}P_\vk(::,\tau'L_2L_3)\Sigma^{\tau'}_{L_3 L_2}\right]^{-1}_{ij}.
\label{Causal2}
\end{eqnarray}
This equation must be satisfied for any matrix form of the
self-energy $\Sigma$.  Moreover, it has to be satisfied for each $L$
and $L''$. We will show below that Eq.~(\ref{Causal2}) is satisfied
for a separable projection (see Eq.~\ref{separable} for a
definition), while for a non-separable projection, it likely is
not. One can check explicitely that $P^0$ violates the 
condition Eq.~(\ref{Causal2}). Only after applying an additional trace
over $LL''$ will the two matrices $P\Sigma$ cancel.  However, for
any given choice of $LL''$, $P^0$ does not satisfy the causality
condition.  Instead a pole in the self-energy results in a diverging
$\Delta$, with the imaginary part having the wrong sign. The
projection $P^0$ is implemented in the \textit{qtl} package
\cite{NovakQTL} of Wien2K\cite{Wien2K}.  The LDA+U
implementation within Wien2K~\cite{Shick} also uses $P^0$, but this
does not cause any causality issues since the problem is unique to
DFT+DMFT.  Additionally, simple impurity solvers such as Hubbard-I
(Ref.~\onlinecite{Lichtenstein_new}) do not incorporate a true
hybridization so they also avoid issues with causality.

Finally, let us mention an attractive feature of $P^0$.  Within this
scheme, the self-energy is independent of the radial distance from the
atom $r_\tau$, having only angular dependence in the form
$\Sigma(\hr,\hr')$.  This matches the conceptual fact that the
impurity solver within the DMFT framework can not determine the radial
dependence of the self-energy.  The impurity solver can only be used
to obtain the angular dependence of $\Sigma$ by determining the
expansion coefficeints $\Sigma_{LL'}$.  In the absence of any
knowledge of the radial dependence of $\Sigma$, the natural choice is
a constant function, independent of radius $r_\tau$. Since
$\Sigma(\vr,\vr')$ is a function of two vectors, a radial delta
function would be an obvious choice.  However, issues with causality
preclude the use of this projection.


%

The second projection $P^1$ of Eq.~(\ref{PP01}) takes the following form
in the Kohn-Sham basis:
\begin{multline}
P_\vk^1(ij,\tau LL') = \\\sum_{\kappa\kappa'}\cA_{i L}^{\tau\kappa}(\vk)A_{j L'}^{\tau \kappa' *}(\vk)
\langle u_l^{\tau\kappa}|u_l^0\rangle\langle u_{l'}^0|u_{l'}^{\tau\kappa'}\rangle.
\end{multline}
The partial density of states computed from the correlated Green's
function using $P^1$ is
\begin{eqnarray}
&& D_{\tau L}(\omega) = \label{DOS1}\\
&&\sum_{\kappa\kappa'\vk i}\cA_{i L}^{\tau\kappa}(\vk)A_{i L}^{\tau \kappa' *}(\vk)
\langle u_l^{\tau\kappa}|u_l^0\rangle\langle u_l^0|u_{l}^{\tau\kappa'}\rangle\delta(\omega+\mu-\varepsilon_{\vk i})
\nonumber
\end{eqnarray}
Comparing Eq.~(\ref{DOS1}) with (\ref{DOS0}), we notice that 
$\langle u_l^{\tau\kappa}|u_{l}^{\tau\kappa'}\rangle$ is replaced by
$\langle u_l^{\tau\kappa}|u_l^0\rangle\langle u_l^0|u_{l}^{\tau\kappa'}\rangle$, 
which leads to incorrect spectral weight. In particular, for $\kappa=1$,
the original overlap in Eq.~(\ref{DOS0}) is $\langle
\dot{u}_l^{\tau}|\dot{u}_{l}^{\tau}\rangle$, while the overlap
obtained by $P^1$, vanishes.

Causality is not violated for any projection $P$, which is 
separable, i.e., can be cast into the form
\begin{equation}
P^\vk (ij,\tau LL')=U^{\vk\tau}_{iL} U^{\vk\tau *}_{jL'}.
\label{separable}
\end{equation}
The condition Eq.~(\ref{Causal2}) can then be expressed as
\begin{eqnarray}
1=\sum_\vk U^{\vk\tau\dagger}(U^{\vk\tau}\Sigma^{\tau} U^{\vk\tau\dagger})^{-1}U^{\vk\tau}\Sigma^\tau
\end{eqnarray}
which is clearly satisfied when $U^{\vk\tau \dagger}U^{\vk\tau}$ is
invertible matrix because
$U^\dagger (U\Sigma U^\dagger)^{-1} U \Sigma \, U^\dagger U (U^\dagger U)^{-1}=1$.
This is satisfied when the Kohn-Sham Hilbert space is of larger
dimension than the correlated Hilbert space.
The projection $P^1$ leads to causal DMFT equations, and therefore is
a better choice than $P^0$.  However, some spectral weight is lost at
energies away from the linearization energy $E_\nu$. To this end, we
also implemented an alternative projection within Wien2K package
\cite{Wien2K}, which preserves both causality and spectral weight.
This projector is given by
%
%
%
\begin{widetext}
\begin{eqnarray}
&& P^2(\vr\vr',\tau LL') =
\sum_{ij\vk\kappa\kappa'}\psi_{i\vk}(\vr)\cA^{\tau \kappa}_{iL}(\vk)
\langle u_l^{\tau\kappa}|u_l^{\tau 0}\rangle\langle u_{l'}^{\tau 0}|u_{l'}^{\tau\kappa'}\rangle
\cA_{jL'}^{\tau  \kappa' *}(\vk) \psi^*_{j\vk}(\vr')\times
\nonumber\\
&&\qquad\qquad
\sqrt{
  \left(\frac{\sum_{\kappa_1\kappa_2}\cA_{i  L}^{\tau\kappa_1}\cA_{i L}^{\tau\kappa_2 *}\langle u_l^{\tau\kappa_1}|u_l^{\tau\kappa_2}\rangle}
             {\sum_{\kappa_1\kappa_2}\cA_{i  L}^{\tau\kappa_1}\cA_{i L}^{\tau\kappa_2 *}\langle u_l^{\tau\kappa_1}|u_l^{\tau 0}\rangle\langle u_l^{\tau0}|u_l^{\tau\kappa_2}\rangle}
  \right)
  \left(\frac{\sum_{\kappa_1\kappa_2}\cA_{j  L'}^{\tau\kappa_1*}\cA_{j L'}^{\tau\kappa_2}\langle u_{l'}^{\tau\kappa_1}|u_{l'}^{\tau\kappa_2}\rangle}
             {\sum_{\kappa_1\kappa_2}\cA_{j  L'}^{\tau\kappa_1*}\cA_{j L'}^{\tau\kappa_2}\langle u_{l'}^{\tau\kappa_1}|u_{l'}^{\tau0}\rangle\langle u_{l'}^{\tau0}|u_{l'}^{\tau\kappa_2}\rangle}
  \right).
}
\label{P2P}
\end{eqnarray}
\end{widetext}
%
%
%
%
%
%
%
Here index $L$ runs over the local basis in which the green's function
is minimally off-diagonal (cubic harmonics or relativistic harmonics).

The projector is separable, as postulated in
Eq.~(\ref{separable}), and the transformation $U$ is
\begin{eqnarray}
U_{i L}^{\vk\tau}=\sum_\kappa \cA_{i L}^{\tau\kappa}(\vk) \langle
u_l^{\tau\kappa}|u_l^{\tau0}\rangle S_{iL}^\tau
\end{eqnarray}
with
\begin{equation}
  S_{i L}^\tau = \sqrt{\frac{\sum_{\kappa_1\kappa_2}\cA_{i  L}^{\tau\kappa_1}\cA_{i L}^{\tau\kappa_2 *}\langle u_l^{\tau\kappa_1}|u_l^{\tau\kappa_2}\rangle}
             {\sum_{\kappa_1\kappa_2}\cA_{i  L}^{\tau\kappa_1}\cA_{i L}^{\tau\kappa_2 *}\langle u_l^{\tau\kappa_1}|u_l^{\tau 0}\rangle\langle u_l^{\tau0}|u_l^{\tau\kappa_2}\rangle}
}
\end{equation}
Hence the DMFT equations are causal. Moreover, $P^{2}_\vk(ii,LL)$ is
identical to $P^{0}_\vk(ii,LL)$ and hence the partial density of
states $D_L(\omega)$, obtained by $P^2$, is identical to
Eq.~(\ref{DOS0}). Hence the projection correctly captures the partial
spectral weight. 
Knowledgeable reader would notice that the projection
is slightly non-local because $S^\tau_{i L}$ is weakly momentum
dependent. At energies where $\dot{u}$ or local orbital substantially
contribute to the spectral weight (away from the Fermi level), we give
up locality in expense of correctly capturing the spectral weight.

All projection schemes lead to slightly non-orthonormal correlated
Green's function. This is because the interstitial weight is not taken
into account and because the full potential basis is overcomplete.  To
have an orthonormal impurity problem, we compute the overlap
$\sum_{ii} P^2(ii,\tau LL')=O^\tau_{LL'}$ and renormalize $P^2(ij,\tau
LL')\rightarrow \sum_{L_1L_2}(\frac{1}{\sqrt{O}})_{LL_1} P^2(ij,\tau
L_1 L_2) (\frac{1}{\sqrt{O}})_{L_2 L'}$.


Finally, we remark that the segment of our code which builds
projections $P^0$, $P^1$ and $P^2$ within Wien2K~\cite{Wien2K} is
based on the \textit{qtl} package of Pavel Novak~\cite{NovakQTL}.

Similar projections within LDA+DMFT method were proposed before. In
particular the method by B. Amadon et.al.~\cite{Wannier2} proposed to
construct the Wannier functions for the correlated subset only, while
the DMFT equations were solved in the Kohn-Sham basis, restricted to
some subset of low energy bands. The local orbitals used for the
projection were either all-electron atomic partial waves in the PAW
framework, or pseudo-atomic wave functions in mixed-basis
pseudopotential code. Hence, in the language of projectors, the method
was similar to choosing the projector to be $P= |\chi_{\vk
  m}^\vR><\chi_{\vk m}^\vR|$, where $\chi_{\vk m}^\vR$ is the the
partial waves or pseudo-atomic wave function. While this method is
clearly causal, it looses spectral weight of the correlated angular
momentum character. Moreover, the implementation of the method did not
allow the self-consistent evaluation of the electronic charge.  The
method of Anisimov \textit{et.al.}~\cite{Anisimov2} also proposed a
construction of the Wannier functions using an arbitrary set of
localized orbitals. In their work, the LDA Hamiltonian was truncated
to Wannier representation for the purpose of obtaining the DMFT
self-energy. This simplifies the self-consistent DMFT problem, but
makes it impossible to implement the charge self-consistency. Finally,
Savrasov et.al.~\cite{Savrasov04} proposed a projector particular to
LMTO basis set, for which causality was not proven.

\section{DFT+DMFT Formalism}
\label{Sfunctional}

To derive the DFT+DMFT equations, we define a functional of the
correlated Green's function $\cG(\vr,\vr')$ and extremise it.
The correlated Green's function $\cG(\vr,\vr')$ is defined by
Eq.~(\ref{corrG}), and the functional to be extremise is
\begin{equation}
\Gamma[\cG,\rho] = -\Tr\ln(G^{-1}) - \Tr[\Sigma^{tot} G] + \Phi[\cG,\rho],
\label{functional}
\end{equation}
where $\Tr$ runs over all space (orbitals,momenta) and time
(frequency).
The quantities apprearing in the above functional are
\begin{widetext}
\begin{eqnarray}
&& G^{-1}_\omega(\vr,\vr') =
  \left[i\omega+\mu+\nabla^2-V_{ext}(\vr)\right]\delta(\vr-\vr')-\Sigma^{tot}_\omega(\vr,\vr')
  \label{G1}\\
&& \Sigma^{tot}_\omega(\vr,\vr') =
  \left[V_H(\vr)+V_{xc}(\vr)\right]\delta(\vr-\vr') +
  \left[\Sigma_\omega(\vr,\vr')-E_{DC}\delta(\vr-\vr')\right]\Theta(r<S)
  \label{S1}\\
&& \Phi[\cG,\rho] = \Phi_H[\rho] + \Phi_{xc}[\rho] + \Phi_{DMFT}[\cG] - \Phi_{DC}[\cG]
  \label{P1x}\\
&&  \rho = \widetilde{\Tr}[G]
\nonumber
\end{eqnarray}
\end{widetext}
where $\widetilde{\Tr}$ is trace over time only (not space), $V_{ext}$
is the potentials due to ions, $V_H, V_{XC}$ are the Hartree, and
exchange-correlation potential, respectively. $\Phi_{DMFT}[\cG]$ is
the sum of all local two particle irreducible skeleton diagrams
constructed from $\cG$, and the Coulomb repulsion $\hat{U}$ (screened by
orbitals not contained in $\cG$), and $\Phi_{DC}$ is the double
counting functional.

We assume that the Coulomb interaction $\hat{U}$ has the same form as
in the atom, i.e.,
\begin{widetext}
\begin{eqnarray}
&& \hat{U}=
\sum_{L_a,..L_d,,m,\sigma\sigma'}\sum_{k=0}^{2 l}\frac{4\pi F^{k}_{\{l\}} }{2 k+1}
\langle Y_{L_a}|Y_{k m}|Y_{L_c}\rangle
\langle Y_{L_b}|Y^*_{k m}|Y_{L_d}\rangle
f^\dagger_{L_a\sigma}
f^\dagger_{L_b\sigma'}f_{L_d\sigma'}f_{L_c\sigma}
\label{CoulombUU}
\end{eqnarray}
\end{widetext}
however, the Slater integrals are reduced due to screening
effects. Typically, we renormalize $F^2\cdots F^6$ by 30\%, from their
atomic values, while $F^0$, being renormalized more, can be estimate
by constraint LDA or constraint RPA \cite{Ferdi}.

To extremize the functional Eq.~(\ref{functional}), we take $\cG$
and $\rho$ as independent variables, and use the following functional
dependence:
$\Sigma[\cG]$, $\Phi_{DMFT}[\cG]$, $E_{DC}[\cG]$,
$\Phi_{DC}[\cG]$ are functionals of $\cG$. Consequently, $G$ is also a
functional of $\cG$, i.e., $G[\Sigma[\cG]]$. On the other hand,
$V_{H}[\rho]$, $V_{xc}[\rho]$, $\Phi_{H}[\rho]$, $\Phi_{xc}[\rho]$ are
functionals of the total electron density, hence $G$ is also a
functional of $\rho$ since $G[V_{H}[\rho]+V_{xc}[\rho]]$. Finally it
is easy to check that
$$\Tr[\Sigma^{tot} G]=\Tr[(V_{H}+V_{xc})\rho]+\Tr[(\Sigma-E_{DC})\cG].$$
With the above functional dependence in mind, minimization with
respect to $\cG$ gives
$$ \Sigma-E_{DC}=\frac{\delta\Phi_{DMFT}[\cG]}{\delta\cG} -\frac{\delta\Phi_{DC}[\cG]}{\delta\cG},$$
and minimization with respect to $\rho$ leads to
$$V_{H}+V_{xc} = \frac{\delta\Phi_{H}[\rho]}{\delta\rho}+\frac{\delta\Phi_{xc}[\rho]}{\delta\rho}.$$
Hence the Hartree and exchange-correlation potential are computed in
the same way as in DFT method (note however $\rho$ is electron density
in the presence of DMFT self-energy), while the DMFT self-energy is
the sum of all local Feynman diagrams, constructed from $\cG$ and
Coulomb interaction $\hat{U}$.

%
%

To sum up all local diagrams, constructed from $\cG$ and screened
Coulomb interaction $\hat{U}$, we solve an auxiliary quantum impurity
problem, which has $\cG=G_{imp}$ as the impurity green's function, and
$\Sigma$ as the impurity self-energy $\Sigma_{imp}=\Sigma$. The
impurity Green's function is
$G_{imp}=1/(i\omega-E_{imp}-\Sigma_{imp}-\Delta)$, hence the DMFT
self-consistency condition reads
\begin{equation}
\hat{P}(i\omega+\mu-H^{DFT}-\hat{E}\overline{\Sigma}))^{-1} = (i\omega-E_{imp}-\Sigma_{imp}-\Delta)^{-1}.
\end{equation}
where $\overline{\Sigma}\equiv\Sigma-E_{DC}$, and $E_{DC}$ is the
interaction included in DFT (double counting).
The self-consistency condition takes the explicit form
\begin{widetext}
\begin{eqnarray}
\int_{(r,r')<S_{\tau}}
d\vr d\vr' P(\vr\vr',\tau L L')
\left\{\left[i\omega+\mu+\nabla^2-V_{KS}(\vr)\right]\delta(\vr-\vr')
  - \sum_{L_1 L_2\in H}
  P(\vr'\vr,\tau L_1L_2)
  \overline{\Sigma}^{\tau}_{L_2  L_1}
  \right\}^{-1}
  \nonumber \\
=\left[\left(i\omega-E^{\tau}_{imp}-\Sigma^{\tau}-\Delta^{\tau}\right)^{-1}\right]_{LL'}
\label{SCC}
\end{eqnarray}
\end{widetext}
where $V_{KS}=V_{ext}+V_{H}+V_{xc}$ and $S$ is the muffin-tin radius.

For efficient evaluation of the DMFT self-consistency condition
Eq.~(\ref{SCC}), we choose to work in the Kohn-Sham (KS) basis. At
each DFT+DMFT iteration, we first solve the KS-eigenvalue problem
\begin{equation}
\left[-\nabla^2+V_{KS}(\vr)\right]\psi_{\vk i}(\vr) = \epsilon_{\vk i}\psi_{\vk i}.
\end{equation}
Then we express the projection $\hat{P}$ in KS basis, $P_k(ij,\tau LL')$,
where $i,j$ run over all bands.
%
We then perform the embedding of the self-energy, i.e., transforming
it from DMFT base to the KS base
\begin{eqnarray}
\overline{\Sigma}_{\vk, ij}(\omega)=\sum_{\tau,L_1 L_2}
P_{\vk\tau}(j i,\tau L_2 L_1)\;\overline{\Sigma}^{\tau}_{L_1 L_2}(\omega)
\label{Sigdef}
\end{eqnarray}
In KS-base, we can invert the Green's function Eq.~(\ref{SCC}), to
obtain the practical form of the self-consistency condition
\begin{eqnarray}
\cG^{\tau}_{LL'}&=&\sum_{\vk ij}
P_{\vk\tau}(ij,LL')\left[\left(i\omega+\mu-\epsilon_{\vk}-\overline{\Sigma}_{\vk}(\omega)\right)^{-1}\right]_{ji}
\label{Gdef}\\
\cG^{\tau}_{LL'}&=&\left[\frac{1}{i\omega-E^{\tau}_{imp}-\Sigma^{\tau}(\omega)-\Delta^{\tau}(\omega)}\right]_{LL'}
\label{SCCdef}
\end{eqnarray}
This is of course equivalent to Eq.~(\ref{SCC}).  Finally we solve
this self-consistency equation for a given self-energy
$\Sigma(\omega)$ to obtain the hybridization function $\Delta^{\tau}$
and the impurity
levels $E^{\tau}_{imp}$.

We note in passing that the self-energy $\Sigma(\omega)$ is a complex
function, and its imaginary part is related to the electron-electron
scattering rate, which is very large in correlated materials. In Mott
insulators, it is even diverging. Hence the DMFT "effective
Hamiltonian" $\epsilon_{\vk}+\overline{\Sigma}_{\vk}(\omega)$ can not
be diagonalized by standard methods to obatin a set of eigenvalues,
i.e., bands. The eigenvalues are complex and hence only the spectral
weight $A(\vk,\omega)=(G_\vk^\dagger(\omega)-G_\vk(\omega)/(2\pi i)$
is a well defined quantity.  The absence of well defined bands in
correlated materials makes computational techniques more
challenging. For example, the calculation of the chemical potential is
far more demanding because one can not assign a unity of charge to
each fully occupied band. Rather all complex eigenvalues, even those
which are far from the Fermi level, need to be carefully
considered. This point will be addressed below in
section~\ref{algorithm}, item 5. Further, the tetrahedron method
\cite{Tetra}, a very useful technique to reduce the number of
necessary momentum points in practical calculation, is not applicable
since it needs real eigenvalues.  We address the necessary
generalization of this method is chapter~\ref{complex-tetrahedra}.

Note that generalization of the projector and the LDA+DMFT formalism
to cluster-DMFT is very straightforward. One needs to increase the
unit cell to include more sites of the same atom type. The self-energy
and the Green's function become matrices in index $\tau$, i.e.,
$\Sigma^{\tau\tau'}_{LL'}$, $\cG^{\tau\tau'}_{LL'}$. The
transformation $\hat{P}$ is also straightforwardly generalized to
matrix form $P_{\vk}(ij;\tau L \tau' L')$. The only difference in the
definition of the projector Eq.~(\ref{P2P}) is that $\cA_{L'}^\tau$ is
replaced by $\cA_{L'}^{\tau'}$ ($\cA_L^\tau$ remains unchanged), which
amounts to the integral over two different spheres around two atoms of
the same type.
Finally, in cluster-DMFT case, the self-energy in KS-basis
Eq.~(\ref{Sigdef}) has to be summed over both $\tau$ and $\tau'$, and
self-consistency condition Eq.~(\ref{SCCdef}) becomes a matrix
equation in $\tau,\tau'$. The challenging part of the cluster-DMFT
formalism is in solving the cluster-impurity problem. In combination
with impurity solvers based on the hybridization expansion (discussed
below) the computational effort grows exponentially with the number of
correlated sites. In the weak coupling impurity solvers, the
computational effort grows as a power-law, however, these techniques
usually can not reach the interesting regime of strong correlations
and low temperatures.

The major bottleneck in evaluating the DMFT self-consistency condition
in our method is the multiplication of the projector
$P_{\vk\tau}(ij,LL')$ with $\Sigma$ in Eq.~(\ref{Sigdef}) and
multiplication of projection with Green's function $G_{\vk,ji}$ in
Eq.~(\ref{Gdef}). Since projection $P^2$ is separable, one can write
the operation in terms of matrix products. Still, these sums run over
all $\vk$-points (typically few thousands) and all frequency points
(typically few hundreds).

For the efficient implementation of the set of Eqs.~(\ref{Sigdef}) and
(\ref{Gdef}), we first notice that the transformation $P$ (or its
separable part $U$) is very large and is not desirable to be written
to the computer hard disc.  Hence we generate it only for one $\vk$-point
at a time, and evaluate both products at this particular
$\vk$-point. Non-negligible amount of time is necessary to generate
the transformation Eq.~(\ref{P2P}), and because this transformation
does not depend on frequency, it needs to be used for all frequencies
in Eqs.~(\ref{Sigdef}) and (\ref{Gdef}).  Hence paralization over
frequency is not implemented, while paralization over $\vk$-points is.

Note that because of the sum over atoms ($\tau$) in
Eq.~(\ref{Sigdef}), the transformation for all atoms needs to be
computed first, and only then the sum in Eq.~(\ref{Sigdef}) can be
evaluated and the self-consistency condition Eq.~(\ref{SCCdef}) can be
inverted.

To optimize the sum in Eqs.~(\ref{Sigdef}) and (\ref{Gdef}), one can
notice that local quantities like self-energy and local green's
function possess a large degree of symmetry when written in proper basis
(real harmonics, relativistic harmonics): many off-diagonal matrix
elements vanish, and many matrix elements are equivalent.  For
example, in a $d$ system with cubic symmetry, one has only two types of
self-energy $t_{2g}$ and $e_g$. Hence, instead of summing over $10\times
10$ matrix elements in Eq.~(\ref{Sigdef}), one can rewrite the sum
over two matrix elements $t=(0,1)$, i.e.,
\begin{eqnarray}
\Sigma_{\vk, ij}(\omega)=\sum_{\tau,t}\Sigma^{(\tau)}_{t}(\omega)P_{\vk\tau}(j i,t)
\label{Sigdef_optimize}
\end{eqnarray}
where 
$P_{\vk\tau}(j i, t) =\sum_{\Sigma(L_1,L_2)=\Sigma(t)}P_{\vk\tau}(ji,L_2L_1)$
and the indices $L_1,L_2$ here stand for the real harmonics rather
than spheric harmonics.
The later transformation is independent of frequency, while the sum
Eq.~(\ref{Sigdef_optimize}) needs to be performed for all frequencies,
hence the compact form of the transformation saves a lot of computer
time.


\section{The algorithm}
\label{algorithm}

The implementation of the DFT+DMFT algorithm is done in the following
few steps:
\begin{itemize}
\item[1)] $\rho(\vr)$: We converge the LDA/GGA equations to get
  the starting electronic charge $\rho(\vr)$. We use the non-spin
  polarized solution as starting point. In the ordered state, the DMFT
  self-energy is allowed to break the symmetry, while typically the
  exchange-correlation potential is not allowed to break the symmetry
  (LDA rather than LSDA).
  
  In this preparation step we also obtain good estimates for the
  Coulomb repulsion $U$ (which is represented by Slater integrals
  $F^0$, $F^2$, $F^4$ and $F^6$). Slater integrals are computed by
  the atomic physics program of Ref.~\onlinecite{Cowan}, and they are
  scaled down by 30\% to account for the screening in the solid. The
  $F^0$ terms is very different from the atomic $F^0$ and is obtained
  by constraint LDA calculation, or constraint RPA calculation
  \cite{Ferdi}.

\item[2)] $\psi_{\vk i}(\vr)$: We solve the DFT KS-eigenvalue problem
  $$ (-\nabla^2 + V_{KS}(\vr))\psi_{\vk i}(\vr) =  \psi_{\vk i}(\vr)\varepsilon_{i\vk}^{DFT}$$
  to obtaine KS eigenvectors, core, and semicore charge, and
  linearization energies $E_{\nu}$.
  
\item[3)] $\overline{\Sigma}_{LL'}:$ We start with a guess for the
  lattice self-energy correction
  $\overline{\Sigma}(\omega) =
  \widetilde{\Sigma}(\omega)+\Sigma_{\infty}-E_{dc}$ (here
  $\widetilde{\Sigma}$ is the dynamic part of the self-energy with the
  property $\widetilde{\Sigma}(\infty)=0$).  A reasonable starting
  point is $\widetilde{\Sigma}(\omega)=0$ and
  $E_{dc}=\langle\Sigma_{\infty}\rangle$. The potential in the first DMFT iteration is
  thus the DFT potential.

\item[4)] $\overline{\Sigma}_{\vk, ij}$: Next we embed the DMFT
  self-energy $\overline{\Sigma}_{LL'}^{(\tau)}(\omega)$ (shifted by
  double counting)
  to Kohn-Sham base by the
  transformation Eq.~(\ref{Sigdef}) to obtain $\overline{\Sigma}_{\vk,ij}(\omega)$.
  
\item[5)] $\mu$: Using the current DMFT self-energy $\overline{\Sigma}(\omega)$, and the
  current DFT KS-potential $V_{KS}$, we compute the current chemical
  potential. This is done in the followin steps:
  \begin{itemize}
  \item Complex eigenvalues $\varepsilon_{\vk l}(\omega)$ of the full
    Green's function are found in the large enough energy interval (at
    least [$-2U$,$2U$]) by solving
    $$\sum_j [\varepsilon_{\vk i}^{DFT}\delta_{ij}+\overline{\Sigma}_{\vk ij}(\omega)]C_{jl}^\vk(\omega)=C^\vk_{il}(\omega)\varepsilon_{\vk l}(\omega).$$
    Here $C_{ji}$ are DMFT eigenvectors expressed in KS base.
    The DMFT eigenvalues outside this interval are set to DFT
    eigenvalues. We need only eigenvalues in this step, but not eigenvectors.
    
  \item The chemical potential is determined using precomputed complex
    and frequency dependent eigenvalues $\varepsilon_{\vk l,\omega}$.
    On imaginary axis we solve
    $$N_{val}=T\sum_{\vk l, \omega_n}\frac{1}{i\omega_n+\mu-\varepsilon_{\vk l}(i\omega_n)}$$
    and on real axis we solve
    $$N_{val}=-\frac{1}{\pi}\Im\sum_{\vk l} \int\frac{f(\omega)d\omega}{\omega+\mu-\varepsilon_{\vk l}(\omega)}$$
    If enough $\vk$-points can be afforded, we use special point
    method, otherwise the ``complex tetrahedron method'' can be used
    (see chapter~\ref{complex-tetrahedra}).
    
    For numerical evaluation of the real axis density, we discretize
    the integral
    $$N_{val}=-\frac{1}{\pi}\Im\sum_{i} f(\omega_i) \sum_{\vk l} \int_{a_i}^{b_i} \frac{d\omega}{\omega+\mu-\varepsilon_{\vk l}(\omega_i)}$$
    with $a_i=(\omega_i+\omega_{i-1})/2$ and
    $b_i=(\omega_{i+1}+\omega_{i})/2$.
    When using the special point method, the integral over frequency
    is evaluated analytically, and the terms of the form
    $\log({a_i+\mu-\varepsilon_{\vk l}(\omega_i)})$ are summed
    up. Alteratively, we sometimes use the complex tetrahedron method,
    where the four-dimensional integral is evaluated analytically
    (see chapter~\ref{complex-tetrahedra})
    
    When DMFT is done on imaginary axis (using imaginary time
    impurity solvers), we evaluate
    \begin{eqnarray}
    &&N=
    \sum_{\vk l}f(\varepsilon_{\vk l}^0-\mu)+\nonumber\\
    &&2T\sum_{0<\omega_n<\omega_N}
    \sum_{\vk l}[
      \frac{1}{i\omega_n+\mu-\varepsilon_{\vk l}(i\omega_n)}-
      \frac{1}{i\omega_n+\mu-\varepsilon_{\vk l}^0}]\nonumber\\
     &&-\frac{1}{\pi}\arctan\left(\frac{\varepsilon_{\vk l}^{\infty}-\mu}{\omega_N}\right)+
     \frac{1}{\pi}\arctan\left(\frac{\varepsilon_{\vk l}^0-\mu}{\omega_N}\right)
    \end{eqnarray}
    Here $\varepsilon_{\vk l}^0$ is the real part of the eigenvalue at
    arbitrary frequency. We choose the lowest or the last Mastubara point.
    Again, the tetrahedron method can be used for momentum sum.
  \end{itemize}
  For Mott insulators, the above described method is not very
  efficient, because even a small numerical error in computing
  $N_{val}$ places chemical potential at the edge of the Hubbard
  band, either upper or lower. This instability usually does not allow
  one to reach a stable self-consistent solution. We devised the
  following method to remove this instability:
  \begin{itemize}
    \item The diagonal components of the self-energy were fitted
      by a pole-like expression
      $\overline{\Sigma}_{L L'}=\overline{\Sigma}_{\infty} + \frac{W_{L}}{i\omega-P_L+i\Gamma_L}$.
    \item Next, we neglected broadening of the pole ($\Gamma_L$), which
      should be small in the Mott insulating state. We computed a
      quasiparticle approximation for the 
      Green's function $G_\vk^{qp}$, i.e.,
      \begin{equation}
        (G^{qp}_\vk)^{-1}_{ij}=i\omega-\varepsilon_{\vk i}^{DFT}-\overline{\Sigma}_{\infty,ij}-U^{\vk\tau}_{iL}\sqrt{W_L}\frac{1}{i\omega-P_L} \sqrt{W_L}U^{\vk\tau *}_{jL}
      \end{equation}
      where $U^{\vk\tau}_{iL}$ is part of the projector $P^\vk
      (ij,\tau LL')=U^{\vk\tau}_{iL} U^{\vk\tau *}_{jL'}$ defined above.
    \item  The above Green's function formulae can be cast into a block form
      \begin{equation}
        G^{qp}_{\vk}=\left[i\omega-\left(\begin{array}{cc}
         \varepsilon_{\vk}^{DFT}+\overline{\Sigma}_{\infty} &
         U^{\vk\tau}\sqrt{W}\\
         \sqrt{W}U^{\vk\tau\dagger} & P
       \end{array}\right)\right]^{-1}\equiv \left(i\omega-H_{\vk}^{qp}\right).
      \end{equation}
      Here $H^{qp}_\vk$ is the quasiparticle Hamiltonian which can be
      diagonalized to obtain the quasiparticle bands. We notice that
      the number of quasiparticle bands of the Mott insulator is
      larger then the number of Kohn-Sham bands because Mott
      insulators have at least two Hubbard bands.
      The quasiparticle
      bands are not very accurate away from the Fermi level, however
      they are sufficiently acurate at low energy and allow one to
      identify gaps at the Fermi level. Once a gap in the spectra of
      $H^{qp}_\vk$ is identified, the charge is computed using the full
      DMFT density matrix to verify the neutrality of the solid. If
      the solid is neutral when chemical potential is in the gap, the
      chemical potential is set to the middle of the gap.
  \end{itemize}
\item[6)] $\Delta$: Impurity hybridization function $\Delta(\omega)$ and
  impurity levels $E_{imp}$ are computed in this step.
  
  We use equation (\ref{Gdef}) to get $\cG_{LL'}^{(\tau)}$ and we use
  the high frequency expansion of both equations
  (\ref{Gdef}) and
  (\ref{SCCdef})  to determin impurity levels
  $${E_{imp}}_{LL'}=-E_{DC}\delta_{LL'}+\sum_{\vk i}P_{\vk\tau}(ii,LL')\varepsilon_{\vk i}^{DFT}$$
   
\item[7)] $\Sigma_{imp}$: Impurity solver uses $\Delta_{LL'}(\omega)$, $E_{imp}$, and
  Coulomb repulsion $U$ (which is represented by Slater integrals
  $F^0$, $F^2$, $F^4$ and $F^6$) as the input and gives the new
  self-energy $\Sigma_{LL'}(\omega)$ as the output.

  Currently we integrated the following impurity solvers: OCA (see
  chapter \ref{OCA}), Non-crossing approximation (NCA), Continuous
  time quantum Monte Carlo (CTQMC)~\cite{CTQMC}. The latter is
  implemented on imaginary axis, and the former two on real axis.
  
  Before the impurity solver is run, we exactly diagonalize the atomic
  problem in the presence of crystal fields, to obtain all atomic
  energies $E_m$ and the matrix elements of electron creation operator
  in the atomic basis $\langle m|f^{\dagger\alpha}|n\rangle$.  Since
  the impurity levels can change during the iteration, the crystal
  field of the atomic problem can change as well. In case of
  $f$-systems, the crystal field splittings are small and one can
  assume that they do not change substantially from their DFT
  value. Hence the exact diagonalization can be done only once at the
  beginning. For the $d$-systems, the crystal field splittings are
  larger, and this approximation is in general not necessary
  satisfactory, hence the exact diagonalization needs to be repeated
  in the charge self-consistent cycle. A special care needs to be
  taken here when using CTQMC. To speed up the convergence of CTQMC
  solver, we typically start simulation with the status of the kink
  distribution from previous DMFT step. Since exact diagonalization
  can reorder eigenstates, these kinks need to be properly renumbered,
  to efficiently restart simulation.
  
\item[8)] $\Sigma_{\infty}$: It is very hard to achieve reasonably
  precise self-energy at high frequency with impurity solvers based on
  hybridization expansion. 
  However, to correctly compute electronic
  charge, it is crucial that the self-energy at high frequency
  approaches its Hartree-Fock value and the impurity Green's function
  and self-energy at large frequency properly behave. Hence we correct
  $\Sigma_\infty$ at each iteration.
  This is quite straighforward,
  given the fact that impurity solvers determine the impurity density
  very precisely. This steps only corrects the high energy tails of
  the impurity green's function and impurity self-energy, while we make sure
  that the low energy part, which is computed very precisely by these
  methods, is not altered.

  In the case of CTQMC solver, we compute the atomic Green's function
  using CTQMC probabilities for each atomic state (see
  Ref.~\onlinecite{CTQMC} for details). The high-frequency tails of
  the self-energy can then be computed. These analytic tails are then
  used instead of noisy QMC data.

  In OCA and NCA impurity solvers, we project out very high excited
  atomic states. This has negligible effect on the low energy physics,
  however, it results in a missing weight at high frequency, and hence
  wrong self-energy at infinity. To correct for this deficiency, we
  add two lorentzians to the impurity Green's function
  $$\cG(\omega) = \int\frac{A(x)dx}{\omega-x} + \frac{a_1}{\omega-\eps_1+i\Gamma}  + \frac{a_2}{\omega-\eps_2+i\Gamma}$$
  typically with $\eps_1 < -U$ and $\eps_2>U$.
  Here we omitted the subscript $LL'$ for the impurity Green's
  function $\cG_{LL'}$ for clarity.
  The parameters $a_1, a_2, \eps_1, \eps_2$ are determined by the
  following constraints:
  \begin{itemize}
  \item \textit{normalization}:    $m_0 + a_1 + a_2 = 1$, where $m_0$ is the
    integral of $A(x)$.
  \item \textit{density}: $n + a_1 = n_{exact}$, where $n=\int A(x)f(x) dx$ and
    $n_{exact}$ is the impurity density determined by the impurity
    solver in an alternative, more precise way (from pseudo-particle
    density).
  \item $\Sigma_\infty$: $m_1 + a_1 \eps_1 + a_2 \eps_2 = E_{imp} + \Sigma_\infty$,
    where $m_1$ is the first moment $m_1=\int x A(x) dx$.
  \end{itemize}
  Once the following three constrains are satisfied, the self-energy
  at high frequency approaches its Hartree-Fock value, and the spectral
  function respects the total impurity density.
  
\item[9)] $\overline{\Sigma}$:  Using the new impurity self-energy, we
  determine the new lattice self-energy
  $\overline{\Sigma}(\omega) = \widetilde{\Sigma}(\omega)+\Sigma_{\infty}-E_{DC}$, where
  $E_{DC}= U(n-1/2) - J(n/2-1/2)$, with $n$ the correlated nominal occupancy. 
  
\item[10)] \textit{goto 4}: If the convergence of charge is hard to
  achieve, we iterate the DMFT loop a few times. We call this loop the
  DMFT loop. If the DMFT loop is to be iterated, jump to 4.

\item[11)] $\mu,\rho(\vr)$: The eigevalue problem is solved for all
  momentum and frequency points,
  $$\sum_j [\varepsilon_{\vk
      i}^{DFT}\delta_{ij}+\overline{\Sigma}_{\vk
      ij}(\omega)]C^{\omega,R}_{\vk jl}=C^{\omega,R}_{\vk il}\varepsilon_{\vk
    l\omega}.$$ Here we evaluate both, eigenvalues and eigenvectors.
  Since this is a non-hermitian eigenvalue problem, the left and right
  eigenvectors are not complex conjugates of each other. We use notation
  $C^{\omega R}_{\vk il}$  for the right and 
  $C^{\omega L}_{\vk il}$  for the left eigenvector.
  
  Using the DMFT eigenvalues, we recompute the chemical potential as in 5.

  We then recompute the electronic charge from the DMFT eigenvectors
  $$\psi_{\vk l\omega}(\vr) = \sum_i \psi_{\vk i}(\vr) C^\omega_{\vk i l}$$
  where $\psi_{\vk i}$ are Kohn-Sham eigenvectors (solutions of the
  LDA eigenvalue problem).
  The electronic valence charge on real axis is
  $$\rho_{val}(\vr) = -\frac{1}{\pi}\Im\sum_{\vk l}\int \psi^R_{\vk l\omega}(\vr) \frac{f(\omega)d\omega}{\omega+\mu-\varepsilon_{\vk l\omega}}\psi^L_{\vk l\omega}(\vr)$$
  and on imaginary axis is
  $$\rho_{val}(\vr) = T\sum_{\vk l,\omega_n} \psi^R_{\vk l\omega_n}(\vr) \frac{1}{i\omega_n+\mu-\varepsilon_{\vk l\omega_n}}\psi^L_{\vk l\omega_n}(\vr).$$

  We compute the electronic charge using similar technique as used
  above to compute the chemical potential.
  The electronic charge is
  $$\rho_{val}(\vr) = \sum_{\vk ij} \psi_{\vk i}(\vr)\psi^*_{\vk j}(\vr) W^{DMFT}_{\vk ij}.$$
  The weights $W^{DMFT}_{\vk, ij}$ on real axis are combuted as
  $$W^{DMFT}_{\vk, i j} = \sum_{l p} C^{\omega_p R}_{\vk i l}C^{\omega_p L}_{\vk j l} w_{\vk l p}$$
  with
  $$w_{\vk l p} = -\frac{1}{\pi}f(\omega_p)\Im\int_{a_p}^{b_p}d\omega\frac{1}{\omega+\mu-\varepsilon_{\vk l \omega_p}}$$
  and $a_p=(\omega_p+\omega_{p-1})/2$,
  $b_p=(\omega_{p+1}+\omega_{p})/2$.

  On imaginary axis we evaluate the weights by the following expression
  \begin{eqnarray}
    W^{DMFT}_{\vk, ij}&=& T \sum_{\omega_n,l}
    \left(\frac{C^{\omega_n R}_{\vk i l}C^{\omega_n L}_{\vk j l}}{i\omega_n+\mu-\varepsilon_{\vk l\omega_n}}-
    \frac{C^{\omega_0}_{\vk i l}C^{\omega_0 *}_{\vk j l}}{i\omega_n+\mu-\varepsilon_{\vk l\omega_0}}\right)\nonumber\\
    &+& \sum_l C^{\omega_0}_{\vk i l}C^{\omega_0*}_{\vk j l}\;f(\varepsilon_{\vk l\omega_0}-\mu)\nonumber
  \end{eqnarray}

  Note that the DMFT density matrix $W^{DMFT}_{\vk, i j}$ is a hermitian matrix in Kohn-Sham band
  indeces $i$ and $j$. Hence, we can use eigenvalue techniques for
  hermitian matrices to decompose $W$ into
  $$ W^{DMFT}_{\vk, ij} = \sum_l U_{\vk, i l}w_{\vk,l}U^*_{\vk,j l}. $$
  The LDA+DMFT electronic charge can then be evaluated
  by rotated Kohn-Sham vectors, and DMFT weights $w_{\vk,l}$ by
  $$\rho_{val}(\vr)=\sum_{\vk,l} \left[\sum_i U_{\vk,il} \psi_{\vk i}(\vr)\right] w_{\vk,l}\left[\sum_j \psi^*_{\vk j}(\vr)U^*_{\vk,jl}\right].$$
  Hence, the code to compute the LDA charge can be simply converted to
  compute the DMFT charge by just replacing the Kohn-Sham LDA weight
  by DMFT weight $w_{\vk,l}$, and by rotating the Kohn-Sham
  eigenvectors by
  the above computed eigenvectors $U_\vk$.

  Finally, the DFT core and DFT semicore charge is added to the
  valence charge, and the resulting total charge is renormalized in
  the standard way, such that the charge neutrality is satisfied to
  high accuracy.
  
\item[12)] $E_{tot}$:The total energy is computed on the output
  density $\rho(\vr)$, using the low temperature limit of the
  functional Eq.~(\ref{functional}) evaluated on the DFT+DMFT
  solution:
\begin{equation}
E_{total} = \Tr[(-\nabla^2+V_{ext})G] + \frac{1}{2}\Tr[\Sigma G] +
E_{H}+E_{xc}-\Phi_{DC}
\nonumber
\end{equation}
For computation, the formula is cast into the following form
\begin{eqnarray}
E_{total} &=& \Tr[(-\nabla^2+V_{KS})G] -
\int(V_{H}(\vr)+V_{xc}(\vr))\rho(\vr)d\vr\nonumber
\\
&+& E_{H}+E_{xc}+\frac{1}{2}\Tr[\Sigma G]-\Phi_{DC}
\nonumber
\end{eqnarray}
and evaluated by
\begin{eqnarray}
E_{tot} &=&  \sum_{i}\varepsilon_{\vk i}^{DFT} {W_{\vk,ii}^{DMFT}}
-\int(V_{H}(\vr)+V_{xc}(\vr))\rho(\vr)d\vr\nonumber\\
&+&E_H+E_{xc}+E_{potential}^{imp}-\Phi_{DC}\nonumber
\end{eqnarray}
where $W_{\vk}^{DMFT}$ is the DMFT density matrix defined above, and
\begin{equation}
E_{potential}^{imp}=\frac{1}{2}T\sum_{\omega_n,\tau LL'}\Sigma^{(\tau)}_{LL'}(\omega_n)\cG^{(\tau)}_{L'L}(\omega_n)
\end{equation}
is the impurity potential energy, which can be computed very precisely
by most impurity solvers, such as CTQMC or OCA. For example, in CTQMC
we sample probability for each atomis state $P_m$. Using these
probabilities, we can evaluate $E_{potential}^{imp}=\sum_m P_m
E_m^{atom}-\sum_{L L'}E^{imp}_{L L'}n^{imp}_{L' L}$.

\item[13)] \textit{mix}: The total electronic charge is mixed with the
  charge from previous iterations using multi-secant mixing of Marks
  and Luke \cite{Mixing}.
  
\item[14)] \textit{DFT}: In this step, we recompute the DFT potential
  (hartree, exchange-correlation potential), the Kohn-Sham orbitals
  and linearization energies.

\item[15)] \textit{goto 11}: If the self consistency is hard
  to achieve, jump to 11 and determine the best electronic charge
  $\rho(\vr)$ on the current impurity self-energy $\Sigma$. We call
  this loop the LDA loop.
  
\item[16)] \textit{goto 6} If the electronic charge and self-energy
  are not converged, jump to 6. We call this loop the charge loop.
  
\end{itemize}

\section{Complex tetrahedron method}
\label{complex-tetrahedra}

The calculation of the electronic density, as well as the correlated Green's
function, requires precise evaluation of integrals, which contain diverging
poles. In systems with many atoms per unit cell, one can not afford
enough $\vk$-points to get hybridization function $\Delta(\omega)$
smooth on a scale of temperature $T$ without introducing artifical
broadening larger than $T$. Hence, to avoid artifical broadening
larger than the low energy scale, we need to use alternative summation
over momentum.  The tetrahedron method \cite{Tetra} is used in
this case.  In the context of DFT+DMFT, an aditional complication is
that the eigenvalues are complex numbers. Although the analytic
formulas for the integration over a tetrahedron can straighforwardly
be evaluated, and are given in appendix A, a more severe problem is
the interpolation of the multidimensional complex functions $\epsilon_{i \vk}$
in momentum space. Below we give details on a method to overcome this
difficulty.

Computation of the Green's function requires the evaluation of the
following integral
$$g = \sum_\vk \frac{C_{i\vk}}{\omega-\epsilon_{i\vk\omega}},$$
which can be rewriten
as
$$g = \sum_t \int_t d^3k \frac{C_{i\vk}}{\omega-\epsilon_{i\vk\omega}},$$
where the sum runs over all
tetrahedrons $t$, and integral needs to be performed over the particular
tetrahedron $t$. $i$ is the band index. The linear interpolation of
$C_{i\vk}$ and linear interpolation of $\epsilon_{i\vk\omega}$ in
momentum space leads to analytic formulas for the weight functions
$w(i,\vk,\omega)$ (given in appendix A), which can be used to evaluate
$g$ to higher precision by $g=\sum_\vk w(i,\vk,\omega) C_{i\vk}$.

Similarly, the electron density is computed by
$$N_{val}=\sum_{i\vk}\int \frac{d\omega f(\omega)}{\omega+\mu-\epsilon_{i\vk\omega}}.$$
We take a frequency mesh, which is sufficiently dense at zero freqeuncy that it
can resolve the fermi function $f(\omega)$, and we approximate
\begin{eqnarray}
N_{val}&=&-\frac{1}{\pi}\Im \sum_{t,i,j} f(\omega_j)\int_t d^3k \int_{(\omega_{j}+\omega_{j-1})/2}^{(\omega_{j+1}+\omega_{j})/2} \frac{d\omega}{\omega+\mu-\epsilon_{i\vk\omega}}\nonumber\\
&=&-\frac{1}{\pi}\Im \sum_{\vk,i,j} f(\omega_j) wi_i(\vk,\frac{\omega_{j+1}+\omega_{j}}{2},\frac{\omega_{j}+\omega_{j-1}}{2})
\end{eqnarray}
Here the integral $\int_t$ is the integral over a particular
tetrahedron $t$.
The weights can again be computed analytically and are give in
Appendix A.

To evaluate the integral over a tetrahedron $t$, which has corners in
momentum points $k_1, k_2, k_3, k_4$, we need to interpolate the
eigenvalues $\epsilon_{i_1 k_1}, \epsilon_{i_2 k_2}, \epsilon_{i_3 k_3},
\epsilon_{i_4 k_4}$ inside the volume of the tetrahedron. Since there
are many crossing bands (index $i$), it is not at all simple to find a
good interpolation of $\epsilon_{i k}$ inside the tetrahedron.

In the standard tetrahedron method, where eigenvalues are real
numbers, one sorts the eigenvalues at each $k$-point, to get the
vector of increasing energies $\epsilon_{1,k},\epsilon_{2,k},\cdots$,
and then one linearly interpolates each sorted component of the vector
$\epsilon_{i,k_1},\epsilon_{i,k_2},\epsilon_{i,k_3},\epsilon_{i,k_4}$
inside the tetrahedron. Hence all crossings are avoided. It is however
important that no artifical crossings are obtained in the
interpolation, because a crossing gives a diverging contribution to
the integral.

Complex eigenvalues, which appear in DFT+DMFT, can not be
sorted. Hence the interpolation is not at all simple. A reasonable
attempt would be to sort eigenvalues according to their real parts,
and just neglect their imaginary parts when sorting. It turns out that
in strongly correlated regime, where the self-energy becomes very
large at some frequency points, the error in tetrahedron method is so
large that the hybridization function can become non-causal in such
points. Due to this non-adequate interpolation, the Green's function
has a lot of noise, superimposed on a smooth curve. However,
hybridization function, which is many times more sensitive than the
Green's function, has unbearable large error, which cause enormous
error in the solution of the impurity problem.

To overcome this problem, we implemented a special type of smooth
interpolation, based on the idea that the absolute value of the energy
should not change much from one k-point to its neighboring k-point. For each
tetrahedron, we minimize the following functional
\begin{eqnarray}
\sum_i \sum_{(\alpha,\beta)\in pairs} |\epsilon_{i,k_\alpha}-\epsilon_{i,k_\beta}|^2=min
\end{eqnarray}
where the 6 $pairs$ of the tetrahedron corners are:
$(1,2),(1,3)\cdots(3,4)$, and $i$ runs over all bands. We minimize the
functional with respect to the order of eigenvalues in all corners
of the tetrahedra.

To minimize the above functional, we can choose an arbitrary order of
bands in the first $k$-point $k_1$, and then we have to permute the
components of the other three $k$-points ($k_2$,$k_3$,$k_4$).
Hence the number of all possible trial steps is $(n!)^3$, where $n$ is
the number of bands, and is typically of the order of few
hundred. Obviously, not all arrangements of the eigenvalues can be
tried. Our algorithm for sorting the eigenvalues is
\begin{itemize}
\item[1] Sort the eigenvalues according to their real parts.
\item[2] Use Metropolis Monte Carlo method (for $T=0$) to flip
  components of a vectors $\epsilon_{k,i} \Longleftrightarrow \epsilon_{k,j}$.
  Try to flip
  components in any of the momentum points $k_2,\cdots,k_4$.
\end{itemize}
The trial steps are chosen in such a way that the probability for
flipping two eigenvalues, which have very different real parts, is
very small. We typically choose an exponential distribution function
with probability $P(|i-j|)\propto \exp(|i-j|/5)$.

\section{Transport calculation using DFT+DMFT}
\label{transport}

In this section, we will give the efficient algorithm to compute the DC
conductivity within DFT+DMFT. The higher order transport coefficients
can be computed along the similar lines, although the computation
becomes more technically involved.

The DC-conductivity can in general be expressed by 
\begin{eqnarray}
\sigma^{\mu\nu} = \lim_{\omega\rightarrow 0}\frac{1}{\omega}\chi^{''}_{\mu\nu}(\omega+i\delta)
\label{opt3}
\end{eqnarray}
where the current-current correlation function $\chi$ is expressed
diagrammatically through the electron Green's functions and the
current vertex function by
\begin{widetext}
\begin{eqnarray}
\chi_{\mu\nu}(i\omega_n)=-T \sum_{\vk\sigma\nu_m,p_1,p_2,p_1',p_2'}
v^{\vk \mu}_{p_1 p_2}
G^{p_1' p_1}_{\vk\sigma}(i\nu_m)G^{p_2 p_2'}_{\vk}(i\nu_m-i\omega_n)
\Gamma^{\sigma\nu}_{p_2' p_1'}(\vk\nu_m,\omega_n).
\label{opt_chi}
\end{eqnarray}
Here $\Gamma(\vk\nu_m,\omega_n)$ is the current vertex function,
which satisfies the integral equation
\begin{eqnarray}
\Gamma^{\sigma\nu}_{p_2' p_1'}(\vk\nu_m,\omega_n) = v^{\vk\nu}_{p_2' p_1'} - 
T\sum_{\vk'\sigma'\nu_m',p_3',p_4',p_3,p_4}
I^{\sigma\sigma'}_{p_1' p_2' p_3' p_4'}(\vk\nu_m,\vk'\nu_m';\omega_n)
G^{ p_4 p_4'}_{\vk'\sigma'}(i\nu_m')G^{ p_3' p_3}_{\vk' \sigma'}(i\nu_m'-i\omega_n)
\Gamma^{\sigma'\nu}_{p_3 p_4}(\vk'\nu_m',\omega_n)
\label{vertexFunction}
\end{eqnarray}
\end{widetext}
and $I(\vk\nu_m,\vk'\nu_m';\omega_n)$ is the particle hole irreducible
vertex, whose limit at zero frequency and Fermi momenta is
the Landau interaction function. $v^{\vk\nu}$ are velocities, given by
$$v^{\vk\nu}_{p_1 p_2} = -\frac{i e}{2 m}\langle \psi_{\vk p_1}|\nabla_\nu| \psi_{\vk p_2}\rangle.$$
All quantities are expressed in a Bloch-basis, for example the
Kohn-Sham basis, which diagonalizes the static part of the action.

In general, the two particle vertex function is very difficult to
compute. In some cases, the vertex corrections vanish and the
transport quantities can be computed from the lowest order bubble
diagram.

If self-energy is momentum independent, and the single band
approximation is appropriate, the vertex correction vanish, as shown
by Khurana \cite{Khurana}. In
multiband system, the following set of conditions are sufficient for
the vertex correction to vanish:
\begin{itemize}
\item[1)] The irreducible vertex function is local, i.e.,
$I(\vk\nu,\vk'\nu';\omega_n)$ does not depend on $\vk$ or $\vk'$.
\item[2)] Velocities are odd functions of momentum, i.e., $v^{-\vk}=-v^{\vk}$
\item[3)] Green's function is even functions of momentum, i.e., $G_{-\vk}=G_{\vk}$.
\end{itemize}
Under the above conditions, it is clear from
Eq.~(\ref{vertexFunction}) that only the zeroth order term remains and
vertex is unrenormalized $\Gamma(\vk)=v^\vk$.  Consider the first
order term $\sum_{\vk'} I \; G_{\vk'} G_{\vk'}v^{\vk'}$ in
Eq.~(\ref{vertexFunction}) or the second order term $\sum_{\vk'\vk''}
I \; G_{\vk'} G_{\vk'} I\; G_{\vk''} G_{\vk''} v^{\vk''}$ in
Eq.~(\ref{vertexFunction}). The function being summed is odd in $\vk'$
and $\vk''$, respectively, and hence the terms vanish.

Under which circumstances the above three conditions are met?  The
first condition is exact in the limit of infinite dimensions. Thus in
Dynamical Mean Field Theory, the irreducible vertex is local.  For
many three dimensional systems, it is believed to be an excellent
approximation. However, the velocities are not necessary odd functions
of momentum, in particular, they are obviously nonzero in strict atomic
limit, thus violating the condition (2).
Finally, the third condition is obviously satisfied in single band
theories with local self-energy, where
$G_{\vk}(\omega)=1/(\omega+\mu-\epsilon_\vk-\Sigma(\omega))$ because
$\epsilon_{-\vk}=\epsilon_{\vk}$.  In Dynamical Mean Field Theory the
self-energy operator is approximated by a purely local quantity.
However, the local approximation is made in a localized basis. The
self-energy in the Kohn-Sham basis is given by Eq.~(\ref{Sigdef}), and
is obviously momentum dependent.  In general case, the resulting
self-energy $\varepsilon_\vk+\hat{P}_{\vk\tau}\Sigma$ is not an even
function of momentum, and hence $G_{-\vk}\ne G_\vk$.
%
%

Due to difficulties in computing the two particle vertex function to
high accuracy on real axis, the vast majority of theoretical
calculations ignore the vertex corrections to conductivity. At present
it is not clear how important the vertex corrections to optical
conductivity and transport are in correlated electron materials.  They
are likely small because they vanish at low energy, where an effective
single band approximation is possible. And they are also small at
intermediate energies where the interband transitions give major
contribution to optical conductivity. However, a thorough
investigation of the vertex corrections and consequently appearance of
excitons in correlated materials is a very interesting avenue for
future research.

In the absence of vertex corrections, the current-current corelation
function Eq.~(\ref{opt_chi}) becomes
\begin{eqnarray}
\Im{\chi_{\mu\nu}(\omega)}=\frac{\pi e_0^2}{V_0}\sum_\vk \int
dy[f(y-\omega)-f(y)]\times
\nonumber\\
\Tr\left(\rho_{\vk}(y)v^{\vk\mu}\rho_\vk(y-\omega)v^{\vk\nu}\right)
\label{chin}
\end{eqnarray}
where ${\rho_\vk}=(G^\dagger-G)/(2\pi i)$.  Both spectral density
$\rho_\vk$ and velocity $v_\vk$ are matrices in orbital indices and
trace is taken over the orbitals and spins in Eq.~(\ref{chin}).
Finally, the real part of the DC conductivity is given by
\begin{eqnarray}
{\sigma'}^{\mu\nu}=\frac{\pi e_0^2}{V_0} \sum_\vk \int
dy\left(- \frac{df}{d y}\right)
\Tr\left(\rho_{\vk}(y)v^{\vk\mu}\rho_\vk(y)v^{\vk\nu}\right).
\label{sigmaB}
\end{eqnarray}

The dynamic self-energy is computed by an impurity solver, which is
implemented either on the real or imaginary axis. The most precise
impurity solvers, such as CTQMC, are implemented on imaginary axis,
hence we would like to formulate the method also for the case of
imaginary axis self-energy. Since the DC transport is sensitive to the
behaviour of the self-energy at low frequency, we take the power
expansion for $\Sigma(i\omega)$ and we determine the coefficients
directly on imaginary axis
\begin{eqnarray}
\Sigma(\omega) = \Sigma(0) + (1-Z^{-1})\omega - i \omega^2 B +\cdots .
\label{linearexp}
\end{eqnarray}
For the DC conductivity, the expansion to the quadratic order is quite
accurate. However, for the thermoelectric power, the truncation at
quadratic order is not sufficient since the qubic terms in the
self-energy expansion (the asymmetry of the scattering rate) is
crucial even at low temperature (see Ref.~\cite{Proceedings}).

We first embed the quasiparticle renormalization amplitude $Z$ and
scattering rate $B$ to the Kohn-Sham basis using Eq.~(\ref{Sigdef}),
i.e., $Z^{-1}_\vk = \hat{P}_{\vk\tau} Z^{-1}$ and $B_\vk =
\hat{P}_{\vk\tau} B$. Then we can express the low energy
electron Green's function in the Kohn-Sham basis as
\begin{eqnarray}
G_\vk(\omega) = (\omega Z^{-1}_\vk + \mu -\Sigma(0) -\varepsilon_{\vk}+ i \omega^2 B_\vk)^{-1}
\end{eqnarray}
Here $Z$ and $Z_\vk$ are hermitian matrices, while $\Sigma(0)$ has
both real and imaginary parts and is a complex
non-hermitian matrix.

Next we compute the square root $r_\vk \equiv \sqrt{Z_\vk}$ through
the eigensystem of $Z_\vk$. We thus have
\begin{eqnarray}
G_\vk(\omega) = r_\vk (\omega- r_\vk(-\mu+\Sigma(0)+\varepsilon_{\vk}-i\omega^2 B_\vk)r_\vk)^{-1} r_\vk
\end{eqnarray}
We first solve the non-hermitian eigenvalue problem
\begin{eqnarray}
\left[r_\vk (\varepsilon_{\vk}-\mu + \Sigma(0)) r_\vk \right] A^R_\vk = A^R_\vk E_\vk\\
A^L_\vk \left[r_\vk (\varepsilon_{\vk}-\mu + \Sigma(0)) r_\vk\right] = E_\vk A^L_\vk,
\end{eqnarray}
and compute the scattering rate in the eigenbase
\begin{eqnarray}
A_\vk^L r_\vk B_\vk r_\vk A_\vk^R = \Gamma_\vk.
\end{eqnarray}
to get
\begin{eqnarray}
G_\vk(\omega) = r_\vk A^R_\vk \frac{1}{\omega - E_{\vk\omega}} A^L_\vk
r_\vk
\label{Grn}
\end{eqnarray}
Here we used $E_{\vk\omega} = E_{\vk}-i\omega^2 \Gamma_\vk$.  Next we
insert Eq.~(\ref{Grn}) into (\ref{sigmaB}) and we neglect the
off-diagonal components of the scattering rate ( $(\Gamma_\vk)_{pq}
\sim \Gamma_{\vk p}\delta_{p,q}$), since the scattering between quasiparticles is
subleading at low temperature. We thus obtain
\begin{eqnarray}
{\sigma'}=-\frac{e_0^2}{2\pi V_0}\Re\sum_{\vk pq}\left[ C^\vk_{pq}S^\vk_{qp}-D^\vk_{pq}R^\vk_{qp}\right]
\end{eqnarray}
where
\begin{eqnarray}
C^\vk_{pq} &=& (A^L_\vk r_\vk v^\mu_\vk r_\vk A^R_\vk)_{qp}(A^L_\vk r_\vk v^\nu_\vk r_\vk A^R_\vk)_{pq}\\
D^\vk_{pq} &=& (A^L_\vk r_\vk v^\mu_\vk r_\vk A^{L\dagger}_\vk)_{qp}(A^{R\dagger}_\vk r_\vk v^\nu_\vk r_\vk A^R_\vk)_{pq}\\
S^\vk_{qp} &=& \int dx \left(-\frac{df}{dx}\right)\frac{1}{(x-E_{\vk x p})(x-E_{\vk x q})}\\
R^\vk_{qp} &=& \int dx \left(-\frac{df}{dx}\right)\frac{1}{(x-E^*_{\vk x p})(x-E_{\vk x q})}
\end{eqnarray}
The integrals $S^{\vk}$ and $R^{\vk}$ have multiple poles and need to
be treated by care. We first rewrite $S^{\vk}$ and $R^{\vk}$ in terms of
the following functions
\begin{eqnarray}
P_1(z) &=& \int  dx\left(- \frac{df}{dx}\right)\frac{1}{x-z}\\
P_2(z,\gamma) &=& \int  dx\left(- \frac{df}{dx}\right)\frac{1}{|x-z+i x^2 \gamma|^2}\\
Q_2(z,\gamma) &=& \int  dx\left(- \frac{df}{dx}\right)\frac{1}{(x-z+i x^2 \gamma)^2}
\end{eqnarray}
If $p=q$, we have
\begin{eqnarray}
S^\vk_{pp} = Q_2(E_{\vk p},\Gamma_{\vk p})\\
R^\vk_{pp} = P_2(E_{\vk p},\Gamma_{\vk p})
\end{eqnarray}
and if $p \ne q$ we approximate
\begin{eqnarray}
S^\vk_{qp} = \frac{P_1(E_{\vk p})-P_1(E_{\vk q})}{E_{\vk p}-E_{\vk q}}\\
R^\vk_{qp} = \frac{P_1(E_{\vk p})-P_1(E_{\vk q})}{E_{\vk p}^*-E_{\vk q}}.
\end{eqnarray}
Here we neglected the term proportional to $x^2 \Gamma$ in the
denominator, since the derivative of the fermi function constrains
$|x|\ll 1$ and since the interband transition give subleading
contribution to the Drude peak.

A special care needs to be taken to compute the integrals $P_1$, $P_2$
and $Q_2$ to high enough precision and avoid divergencies. We give
details on their evaluation in Appendix~\ref{AppB}

\section{Impurity solvers based on hybridization expansion}
\label{Impurity-solvers}

The impurity solvers based on the hybridization expansion have a long
history and were often employed to solve the problem of a degenerate
magnetic impurity in a metallic host
\cite{KK,GK,Ku,Grew,Keiter,Piers,Bickers}. In the past, most of
calculations were limited to the lowest order self-consistent
approximation, called the Non-crossing approximation (NCA). Recently,
many generalization of the approach were studied
\cite{Pruschke,CTMA,SUNCA,Anders}, to overcome the difficulty of the
NCA at low temperature, below the Kondo temperature. It is well known
that the NCA approximation fails to recover the Fermi liquid fixed
point at low temperature and low energy. Typically there are three
types of problems with NCA: i) the Kondo temperature is correct when
only one type of charge fluctuations is dominant (like $N\rightarrow
N-1$, which is equivalent to the limit of $U=\infty$). When more than
one charge fluctuation needs to be considered ($N\rightarrow N+1$ and
$N\rightarrow N-1$) the Kondo temperature is severely underestimated
and hence the Kondo peak is too narrow. ii) The asymmetry of the
Kondo-Suhl resonance and its height is exaggerated in NCA. iii) At
very low temperatures $T\ll T_K$ an additional spurious peak at zero
frequency appears.

For DMFT applications, the problem iii) is not very severe, while the
other two are. The first problem can be corrected by a very moderate
computational expense. Adding the first subleading Feynman diagrams
\cite{Pruschke,SUNCA}, named One crossing approximation (OCA)
\cite{Pruschke,our-rmp} cures the problem of the low energy scale. It also
substantially improves the asymmetry of the Kondo peak as well as its
width.  Not surprisingly, in the context of DMFT, the OCA
approximation gives correct critical $U$ of the Mott transition in the
Hubbard model, while NCA severely underestimates it.  In contrast to
other higher order conserving approximations \cite{SUNCA,CTMA}, the
OCA approximation is relatively straighforward to generalized to the
arbitrary impurity problem. Due its attractive features, OCA was used
in many DMFT applications, such as unraveling the mixed valence state
in Pu \cite{Nature}, the coherence-incoherence crossover in Ce-115
materials \cite{Science}, the transport properties in titanides
\cite{V11}, the $\alpha$ to $\gamma$ transition in Ce, etc. Compared
to exact solution, as obtained by CTQMC, the OCA approximations
typically gives very precise probability for all atomic states
\cite{chris} (the histogram), quite precise coherence scale, and the
quasiparticle renormalization amplitude (the width of the Kondo peak),
which is typically only slightly underestimated. At temperatures below
the coherence scale, the OCA method, however, still suffers from
slight overestimation of the height of the Kondo peak, and hence
causality violation in the context of DMFT. Hence, the OCA
approximation has to be used with care, especially in the systems with
high coherence scale, and the systems with only moderate
correlations.

The OCA equations for the one band problem were given by many authors
\cite{Pruschke,SUNCA}, and their generalization to multiband situation
was briefly discussed in the review Ref.~\onlinecite{our-rmp}, the generalized
equations were however, not yet given, hence we will give them for the
general multiorbital impurity problem, as relevant in the electronic
structure calculations in section \ref{OCA}.

Recently, a renewed interest in the hybridization expansion arouse,
once it was shown \cite{Werner,Rubtsov} that the Feynman diagrams can
be efficiently sampled by Monte Carlo importance sampling. The current
implementation of this algorithm, as applied to realistic material
problems, was discussed in plenty of detail recently
\cite{CTQMC,Werner2}, and it will not be repeated here.

Here we will rather outline an alternative Monte Carlo sampling
approach, which was not yet discussed in the literature nor
implemented.
It is natural to ask if there exists an alternative
regrouping of diagrams in Monte Carlo sampling, such that NCA
approximation would be the lowest order contribution in the
hybridization expansion, i.e., the two kinks approximation.
We detail the method below in section \ref{cbctqmc}, and show results
of a simplified implementation, which truncates the sampling at a
finite order (up to fifth order in hybridization).

\subsection{Towards Bold-CTQMC}
\label{cbctqmc}


The CTQMC \cite{Werner,CTQMC} solver is the most efficient exact
solver for electronic structure problems (see for example
Ref.~\onlinecite{chris} and Ref.~\onlinecite{MillisEfficient}). On the
other hand, the OCA impurity solver is very accurate in many
correlated systems with narrow bands. For example, it gives correct
critical $U$ in Hubbard model, correct Kondo scale in Kondo lattice
model, etc.

The current implementation of CTQMC is equivalent to pseudoparticle
formulation of the expansion around the atomic limit, however, with
bare pseudoparticle propagators.  It is thus natural to expect that
the dressed pseudoparticle propagators would make the
algorithm more efficient, since the two kinks approximation is
equivalent to NCA, and the four kinks approximation to OCA.

The basic idea of the \textit{bold CTQMC} algorithm is to sample the
skeleton Feynman diagrams, with propagators being \textit{dressed}
\cite{Prokofev}.  The Monte Carlo importance sampling samples all
such diagrams, with the probability proportional to their
Luttinger-Ward functional $\Phi$.
Hence contributions to all pseudoparticle self-energies can
be straighforwardly sampled within this approach.

%
Although the formalism of hybridization expansion on real axis was
developed long ago (see for example Ref.\onlinecite{Kroha}), its
imaginary axis counterpart was not yet given.  To our knowledge, the
NCA equations have not yet been implemented on imaginary axis, because
of the problems with diverging term in the projected Dyson equation
(see Eq.~(\ref{PDyson}) below).

In the hybridization expansion, the pseudoparticles are introduced to
diagonalize the atomic part of the Hamiltonian. The impurity problem
is cast into the form
\begin{eqnarray}
 H &=& \sum_{m} |m\rangle E_m \langle m| +
 \sum_{k i} \varepsilon_{k i} c^{\dagger}_{k i}c_{k i}
 \\
 &+&\sum_{mn,k\alpha i} V_{ki\alpha}|m\rangle\langle
 m|f^\dagger_\alpha|n\rangle\langle n| c_{k i}+h.c.
\nonumber
\end{eqnarray}
where we used completeness $\sum_m |m\rangle\langle m|=1$ for atomic states $|m\rangle$.
Each atomic state is represented by corresponding
pseudoparticle $a_m^\dagger|vacuum\rangle = |m\rangle$, and the
completness of atomic basis gives a constraint for pseudoparticles
$\rightarrow \sum_m a_m^\dagger a_m\equiv Q=1$. The Hamiltonian is
then given by
\begin{eqnarray}
 H &=& \sum_{m} E_m a_m^\dagger a_m +
 \sum_{k i} \varepsilon_{k i} c^{\dagger}_{k i}c_{k i}
 \label{Hpseudo}\\
 &+&\sum_{mn,k\alpha i} V_{ki\alpha} a_m^\dagger a_n
 \langle m|f^\dagger_\alpha|n\rangle  c_{k i}+h.c. +
 \lambda (Q-1)
 \nonumber
\end{eqnarray}
and the action is
\begin{eqnarray}
&& S = \sum_m\int d\tau a_m^\dagger (\frac{\partial}{\partial\tau}+E_m+\lambda)a_m
\\
&& \qquad +\sum_{nn'mm'}(F^{\alpha\dagger})_{mn}(F^{\beta})_{n' m'}\times\nonumber\\
&& \times \int d\tau
d\tau' a_m^\dagger(\tau)a_n(\tau)\Delta_{\alpha\beta}(\tau-\tau')a_{n'}^\dagger(\tau')a_{m'}(\tau')
\nonumber
\end{eqnarray}
where $(F^{\alpha\dagger})_{mn}=\langle m|f^\dagger_\alpha|n\rangle$.
We also define $H = H_0 + \lambda Q$.

Any physical quantity has to be evaluated in the $Q=1$ subspace. This
is achieved by letting $\lambda\rightarrow\infty$, to separate the spectra
of $Q=0$, $Q=1$, $Q=2$, $\cdots$. Then we use the Abrikosov's trick to
pick out the $Q=1$ subspace.
%
%
The expectation value, which we want to compute is
\begin{equation}
\langle A\rangle_{Q=1} = \frac{\Tr_{Q=1}(A e^{-\beta
    H})}{\Tr_{Q=1}(e^{-\beta H})},
\label{Abrikosov1}
\end{equation}
while accesible quantities are
$\langle A\rangle = \sum_Q\Tr_Q(A e^{-\beta H})/Z$.
If operator $A$ vanishes in the absence of impurity (in $Q=0$
subspace), the physical expectation value can be computed by
\begin{equation}
\langle A\rangle_{Q=1} =\lim_{\lambda\rightarrow\infty}\frac{\langle A \rangle}{\langle Q\rangle}.
\label{Abrikosov}
\end{equation}
This is clear from expansion
\begin{eqnarray}
Z\langle A\rangle &=& \Tr_{Q=1}(A e^{-\beta H_0-\beta\lambda})
+\Tr_{Q=2}(A e^{-\beta H_0-2\beta\lambda}) +\cdots
\nonumber\\
Z\langle Q\rangle &=& \Tr_{Q=1}(e^{-\beta H_0-\beta\lambda})
+\Tr_{Q=2}(2 e^{-\beta H_0-2\beta\lambda}) +\cdots
\nonumber\\
Z &=& \Tr_{Q=0}(e^{-\beta H_0}) + \cdots
\end{eqnarray}
Notice also that in the $\lambda\rightarrow\infty$ limit
\begin{equation}
\langle Q\rangle e^{\beta\lambda}= \frac{\Tr_{Q=1}(e^{-\beta
    H_0})}{\Tr_{Q=0}(e^{-\beta H_0})}=e^{-\beta F_{imp}}
\label{Q0}
\end{equation}
can be used to obtain impurity free energy.

In more general case, when $\langle A\rangle$ does not vanish in $Q=0$
subspace, Eq.~(\ref{Abrikosov}) should be replace by $\langle
A\rangle_{Q=1} =\lim_{\lambda\rightarrow\infty}\frac{\langle A
  Q\rangle}{\langle Q\rangle}$.

The Green's functions for pseudoparticles obey the Dyson equation,
\begin{eqnarray}
G_m = \frac{1}{\omega-\lambda-E_m-\Sigma_m(\omega)}.
\label{Dyson0}
\end{eqnarray}
where the energies of all pseudoparticles are shifted by $\lambda$
compared to atomic energies $E_m$, due to $\lambda Q$ term in the
Hamiltonian.
In general, the Green's functions for pseudoparticles are
off-diagonal. The states which correspond to the same superstate,
defined in Ref.~\onlinecite{CTQMC}, obey a matrix analog of the above Dyson
equation. However, here we will give equations for diagonal case,
since the generalization is less transparent, but straighforward.

The numeric limit of $\lambda\rightarrow \infty$ is very untractable
for computer. Since bold-CTQMC is implemented in imaginary time, we
thus want to analytically project the pseudoparticle equations on
imaginary time axis.

Before the limit $\lambda\rightarrow\infty$ is taken, the
pseudoparticle Green's functions are given by
\begin{eqnarray}
G_m(\tau) =\left\{
\begin{array}{cr}
  \int \frac{dx}{\pi} f(-x) e^{-x\tau}G_m^{''}(x) & \tau>0 \\
 -\int \frac{dx}{\pi} f(x) e^{-x\tau}G_m^{''}(x) & \tau<0 \\
\end{array}
\right.
.
\end{eqnarray}
The poles of the Green's function $G_m$ are at large frequencies,
comparable to $\lambda$, while $G_m^{''}$ vanishes for $x\ll \lambda$.
Hence $G(\tau<0)$ vanishes because $f(x)G_m^{''}(x)$ vanishes. We thus have
\begin{eqnarray}
&& G_m(\tau<0)=0\\
&& G_m(\tau>0) = e^{-\lambda \tau}\int
\frac{dx}{\pi}e^{-x\tau}G_m^{''}(x+\lambda).
\end{eqnarray}
This equations demonstrate the well known fact that the
pseudoparticles can not propagate back in time.

To derive a set of well posed projected equations, we introduce
projected Green's functions, which remain well behaved in the limit
$\lambda\rightarrow\infty$, and are used for numeric implementation
\begin{eqnarray}
\widetilde{G}_m(\tau) = e^{\lambda\tau}G_m(\tau)
\end{eqnarray}
Of course, these projected propagators vanish for $\tau<0$.  The
projected propagators are analogous to the well known projected
functions on the real axis (see Ref.~\onlinecite{Kroha})
$\widetilde{G}_m(x) = G_m^{''}(x+\lambda)/f(-x)$
since
\begin{equation}
\widetilde{G}_m(\tau) = \int \frac{d\omega}{\pi}
e^{-\omega\tau}f(-\omega)\widetilde{G}_m(\omega)
\label{analtc}
\end{equation}
is the usual $\tau \leftrightarrow \omega$ transformation between the
imaginary time and real frequency.

Our goal is to write all equations in terms of projected
$\widetilde{G}$ and analogous $\widetilde{\Sigma}$ functions, which do
not contain $\lambda$ and are numerically well behaved. The problem
however is that the projected quantities do not have fermionic nor
bosonic character, and hence can not be represented on imaginary
frequency axis. The Dyson equation Eq.~(\ref{Dyson0}) can be expressed
in terms of projected functions by 
%
\begin{eqnarray}
\widetilde{G}(\tau) =
T\sum_{i\omega}\frac{e^{-(i\omega-\lambda)\tau}}{i\omega-\lambda-E-
 \int_0^\beta d\tau' e^{(i\omega-\lambda)\tau'}\widetilde{\Sigma}(\tau')}
\label{PDyson}
\end{eqnarray}
but its evaluation is far from straightforward.
For convenience, we drop the index $m$ from $E_m$, $\widetilde{G}_m$, and 
$\widetilde{\Sigma}_m$.

We need to evaluate this formula in the limit $\lambda\rightarrow\infty$.
It is however not possible to perform the limit numerically because the
exponential factors grow as $\exp(\lambda\beta)$ while the poles are
in infinity on the real axis.

For the implementation of the bold-CTQMC, it is crucial to find
numerically tractable form of the projected Dyson equation. To this
end, we perform expansion in powers of $\widetilde{\Sigma}$, to get
\begin{eqnarray}
\widetilde{G}(\tau) = T\sum_{i\omega}
\frac{e^{-(i\omega-\lambda)\tau}}{i\omega-\lambda-E}\left(
1+\frac{S}{i\omega-\lambda-E}
\right.\nonumber\\
\left.
+\frac{S^2}{(i\omega-\lambda-E)^2}+\cdots
\right)
\end{eqnarray}
where $S=\int_0^\beta d\tau' e^{(i\omega-\lambda)\tau'}\widetilde{\Sigma}(\tau')$.
The summation over imaginary frequency can now be performed, to obtain
\begin{widetext}
\begin{eqnarray}
\widetilde{G}(\tau)=-\sum_{n=0}^{\infty}\frac{1}{n!}\frac{d^n}{dE^n}\left[
  \int_0^\tau d\tau_1 \widetilde{\Sigma}(\tau_1)
  \int_0^{\tau-\tau_1} d\tau_2 \widetilde{\Sigma}(\tau_2)
  \int_0^{\tau-\tau_{n-1}-\cdots-\tau_1} d\tau_n \widetilde{\Sigma}(\tau_n)
  e^{-E(\tau-\tau_1-\tau_2-\cdots-\tau_n)}
  \right]
\label{longexp}
\end{eqnarray}
\end{widetext}
Note that the limits of integration are constraint to the phase
space of forward propagating pseudoparticles. Namely, the limit of
$\lambda\rightarrow\infty$ does not allow the time difference in the
exponent to be negative.

To evaluate the projected Dyson equation in a stable way,  we first
evaluate the following moment-functions
\begin{equation}
S_n(\tau) = \frac{1}{n!}\int_0^\tau d\tau' \widetilde{\Sigma}(\tau')e^{-E(\tau-\tau')}(\tau-\tau')^n,
\end{equation}
and then we convolve the moment-functions with
$\widetilde{\Sigma}$. The Eq.~(\ref{longexp}) is hence implemented by
\begin{eqnarray}
\widetilde{G}(\tau) &=& -e^{-E\tau} + S_1(\tau) -
(\widetilde{\Sigma}*S_2)(\tau) \\
&+& (\widetilde{\Sigma}*(\widetilde{\Sigma}*S_3))(\tau)-\cdots
+(\widetilde{\Sigma}*(\widetilde{\Sigma}*\cdots * S_n))(\tau)\nonumber
\label{expansion}
\end{eqnarray}
where
\begin{equation}
(\widetilde{\Sigma} * Q)(\tau) = \int_0^\tau d\tau' \widetilde{\Sigma}(\tau-\tau')Q(\tau')
\end{equation}
Note that all terms in the expansion have the same sign
(note $\widetilde{\Sigma}<0$), hence the expansion converges quite fast, and
we typically need between 30-50 terms for numerically sufficient
precison.

Convolutions can be evaluated by standard method of Fourier
transforms, or, they can be cast into the form of matrix
multiplications, once the matrix
$\widetilde{\Sigma}_{\tau,\tau'}=\widetilde{\Sigma}(\tau-\tau')d\tau'$
is precomputed and used
for all terms in the expansion.

It is instructive to check the formula in two simple limits: i)
$\widetilde{\Sigma}=\Sigma_0\delta(\tau)$, evaluates to
$\widetilde{G}(\tau)=-e^{-(E+\Sigma_0)\tau}$; ii)
$\widetilde{\Sigma}=\sigma_0=const$ and $E=0$ evaluates to
$\widetilde{G}(\tau)=-\cosh(\tau\sqrt{\sigma_0})$.

The latter limit is very instructive because it shows that
$\widetilde{G}$ can exponentially grow at low temperature and finite
$\tau$. This is well known problem from implementing the NCA equations
on real axis. To keep $\widetilde{G}(\tau)$ finite, and peaked around
the origin on real axis ($\widetilde{G}(\tau)$ roughly constant in
$\tau$), one needs to shift all pseudoparticle energies
$E_m\rightarrow E_m+\lambda_0$ to sufficiently positive energies, such
that $\sum_m -\widetilde{G}_m(\beta-0^+) = const$, where $const$ is of
the order unity.  Namely, in grand canonical ensemble, the
pseudoparticle charge $\langle Q\rangle$, defined in Eq.~(\ref{Q0}),
is
\begin{eqnarray}
\langle Q\rangle = \sum_m G_m(\beta-0^+) =
e^{-\beta\lambda}\sum_m\widetilde{G}_m(\beta-0^+)
\end{eqnarray}
indeed vanishes in the physical $Q=1$ subspace. Once the projection is
done, the physical quantities in $Q=1$ subspace are invariant with
respect to shift of all pseudoparticle energies by the same amount. If
we introduce a finite shift $E_m\rightarrow E_m+\lambda_0$ (which is
equivalent to $\lambda\rightarrow\lambda+\lambda_0$), charge $\langle
Q\rangle$ will decrease for $e^{-\beta\lambda_0}$ while the product
$\langle Q\rangle e^{\beta\lambda0}=e^{-\beta F_{imp}}$ will remain
the same. Similarly, all physical quantities are invariant, while the
projected pseudoparticle quantities are not. Hence, for numerical
stable evaluations, it is crucial to choose the shift $\lambda_0$ such
that pseudoparticle propagators are finite. A large $\lambda_0$ will
make them exponentially small, while vanishing $\lambda_0$ will cause
$\widetilde{G}_m$ to diverge at $\beta$. We thus need to fix the value
of $\lambda_0$ properly.  Two possible choices are $\sum_m
G_m(\beta-0^+)= const$ or $G_{m-gs}(\beta-0^+)=const$, where $m-gs$ is
the pseuodoparticle, which corresponds to the ground state of the
atom.


The basic idea for the bold-CTQMC is to sample self-energies for all
pseudoparticles as well as the local Green's function. This is easiest
to achive by defining the probabilty to be proportional to the
absolute value of the Luttinger-Ward functional $|\Phi[G,\Delta]|$,
and the self-energies then become
\begin{eqnarray}
\Sigma_{mm'}  &=& \frac{\delta \Phi[G,\Delta]}{\delta G_{m'm}}\\
G_{\alpha\beta}&=&\frac{1}{\langle Q\rangle}\frac{\delta \Phi[G,\Delta]}{\delta \Delta_{\beta\alpha}}
\end{eqnarray}
where the first equation is contribution to the pseudoparticles
self-energies, and the second is contribution to the real-electron
Green's function (the impurity Green's function).

The second identify might be less obvious, but it follows from the
fact that the impurity Green's function is the T-matrix for the
conduction electrons
\begin{equation}
\left(\frac{1}{g^{-1}_k-\Sigma_c}\right)_{k i, k' j} = g_{k i}\delta_{k i, j k'} + g_{k i} V^*_{k i\alpha}G_{\alpha\beta}V_{k' j\beta}.
\label{T-matrix}
\end{equation}
We have seen above that $G_{mm'}$ carries a factor of
$e^{-\lambda\beta}$, and we will show below that $\Phi$ also carries
the same factor $e^{-\lambda\beta}$, hence the pseudoparticle
self-energy $\Sigma_{mm'}$ is of the order of unity. On the other
hand, the conduction electron self-energy $\Sigma_c$ is proportional
to $\delta\Phi[G,\Delta]/\delta\Delta$, and hence vanishes as
$e^{-\beta\lambda}$. Therefore both $\Sigma_c$ and $G_{\alpha\beta}$
are proportional to $e^{-\beta\lambda}$.  The expansion of the
equation Eq.~(\ref{T-matrix}) in powers of $e^{-\beta\lambda}$ shows
that i) conduction electron propagator $g_{k}$ is unrenormalized in
this theory (or equivalently the bare hybridization $\Delta$ appears
in functional $\Phi[G,\Delta]$); ii) the impurity Green's function,
evaluated in the grand-canonical ensemble $G_{\alpha\beta}$ is equal
to $\delta\Phi[G,\Delta]/\delta\Delta$, which vanishes as
$e^{-\beta\lambda}$. However, the physical quantities like the
electron Green's function must be evaluated in $Q=1$ subspace, using
Eq.~(\ref{Abrikosov}). The resulting ratio is of order unity and is
invariant with respect to shift of $\lambda_0$, as explained above.

The Luttinger-Ward functional $\Phi[G,\Delta]$ for the lowest order
contribution (two kinks), known under the name NCA, is given by
\begin{eqnarray}
\Phi^0[G,\Delta] = \int_0^\beta d\tau
G_{mm'}(\tau)G_{n'n}(\beta-\tau)\Delta_{\alpha\beta}(-\tau)\nonumber\\
\times (F^{\alpha})_{nm}(F^{\beta\dagger})_{m'n'}
\end{eqnarray}
Note that if integration variable is shifted to
$\tau\rightarrow\beta-\tau$, additional minus sign can appear.  In
case of regular fermions and bosons, this minus sign is automatically
taken care of by the antiperiodicity of fermionic Green's functions
$G(\beta-\tau)=-G(-\tau)$.  The pseudoparticle Green's functions
however vanish at negative times, and one needs to add $\beta$ to the
negative argument, and add an overal minus sign when $\beta$ is added
to the fermionic Green's function.

The corresponding pseudoparticle self-energies are
\begin{eqnarray}
 \Sigma_{nn'}(\tau)= (-1)^{f}
 \frac{\delta \Phi^0[G,\Delta]}{\delta G_{n'n}(\beta-\tau)}
\end{eqnarray}
where $(-1)^{f}$ is $+1$ (-1) if $n$ corresponds to pseudo-boson
(pseudo-fermion). Again, this minus sign is because negative times are
not allowed for pseudoparticles.

Each pseudoparticle propagator carries an exponent
$e^{-\lambda\Delta\tau}$, and the sum of exponents is always
$e^{-\beta\lambda}$. This holds for all diagrams composed of
exactly one loop of pseudoparticles.  These are the only diagrams that
give contribution to the physical quantities.

If we take out the exponential factors, the NCA functional takes the form
\begin{eqnarray}
\Phi^0[G,\Delta] = e^{-\beta\lambda}\int_0^\beta d\tau
\widetilde{G}_{mm'}(\tau)\widetilde{G}_{n'n}(\beta-\tau)\Delta_{\alpha\beta}(-\tau)\nonumber\\
\times (F^{\alpha})_{nm}(F^{\beta\dagger})_{m'n'}
\end{eqnarray}
If we denote
$\widetilde{\Phi}[G,\Delta]=e^{\beta\lambda}\Phi[G,\Delta]$, we see
that
$$\Sigma(\tau)=\frac{\delta\Phi}{\delta G(\beta-\tau)} =
\frac{\delta\widetilde{\Phi}}{\delta \widetilde{G}(\beta-\tau)}e^{-\lambda\tau}=
\widetilde{\Sigma}(\tau)e^{-\lambda\tau},$$
hence
\begin{equation}
\widetilde{\Sigma}_{nn'}(\tau) =
\frac{\delta \widetilde{\Phi}[\widetilde{G},\Delta]}{\delta \widetilde{G}_{n'n}(\beta-\tau)}
\label{Sigmadd}
\end{equation}
The projected $\widetilde{\Phi}[\widetilde{G},\Delta]$ has exactly the
same form as $\Phi[G,\Delta]$, we only need to replace
$G\rightarrow\widetilde{G}$. The NCA diagram hence becomes
\begin{eqnarray}
\widetilde{\Phi}^0[\widetilde{G},\Delta] = \int_0^\beta d\tau
\widetilde{G}_{mm'}(\tau)\widetilde{G}_{n'n}(\beta-\tau)\Delta_{\alpha\beta}(-\tau)\nonumber\\
\times (F^{\alpha})_{nm}(F^{\beta\dagger})_{m'n'}
\label{NCAl}
\end{eqnarray}

From Eqs.~(\ref{Sigmadd}) and~(\ref{NCAl}) it is clear that we
achieved the goal of expressing all equations in terms of projected
quantities, which do not depend on variable $\lambda$, and are
numerically well behaved.

The projected second order diagram, which correspond to OCA
approximation, is given by
\begin{widetext}
\begin{eqnarray}
\widetilde{\Phi}^1[\widetilde{G},\Delta] = \int_0^\beta d\tau_4\int_{0}^{\tau_4}
d\tau_3\int_{0}^{\tau_3} d\tau_2\int_0^{\tau_2}d\tau_1
\widetilde{G}_{m_0 m_0'}(\tau_1-\tau_4+\beta)\widetilde{G}_{m_1 m_1'}(\tau_2-\tau_1)
\widetilde{G}_{m_2 m_2'}(\tau_3-\tau_2)
\widetilde{G}_{m_3 m_3'}(\tau_4-\tau_3)
\nonumber\\
(F^{\alpha\dagger})_{m_1' m_0}\Delta_{\alpha\beta}(\tau_1-\tau_3)(F^\beta)_{m_3' m_2}
(F^{\alpha'\dagger})_{m_2' m_1}
\Delta_{\alpha'\beta'}(\tau_2)
(F^{\beta'})_{m_0' m_3}
\end{eqnarray}
\end{widetext}
The projected pseudoparticles vanish at negative times and are well
behaved at positive times. For the purpose of properly evaluating the
Feynman diagrams in time, we can extend them to negative times
without any loss of generality. The pseudo-bosons hence become
periodic, and the pseudo-fermions antiperiodic. The annoying minus
signs $(-1)^f$ can then be eliminated. However, the projected
pseudoparticles can not be Fourier transformed to imaginary frequency,
and they do not obey the usual Dyson equation, but rather a more
complicated type of Dyson equations derived in
Eq.~(\ref{longexp}). The pseudoparticles can be analytically continued
to real frequencies, and all pseudoparticles satisfy fermionic-type of
continuation, given in Eq.~(\ref{analtc}).

Finally, the Monte Carlo algorithm must generate any skeleton diagram
of any order. The probability to accept the diagram is proportional to
its $|\widetilde{\Phi}[\widetilde{G},\Delta]|$. The contribution to
pseudoparticle self-energy is then $\Sigma_{mm'}(\tau) = \langle
\textrm{sign}(\Phi)/G_{m'm}(-\tau) \rangle$, where $\langle\rangle$ means the average in
the Markov proces, where weights are proportional to $|\Phi|$. Similarly,
the impurity Green's function can be sampled by
$G_{\alpha\beta}(\tau)=\langle
\textrm{sign}(\Phi)/\Delta_{\beta\alpha}(-\tau)\rangle/\langle Q\rangle$.
The sampled self-energies will only be proportional to the exact
self-energies. The renormalization factor can easily be found knowing
the probability for NCA diagram, and its value.

%
The requirement to sample the skeleton diagrams
prohibits us to combine many diagrams into determinant of
hybridization functions $\Delta$, as it was achieved in the algorithm
by Werner et.al. \cite{Werner}. Similar type of trick of combining the
diagrams into determinant of $\Delta$'s would substantially improve
the efficiency of the algorithm.  It is however not clear how to
eliminate non-skeleton diagrams from determinant, and keep the
updating formulas efficient.

\begin{figure}[!bt]
\centering{
  \includegraphics[width=0.99\linewidth]{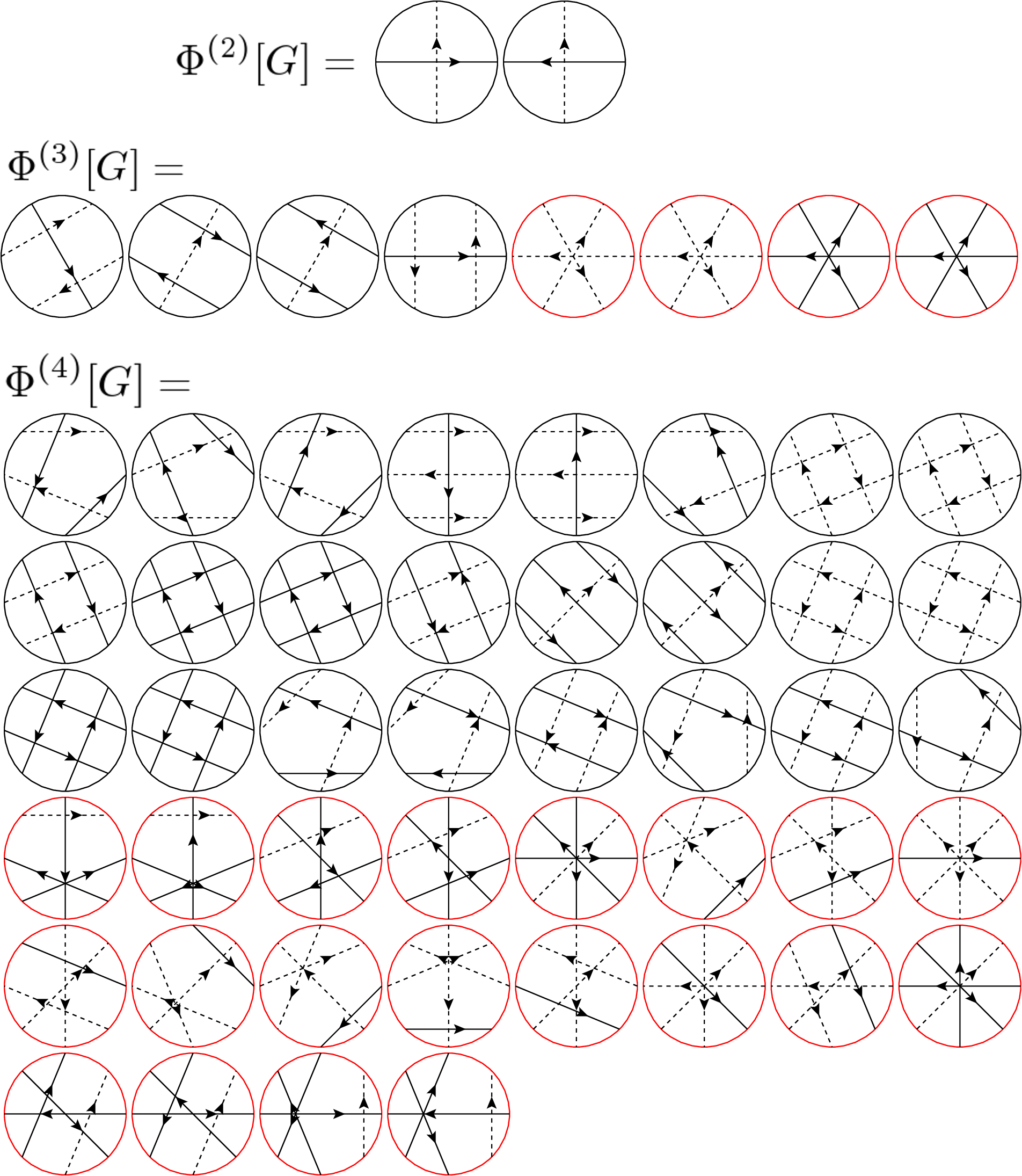}
  }
\caption{ All diagrams of the second, the third, and the fourth order
  in hybridization strength which contribute to the Luttinger-Ward
  functional. The pseudoparticle-propagators run across the ring,
  while the crossing lines stand for the hybridization $\Delta$. The
  full line represents spin-up, and the dashed lines the spin-down
  hybridization. The black diagrams (both diagrams in $\Phi^{(2)}$,
  first four in $\Phi^{(3)}$, and first 24 in $\Phi^{(4)}$) give
  positive contribution to $\Phi$, and the red give negative
  contribution. Some diagrams seems to appear multiple times. This is
  because different pseudoparticles appear in the ring. Since we do
  not use different line for each pseudoparticle, some diagrams seem
  equivalent. However, it is very straighforward to deduce the
  pseudoparticle propagators knowing the type and the direction of the
  conduction electron propagators.
}
\label{bctqmc-diags}
\end{figure}

To test the above described algorithm, and to check its performance
and convergence, we implemented a simplified version of the bold-ctqmc
for the canonical Anderson impurity model. We sampled all diagrams up
to certain order starting with first order (NCA), second order (OCA)
and up to fifth order. The fifth order takes only minutes on a typical
personal computer.  We first found the topology of all diagrams of
certain order, the prefactor and the sign of each diagram. In
Fig.~\ref{bctqmc-diags} we plotted the diagrams for the first few
orders (second - $\Phi^{(2)}$, ... fouth - $\Phi^{(4)}$). We colored
the diagrams according to their sign, positive with black and negative
with red. There are four NCA diagrams, two OCA diagrams, 8 third order
diagrams (4 positive and 4 negative), 44 forth order diagrams (24
positive and 20 negative), 320 fifth order diagrams (128 positive and
192 negative). We evaluated exactly the NCA and OCA diagrams, and we
used Metropolis algorithm to sample the time arguments for higher
order diagrams. The probability for the acceptance of a set of
imaginary times was taken to be proportional to the value of the
total $|\Phi^{(n)}(\tau_1,\tau_2,...\tau_{2n})|$, hence at fifth order
320 diagrams were evaluated at each Monte Carlo step. While this
algorithm can not be used at very high orders in perturbation theory
due to exponential growth in the number of diagrams, its advantage is
in large improvement of the sign problem. Namely, the diagrams of the
same order and the same time arguments tend to cancel at higher
orders. Since we evaluate all of them at each Monte Carlo step, the
sign problem is almost completely eliminated.

The non-interacting limit $U=0$ is the hardest case for the
hybridization expansion algorithm, because the coherence temperature is
infinite. Here we present test of the algorithm in the case of
half-filled non-interacting Hubbard model on the Bethe lattice within
DMFT. We want to emphasize that the algorithm becomes more efficient
and faster converging in strongly interacting limit $U>>0$, a case
which will be presented elsewhere.

\begin{figure}[!bt]
\centering{
  \includegraphics[width=0.99\linewidth]{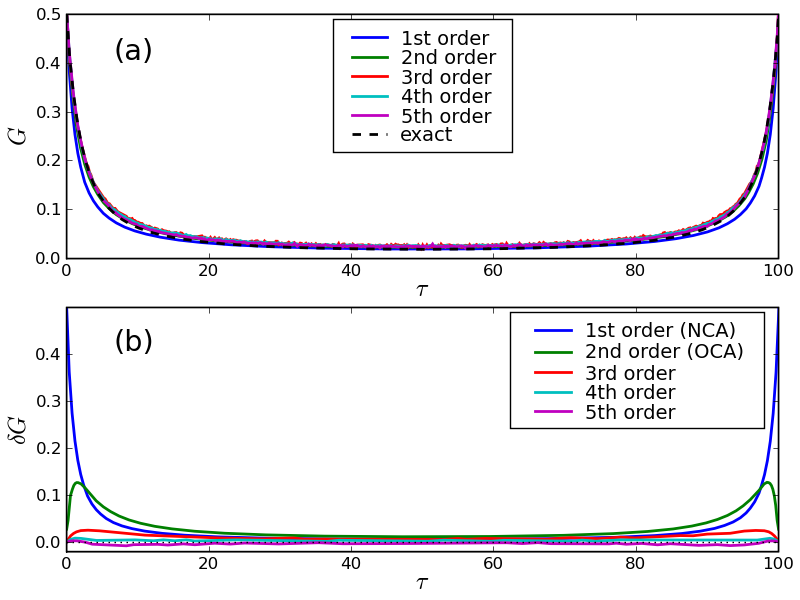}
  }
\caption{
(a) The comparison of
the finite order perturbation theory result with the exact impurity
Green's function.
(b) The contributions to the impurity Green's function up to
the fifth order, plotted separately order by order.
}
\label{bctqmc-g1}
\end{figure}

In Fig.~\ref{bctqmc-g1}(a) we show the impurity Green's function on
imaginary axis (at $1/T=100$) when the perturbation theory is
truncated at certain order. We also display the exact result by the
dashed line. While the NCA curve clearly deviates from the exact
result, the higer order approximations are hardly distinguished from
the exact curve on this plot. In Fig.~\ref{bctqmc-g1}b we show
separately the contributions to the Green's function from different
orders in perturbation theory. As expected the contribution from the
lowest two orders is large, while the higher order contributions are
smaller. This shows why OCA approximation is so successful in many
realistic situations. The fifth order contribution is on average only
$3\times 10^{-3}$, and never exceeds $6\times 10^{-3}$.

\begin{figure}[!bt]
\centering{
  \includegraphics[width=0.99\linewidth]{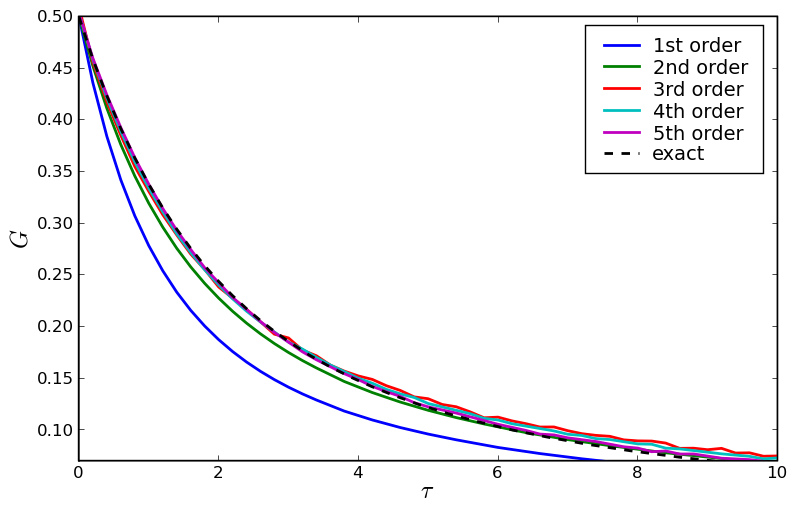}
  }
\caption{ The same as in Fig.~\ref{bctqmc-g1}, but we zoom in the
  short time behaviour.
}
\label{bctqmc-g2}
\end{figure}

In Fig.~\ref{bctqmc-g2} we zoom-in the exponential drop of the Green's
function at short times. We see that the convergence with the
perturbation order is very encouraging.
%

\begin{figure}[!bt]
\centering{
  \includegraphics[width=0.99\linewidth]{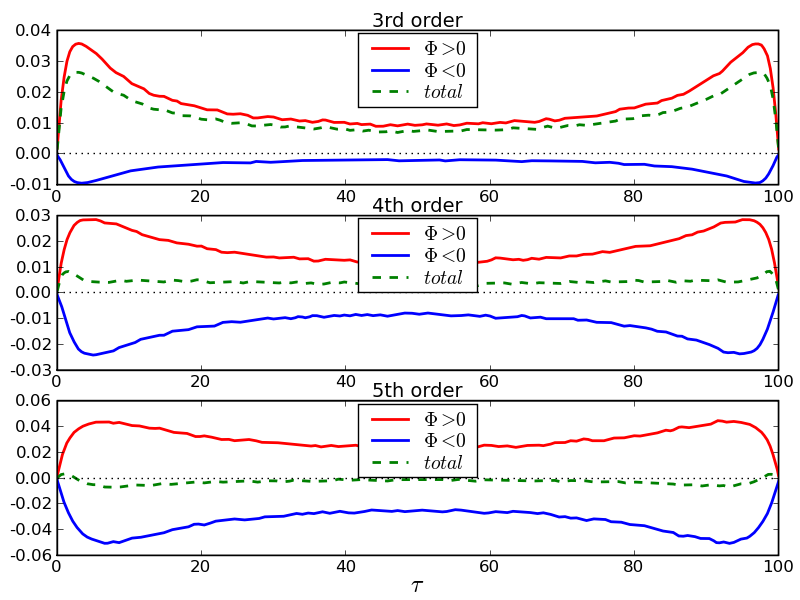}
  }
\caption{
The three panels show the contribution to the impurity Green's
function at 3rd, 4th and 5th order in perturbation theory. We show
separately the contribution from the terms with positive $\Phi$ and
the terms with negative $\Phi$.
}
\label{bctqmc-sign}
\end{figure}

For efficiency of the bold-ctqmc, it is important to monitor the sign
of each individual diagram. In Fig.~\ref{bctqmc-sign} we show
separately the contribution to the impurity Green's function from the
diagrams with positive $\Phi$ and those with negative $\Phi$, together
with the sum of the two. At the third order, the sum is around 70\% of
the positive contribution, while at the forth and fifth order, the
sign drops to 0.2 and 0.07, respectively.
As explained above, the current implementation of the method, which
groups together all diagrams of a certain order in perturbation
theory, does not have a substantial minus sign problem.
%
%
However, this method becomes expensive at high orders, and thus one
needs to resort to sampling of individual diagrams, which can be
performed to arbitrary high order. In the latter case, there will be a
minus sign problem, as estimated here.

%

\subsection{The One crossing approximation}
\label{OCA}

In this section we will give the most general formulas for the One
crossing approximation, and we will explain the crucial steps in
implementing the algorithm.

We start with lowest order approximation, which is the Non-crossing
approximation. When evaluating these diagrams, we have to consider
only two Hilbert subspaces of constant $N$ at once, i.e., $N$ and
$N+1$. The first step is to compute all eigenvalues and eigenvectors
of the atom in the subspace $N$ and $N+1$. We then group together the
atomic eigenstates, which are degenerate, i.e., have the same
atomic energy $E_m$. In the next step we check which of these
degeneracy's survive in the presence of the crystal field environment
(impurity hybridization $\Delta$), and which off-diagonal propagators
need to be considered. We evaluate the following matrix elements
\begin{eqnarray}
C_{b_2,b_1}^{\alpha\alpha'}=  \sum_{f\in deg, (\alpha,\alpha')\in deg\; and\; \Delta_{\alpha\alpha'}\ne 0, }(F^{\alpha'})_{b_2 f}(F^{\alpha\dagger})_{f b_1}
\end{eqnarray}
Here $b$ runs in the Hilbert subspace of $N$ and $f$ in the Hilbert
supspace of $N+1$.  The matrix elements $(F^{\alpha})_{b f}=\langle
b|f_\alpha|f\rangle$ and $(F^{\alpha\dagger})_{f b}=\langle
f|f^\dagger_\alpha|b\rangle$ where $f_\alpha$ is electron destruction
operator.  The sum runs only over the $f$ states which are degenerate
and over one electron states $\alpha$ which are also degenerate and
for which $\Delta_{\alpha\alpha'}$ is nonzero in the considered
crystal field symmetry.  The resulting matrix elements $C_{b_2,b_1}$
have the same symmetry as the propagators of the pseudoparticles
$G_{b_2 b_1}$. Clearly, in high symmetry crystal environment, most of
the off-diagional matrix elements vanish and the degeneracy of
$G_{bb}$ is high, but in low symmetry environment and in the broken
symmetry state, many of the off-diagonal propagators become crucial.

\begin{figure}[!bt]
\centering{
  \includegraphics[width=0.99\linewidth]{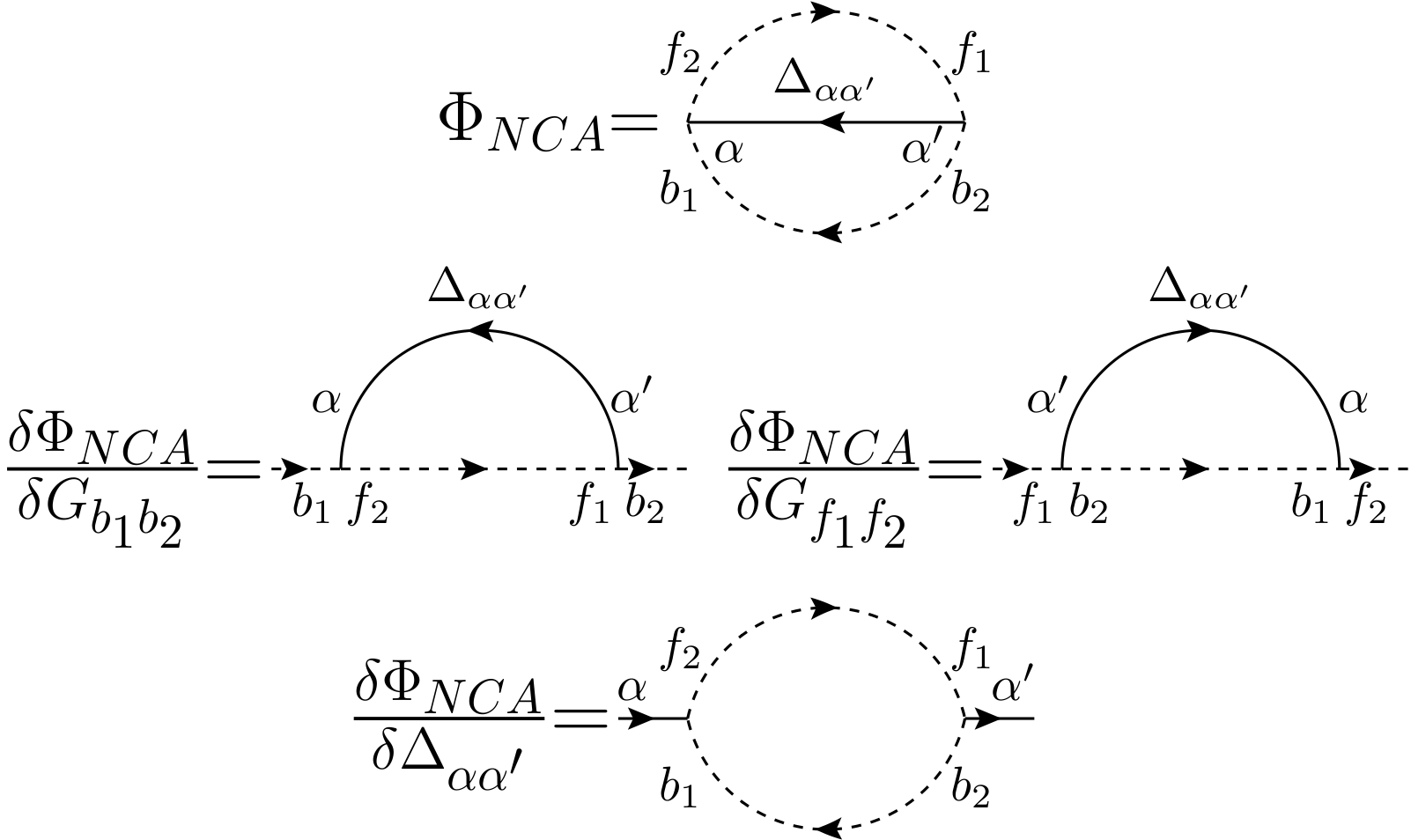}
  }
\caption{
  The NCA Luttinger-Ward functional and the self-energies within NCA.
}
\label{NCALW}
\end{figure}
Once the symmtry of the propagators is known, we determine all
nonvanishing bubbles (NCA diagrams) and the matrix elements for each
bubble. The NCA matrix elements are
\begin{eqnarray}
C_{b_1 b_2 f_1 f_2}^{\alpha\alpha'}=
\sum_{(f_1,f_2), (b_1,b_2), (\alpha,\alpha')\in deg}(F^{\alpha'})_{b_2  f_1}(F^{\alpha\dagger})_{f_2 b_1},
\end{eqnarray}
where we sum only over degenerate states $f,b$ and degenerate crystal
field components $\alpha$.
The Luttinger-Ward functional and the self-energy corrections are
depicted in Fig.~\ref{NCALW}. We associate a factor
$(F^{\alpha\dagger})_{fb}$ to each vertex that marks the creation of
electron in bath $\alpha$. Accordingly, we add a factor 
$(F^{\alpha})_{bf}$ for each vertex of electron anhilation.
\begin{figure}[!bt]
\centering{
  \includegraphics[width=0.5\linewidth]{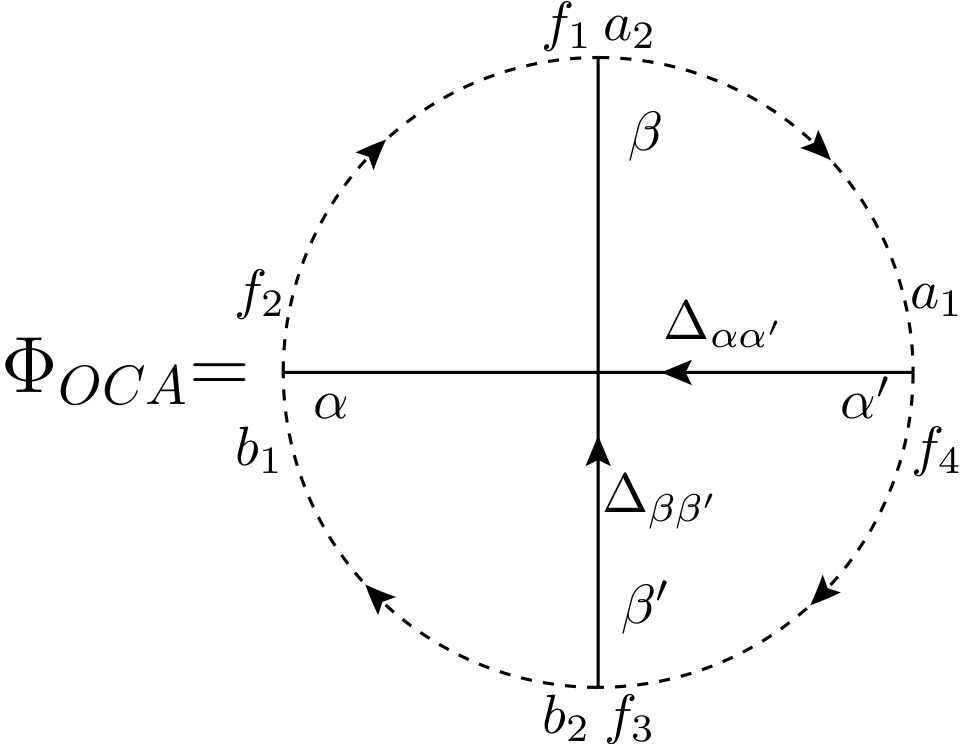}
  }
\caption{
  The Luttinger-Ward functional for the One Crossing Approximation (OCA).
}
\label{OCALW}
\end{figure}

In the next step, we precompute the matrix elements of the
one-crossing diagrams, which are depicted in 
Fig.~\ref{OCALW}.
Here we need to select three different Hilbert subspaces: $N-1$, $N$,
and $N+1$ to compute
\begin{eqnarray}
&& D^{\alpha\alpha'\beta\beta'}_{f_1 f_2 f_3 f_4 b_1 b_2 a_1 a_2} =\nonumber\\
&& \qquad \sum_{deg}(F^{\beta'})_{b_2 f_3}(F^{\alpha'})_{f_4 a_1}(F^{\beta\dagger})_{a_2 f_1}(F^{\alpha\dagger})_{f_2 b_1}
\end{eqnarray}
Here $b$, $f$, $a$ run over the states with $N-1$, $N$ and $N+1$
number of particles, respectively.  We add only the most important
crossing corrections, for which the particle number $N$ is in the
Hilbert subspace of the ground state of the atom. We also select $f_i$
to be only the ground state multiplet of the atom, or the atomic
states with energy very close to the ground state energy.
%
%
We compute the matrix elements $C$ and $D$ only once in the DMFT
self-consistent loop and we save them into the input file for OCA
impurity solver. The matrix elements $C$, $D$ do need to be updated in
the outer LDA+DMFT charge loop. We typically update them every three
to four charge steps, since the relative crystal field splittings
usually change very little during LDA+DMFT iterations. The atomic
energies $E_m$ change much more (due to the chemical potential shift),
and need to be updated at every step.
%

The NCA diagrams on the real axis can be evaluated with conventional techniques, 
and after the projection, they take the following form
\begin{widetext}
\begin{eqnarray}
\Sigma_{b_2 b_1}(\omega) =\sum_{f_1 f_2 \alpha\alpha'}
-(F^{\alpha'})_{b_2 f_1}(F^{\alpha\dagger})_{f_2 b_1}\int\frac{dy}{\pi}f(y)\Delta^{''}_{\alpha\alpha'}(y)
G_{f_1 f_2}(\omega+y)
\end{eqnarray}
\begin{eqnarray}
\Sigma_{f_2 f_1}(\omega) =\sum_{b_1 b_2 \alpha\alpha'}
-(F^{\alpha'})_{b_2 f_1}(F^{\alpha\dagger})_{f_2 b_1}\int\frac{dy}{\pi}f(-y)\Delta^{''}_{\alpha\alpha'}(y)
G_{b_1 b_2}(\omega-y)
\end{eqnarray}
\begin{eqnarray}
A^{imp}_{\alpha'\alpha}(\epsilon) = 
\frac{1}{e^{\beta\lambda}\langle Q\rangle f(-\epsilon)}
\sum_{b_1 b_2 f_1 f_2}\int dy
e^{-\beta y}
(F^{\alpha'})_{b_2 f_1}(F^{\alpha\dagger})_{f_2 b_1}
G_{b_1 b_2}^{''}(y)G_{f_1 f_2}^{''}(y+\epsilon)
\nonumber
\end{eqnarray}
\end{widetext}
where $G^{''}=\Im G$. The pseudoparticle propagators $G$ and the
pseudoparticle self-energies are related by the Dyson equation.
The Eq.~\ref{Hpseudo} shows that $G = 1/(omega-E-\lambda-\Sigma)$.

Many of the pseudoparticle propagators and hybridization functions are
degenerate, hence in practice we do not need to sum over all possible $b$, $f$
and $\alpha$ indices, but we rather use the precomputed matrix
elements $C_{b_1 b_2 f_1 f_2}^{\alpha\alpha'}$, which make sure that no
equivalent diagram (a diagram which has the same frequency dependence)
is not computed multiple times.

To take care of the diverging exponential factors, we work with the
projected quantities $\widetilde{G}(\omega)=G^{''}(\omega)/f(-\omega)$
and $\widetilde{\Sigma}(\omega)=\Sigma^{''}(\omega)/f(-\omega)$, as
explained above. The pseudoparticles have typically very sharp almost
diverging structure near the treshold energy, which is not easy to
Fourier transform. Hence we can not use the Fourier transform for
convolutions. We rather cast the above equation into the form for
matrix multiplication, for which fast linear algebra packages such as
BLAS, exist. We use the logarithmic mesh to resolve the fine structure
of the pseudoparticle green's functions.

It is important to realize that the number of baths $\alpha$ is quite
small (of the order of $2(2L+1)$ for correlated orbital of angular
momentum $L$), while the number of atomic states is much bigger. Hence
we precompute the integral and the first moment of functions
$\Delta^{''}(\omega)f(\omega)$ and of $\Delta^{''}(\omega)f(-\omega)$
for all $\alpha\alpha'$. Within trapezoid rule, the values and the
first moments of these quantities are enough to compute the above
convolutions with matrix multiplications on any given mesh.

To see that, lets consider an arbitrary convolution
\begin{equation}
  C(z)=\int g(x)f(x-z)dx
\end{equation}
Here the function $g(x)$ is defined on a certain mesh $\{x_i\}$, on
which it is well resolved, i.e., $g(x_i)\equiv g_i$. The function
$f(y)$ is defined on another mesh $\{y_i\}$, i.e., $f(y_i)\equiv f_i$.
The convolution can be safely calculated on the union of both mashes
$\{x_i,y_j+z\}$. One of the meshes should be shifted for $z$, thus for
each outside frequency, a different union of the two meshes should be
formed and only then the convolution can be safely evaluated. This is
very time consuming and not done in practice.

When a certain $f$ function needs to be convolved with many other
functions (like $\Delta^{''}(\omega)f(\omega)$ in our example above),
we use the followin trick. We first precompute the integral and the
first moment of the function
\begin{eqnarray}
  F_1(\epsilon_i) = \int_{-\infty}^{\epsilon_i}f(u)du\\
  F_2(\epsilon_i) = \int_{-\infty}^{\epsilon_i}u f(u)du\\
\end{eqnarray}
We then calculate the convolution without building a new inside mesh.
Let's use the mesh $\{x_i\}$ which resolves function $g$. Then, in the
spirit of trapezoid rule, we can linearly interpolate $g$ between the
points
\begin{equation}
  C(z)=\sum_i \int_{x_i}^{x_{i+1}}
  \left[g_i+\frac{g_{i+1}-g_i}{x_{i+1}-x_i}(x-x_i)\right]f(x-z)dx.
\end{equation}
This integral can be expressed by the above defined functions. To show
that, let us rewrite the convolution and expressed it by the new
function $\langle f\rangle_i$ which is defined on the same mesh as $g$
and with which the covolution is a simple scalar product
\begin{eqnarray}
  C(z)=\sum_i g_i\left[
   \int_{x_i-z}^{x_{i+1}-z}\frac{x_{i+1}-z-u}{x_{i+1}-x_i}f(u)du
   \right.\nonumber\\
   \left.   +
   \int_{x_{i-1}-z}^{x_i-z}\frac{z+u-x_{i-1}}{x_{i}-x_{i-1}}f(u)du
 \right] \nonumber\\
  \equiv \sum_i g_i \langle f\rangle_{i z} dh_i .
\end{eqnarray}
Thus $\langle f\rangle_{i z}$ is
\begin{widetext}
\begin{eqnarray}
\langle f\rangle_{i z}=
2\left[
\frac{(x_{i+1}-z)[F_1(x_{i+1}-z)-F_1(x_i-z)]-F_2(x_{i+1}-z)+F_2(x_i-z)}{(x_{i+1}-x_{i-1})(x_{i+1}-x_i)}-
\right.\\
\left.
-\frac{(x_{i-1}-z)[F_1(x_i-z)-F_1(x_{i-1}-z)]-F_2(x_i-z)+F_2(x_{i-1}-z)]}
     {(x_{i+1}-x_{i-1})(x_{i}-x_{i-1})}
\right]
\end{eqnarray}
\end{widetext}
Hence the convolution of $f$ with many functions $g_m$ can be computed
at once $C(m,z) = g_m*f$ by the following matrix product
$C(m,z) = \sum_i g_{m i} \langle f\rangle_{i z}  dh_i$.

Once the NCA contributions are evaluated, we add the second order diagrams,
which correspond to OCA approximation and are depicted in
Fig.~\ref{OCALW}. They take the explicit form
\begin{widetext}
\begin{eqnarray}
&&\Sigma_{b_2 b_1}(\omega) =
-\sum_{f_1 f_2 f_3 f_4 a_1 a_2 \alpha\beta\alpha'\beta'} (F^{\beta'})_{b_2 f_3}(F^{\alpha'})_{f_4 a_1}(F^{\beta\dagger})_{a_2 f_1}(F^{\alpha\dagger})_{f_2 b_1}\times\\
&&\qquad\qquad\qquad\qquad\qquad\qquad
\times \int\frac{dy}{\pi}f(y)\Delta^{''}_{\beta\beta'}(y)G_{f_3 f_4}(\omega+y)
\int \frac{dx}{\pi}f(x)\Delta^{''}_{\alpha\alpha'}(x)G_{f_1 f_2}(\omega+x)G_{a_1 a_2}(\omega+x+y)
\nonumber
\end{eqnarray}
\begin{eqnarray}
&&\Sigma_{a_2 a_1}(\omega) =
-\sum_{f_1 f_2 f_3 f_4 b_1 b_2 \alpha\beta\alpha'\beta'}(F^{\beta'})_{b_2 f_3}(F^{\alpha'})_{f_4 a_1}(F^{\beta\dagger})_{a_2 f_1}(F^{\alpha\dagger})_{f_2 b_1}\times\\
&&\qquad\qquad\qquad\qquad\qquad\qquad
\times \int\frac{dy}{\pi}f(-y)\Delta^{''}_{\alpha\alpha'}(y)G_{f_3 f_4}(\omega-y)
\int \frac{dx}{\pi}f(-x)\Delta^{''}_{\beta\beta'}(x)G_{f_1 f_2}(\omega-x)G_{b_1 b_2}(\omega-x-y)
\nonumber
\end{eqnarray}



\begin{eqnarray}
\Sigma_{f_2 f_1}(\omega) &=&
-\sum_{f_3 f_4 a_1 a_2 b_1 b_2 \alpha\beta\alpha'\beta'}\left[(F^{\beta'})_{b_2 f_3}(F^{\alpha'})_{f_4 a_1}(F^{\beta\dagger})_{a_2 f_1}(F^{\alpha\dagger})_{f_2 b_1}
+(F^{\alpha})_{b_2 f_1}(F^{\beta'})_{f_2 a_1}(F^{\alpha'\dagger})_{a_2 f_3}(F^{\beta\dagger})_{f_4 b_1}\right]\times
\nonumber\\
&\times&\int\frac{dy}{\pi}f(-y)\Delta^{''}_{\alpha\alpha'}(y)G_{b_1 b_2}(\omega-y)
\int \frac{dx}{\pi}f(x)\Delta^{''}_{\beta\beta'}(x)G_{a_1 a_2}(\omega+x)G_{f_3 f_4}(\omega+x-y)
\end{eqnarray}


\begin{eqnarray}
A^{imp}_{\beta'\beta}(\epsilon) &=& 
-\sum_{\alpha\alpha'f_1 f_2 f_3 f_4 b_1 b_2 a_1 a_2}\left[
  (F^{\beta'})_{b_2 f_3}(F^{\alpha'})_{f_4 a_1}(F^{\beta\dagger})_{a_2 f_1}(F^{\alpha\dagger})_{f_2 b_1}+
  (F^{\alpha'})_{b_2 f_1}(F^{\beta'})_{f_2 a_1}(F^{\alpha\dagger})_{a_2 f_3}(F^{\beta\dagger})_{f_4 b_1}
  \right]\times\\
&\times&\frac{1}{e^{\alpha\lambda}\langle Q\rangle f(-\epsilon)}\int dy e^{-\alpha y}
\int\frac{dx}{\pi}f(x)\Delta^{''}_{\alpha\alpha'}(x)\Im\left\{G_{b_1 b_2}(y)G_{f_1 f_2}(x+y)\right\}\Im\left\{G_{f_3 f_4}(\epsilon+y)G_{a_1 a_2}(\epsilon+x+y)\right\}
\nonumber
\end{eqnarray}

\end{widetext}
In practice, we do not sum over all $f$, $b$ and $a$ indices. As
explained above, we precompute the matrix elements
$D^{\alpha\alpha'\beta\beta'}_{f_1 f_2 f_3 f_4 b_1 b_2 a_1 a_2}$ for
the most important processes. We take only the low lying atomic states
into account (only $f$'s which are part of the ground state multiplet
or with energy very close to the ground state). We also take into
account the degeneracy of all atomic states and the degeneracy of
baths $\alpha$, in order to avoid computing the equivalent diagram
mutliple times.
Finally, the convolutions for the OCA approximation can also be cast
into the form of matrix multiplication, once the first moment and
integrals of a few functions are precomputed.

%

\section{The analytic continuation method}
\label{ancont}

The Monte Carlo impurity solvers are implemented on imaginary axis
where the quantity being sampled is real and many times even
``sign-free''. The results obtained in this way are exact, except for
the statistical noise. However, even a tiny statistical error on
imaginary axis precludes the analytic continuation by Pade type of
methods. The standard method, to overcome the difficulty of the
singularity of the kernel, is the Maximum Entropy Method (MEM). The
basic idea of this method is to find a function on the real axis,
which is very close to Monte Carlo data on imaginary axis (within
statistical error), and is smooth function on real axis, locally not
very different from a chosen model function. This approach works very
well for analytical continuation of the Green's function $G(\tau)$ to
obtain spectral function on real axis, i.e., to solve the integral
equation
$$G(\tau) = -\int f(-x)e^{-\tau x}A(x)dx$$ for $A(x)$.

Knowing the spectral function, it is however not possible to obtain
the momentum resolved spectra, or optical conductivity, or transport
coefficients. To compute these properties, it is essential to
analytically continue the self-energy, rather then the Green's
function. The self-energy of correlated materials is however very hard
to analytically continue with maximum entropy method, because the
self-energy typically has very sharp feature or even poles, which
separate the low energy part of the spectra (the quasiparticle peak)
from the high energy part of the spectra (the Hubbard bands). Due to
the maximum entropy method requirements of smoothness, the
analytically continued self-energy at low energy is typically
polluted with the near-by poles, which appear in the self-energy at
the intermediate energy.

A successful analytic continuation method for self-enery needs to
met the following conditions:
\begin{itemize}
\item imaginary axis self-energy is equal to Monte Carlo data within the statistical error
\item real axis self-energy function must be locally smooth
\item the power-expansion around zero frequency should match the
  quantum Monte Carlo data on both, real and imaginar axis.
\end{itemize}
While the first two conditions are met by MEM, the last is not.

We developed an alternative method, which mets the above conditions
and was very successfully used in combination with CTQMC for pnictides
\cite{FeAs}, cuprates \cite{Cedric}, VO$_2$ and other
materials. Although the method has many parameters, which needs to be
choosen appropriately, we can always check its accuracy by recomputing
the spectral function of the lattice, using analytically continued
self-energy, and comparing the spectral function to the maximum
entropy continued spectra.

We expand the self-energy in terms of modified Gaussians $\cL$, and
we add a polynomial function around zero frequency
\begin{equation}
\Sigma_M(z) = \sum_{n} c_n \cL(E_n,z) + f_0(z)
\label{SigM}
\end{equation}
The modified Gaussians
\begin{equation}
\cL^{''}(E_n,\omega) = \frac{1}{b |E_n| \sqrt{\pi}}e^{-b^2/4-(\log(\omega/E_n)/b)^2}
\end{equation}
have a unique peoperty that they are peaked at $E_n$, with the width
of approximately $E_n$, while they exponentially vanish at zero
frequency.  They are asymmetric with slow decay away from zero and
very fast decay towards zero frequency.  We choose the modified
Gaussians centered on a logarithmic mesh of $E_n = \pm \pi T w^n$ with
$w\sim 1.5$.
The modified Gaussians functions were used in connection with
constructing the NRG spectral function \cite{Bulla}. We typically take the
parameter $b$ to be $\sim 0.8$.

Since the modified Gaussians all vanish at zero frequency, we add a
polynomial function around zero frequency. The coefficients of the
polinomial are determined by fitting the imaginary axis self-energy,
i.e.,
\begin{eqnarray}
\Sigma(\omega_n) = \Sigma_0 + (-b_1 +i a_1) \omega_n  +(-a_2-i b_2) \omega_n^2
\end{eqnarray}
which can be analytically continued to
\begin{equation}
\Sigma(\omega)   = \Sigma_0 + (a_1 + i b_1) \omega  + (a_2 + i b_2) \omega^2 
\end{equation}
The polynomial has to drop-off sufficiently fast at high frequency,
hence we choose the following function
\begin{equation}
f_0^{''}(\omega) =
\left\{
\begin{tabular}{l}
$\left(\Sigma_0^{''} + \omega b_1  + \omega^2 b_2\right)/\left(1  + (\omega^2 b_2/\Gamma^2)^2 \right)$\\
$\Sigma_0^{''}\; \Gamma^2/(\omega^2 + \Gamma^2)$
\end{tabular}
\right.
\end{equation}
where the upper choice is made for metals and the lower choice for
insulators and very bad metals. In the FL regime, we have $b_1 \ll 1$,
$|\Sigma_0^{''}|\propto Z^2 \pi^2 T^2 $.  The coefficient $\Gamma$ is
determined by the condition $f_0(\omega=1) \ll 1$

For speed, we precompute $\cL(E_n,i\omega)$ and $\cL^{'}(E_n,\omega)$
by 
\begin{equation}
  \cL(E_n,z)  = -\frac{1}{\pi}\int \frac{dx \cL^{''}(E_n,x)}{z-x}.
\end{equation}
Similarly, we also precompute $f_0(i\omega)$ and $f_0^{'}(\omega)$.
Also the integral of the functions
$I_n  = \int dx \cL^{''}(E_n,x)$ and $I_0 = \int dx f_0^{''}(x)$ are
precomputed.

The coefficients $c_n$ in expansion Eq.~(\ref{SigM}) are determined by
minimizing the following functional
\begin{widetext}
\begin{eqnarray}
\chi = \sum_{\omega_n \in sampled}
|\Sigma_{M}(i\omega_n)-\Sigma_{QMC}(i\omega_n)|^2 +
\alpha_1 |I(\Sigma_M)-I_{D}|^2 +
\alpha_2 |\Sigma_M(0)-\Sigma_0|^2 \\
+\alpha_3 \left|\frac{d\Sigma_M(0)}{d\omega} -(a_1+ib_1)\right|^2+
\alpha_4 \left|\frac{d^2 \Sigma_M(0)}{d\omega^2} -2(a_2+ib_2)\right|^2
\end{eqnarray}
\end{widetext}
Here $\omega_n$ in the first term runs over the imaginary frequencies
which are sampled by QMC (and not over the analytically added
tail). The second terms imposes the correct value of the integral of
the self-energy. The integral of the expansion (\ref{SigM}) is
$$I(\Sigma_M) = \sum_{n} c_n I_n + I_0,$$ which needs to match the
$1/(\omega_n)$ tail of the QMC data 
$$I_{D} = \pi \lim_{\omega_n\rightarrow\infty}\omega_n
\Sigma^{''}_{QMC}(\omega_n)$$ Finally, the last three terms ensure
that the value, and the first two derivatives of the analytically
continued self-energy at zero frequency match the derivatives on
imaginary axis.  For minimization, we use the L-BFGS-B alrorithm of
Ref.~\onlinecite{Byrd}.

\section{Cerium $\alpha$-$\gamma$ transition}
\label{Ce-test}

\begin{figure}[hbt]
\centering{
  \includegraphics[width=1.0\linewidth]{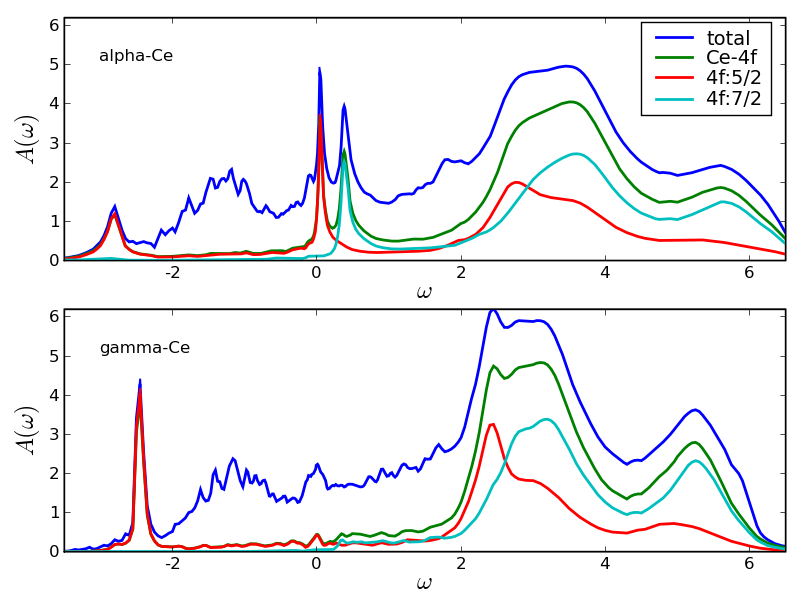}
}
\caption{Total and partial density of states of elemental cerium metal
  in both phases, $\alpha$ and $\gamma$ phase. We used OCA impurity solver.
}
\label{CeDOS}
\end{figure}

\begin{figure}[hbt]
\centering{
  \includegraphics[width=1.0\linewidth]{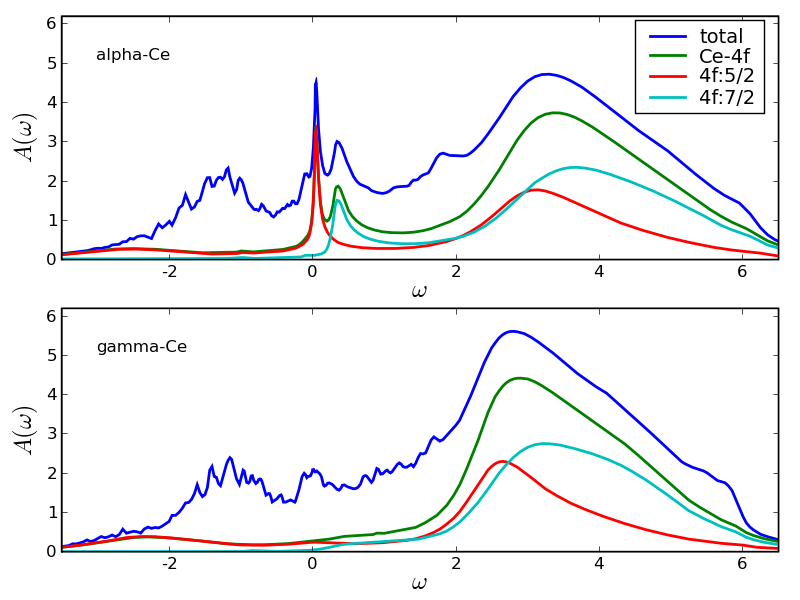}
}
\caption{The same data as in Fig.~\ref{CeDOS}, but obtained by
  continuous time quantum Monte Carlo solver, and analytical continuation
  method.
}
\label{CeDOS2}
\end{figure}

\begin{figure}[hbt]
\centering{
  \includegraphics[width=1.0\linewidth]{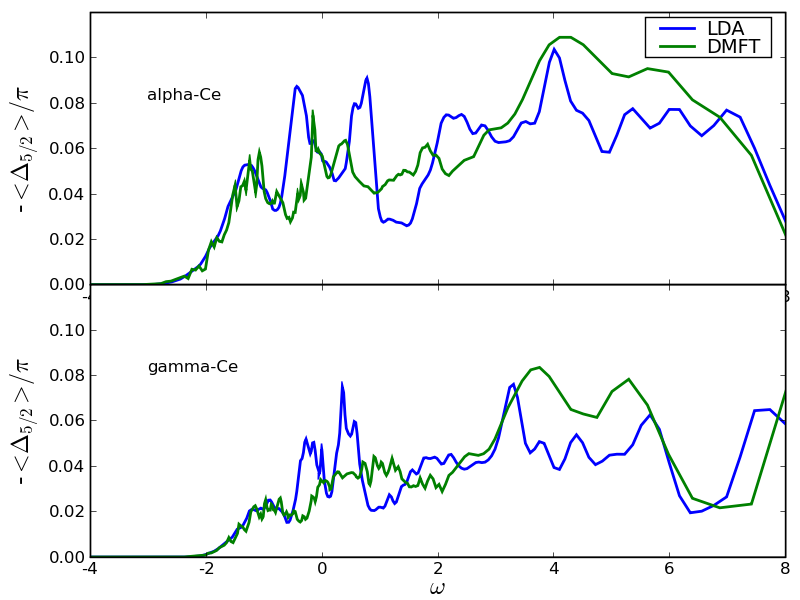}
}
\caption{The hybridization function of the $j_z=5/2$ subshell within
  LDA and within DMFT in both phases.
}
\label{CeHybrid}
\end{figure}

\begin{figure}[hbt]
\centering{
  \includegraphics[width=0.8\linewidth]{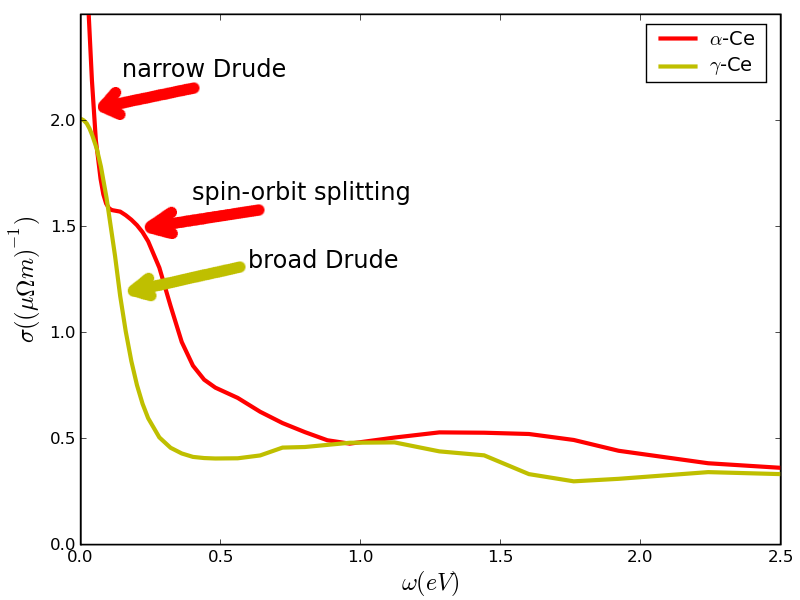}
}
\caption{
Optical conductivity of $\alpha$-Ce and $\gamma$-Ce within
LDA(Wien2K)+DMFT method. Note the shoulder in $\alpha$-Ce
conductivity, which is due to excitations across the two quasiparticle
peaks ($4f:5/2$ and $4f:7/2$) clearly visible in figure \ref{CeDOS},
and also measured by experiment of Ref.~\onlinecite{Ce-exp}.
}
\label{CeOptics}
\end{figure}

To test our implementation of DFT+DMFT within Wien2K method, we show
in Figs.~\ref{CeDOS} and \ref{CeDOS2} results for cerium $\alpha$ to $\gamma$
transition.

At a temperature less than $600\,$K and pressure less than 20$\,$kbar,
elemental cerium undergoes a transition between two isostructural
phases: a high pressure phase or $\alpha$ phase and a low pressure
$\gamma$ phase. In $\alpha$-Ce the $f$ electron is delocalized while
in $\gamma$-Ce the $f$ electron is localized. The transition is well
accounted for by phenomenological Kondo Volume Collapse picture \cite{Allen1,Allen2,Allen3}.

We treat only the Ce $4-f$ electrons as strongly correlated thus
requiring full energy resolution, while all other electrons such as Ce
$spd$ are assumed to be well described by the GGA.
We choose $U=5.5\,$eV and $J=0.68\,$eV for the Coulomb interaction.
The value of $U$ was obtained by constraint DFT calculation
\cite{McMahan} and $J$ was computed using the atomic physics program
of Ref.~\onlinecite{Cowan} and reduced by 30\% to account for
the screening in the solid. Both phases of Ce have fcc unit cell with
quite different volumes, $V_\alpha=28.06\textrm{\AA}^3$ and
$V_\gamma=34.37\textrm{\AA}^3$.  The results were converged with
5000-$\vk$ points, we use the GGA functional for the DFT part and use
OCA and CTQMC impurity solver to solve the auxliliary impurity
problem.

The results in Fig.~\ref{CeDOS} (obtained by OCA) and
Fig.~\ref{CeDOS2} (obtained by CTQMC) are practically identical and
very similar to previous LDA(LMTO)+DMFT results \cite{database}. One
can clearly see the broad quasiparticle peak in $\alpha$-Ce, split by
the spin-orbit coupling $\sim 0.3\,$eV. The lower peak has mostly
$5/2$ character and the upper peak mostly $7/2$-character. The system
is in good Fermi liquid regime at the temperature of $150\,$K used in
the calculation. The second phase with larger volume is in local
moment regime with no visible Kondo peak at the Fermi level, but
enhanced Hubbard bands.

It is instructive to examine the hybridization function
$\Delta=\omega-E_{imp}-\Sigma-1/G$ as computed by LDA and
self-consistent DMFT (see Fig.~\ref{CeHybrid}). It turns out that in
Ce, the low energy hybridization function is substantially reduced
compared to its LDA value. The two large peaks at $-0.4\,$eV and
$0.7\,$eV are absent in DMFT hybridization. Since the coherence scale
is exponential function of hybridization, the coherence scale is lower
in DMFT than it would be in so called one-shot DMFT. It is known from
the early days of the Kondo volume collapse theory \cite{Allen2}, that
the LDA hybridization in a one-shot calculation was too big and had to
be renormalized by phenomenological parameter
\cite{Allen-private}. DMFT reduces the hybridization through the
collective screening effects and hence is able to give correct
coherence scale of the problem.

Finally, let us show optical conductivity, as implemented in
LDA(Wien2K)+DMFT method. The overal agrement with previous LDA+DMFT
results~\cite{V9} is very good. The new computational results are in
even slightly better agreement with experiment of
Ref.~\onlinecite{Ce-exp}, since they both clearly display a shoulder
around $0.3\,$eV in $\alpha$-Ce, which we can now clearly identify as
excitations across the split quasiparticle peak. The splitting is 
due to spin-orbit coupling in Ce.

\section{Heavy fermion 115 materials}
\label{115}

\begin{figure}[!bt]
\centering{
  \includegraphics[width=0.3\linewidth]{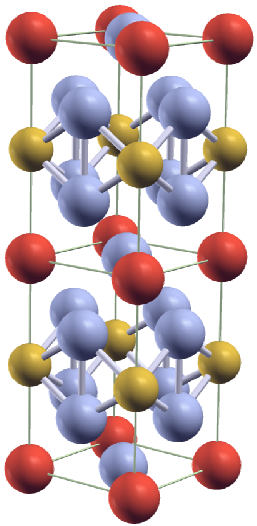}
}
\caption{Crystal structure of Ce$X$In$_5$. Red, yellow and gray
  spheres correspond to Ce, $X$, and In atoms, respectively.
}
\label{structure}
\end{figure}

The heavy fermion 115 materials have a chemical formula Ce$X$In$_5$,
where $X$ is either Co, Rh or Ir. They crystallize in layered
tetragonal structure shown in Fig.~\ref{structure}, composed of Ce-In
layers and $X$-In layers.

At high temperature, the low energy electronic states are composed of
mainly the broad $spd$ bands of In and Ce. The Ce-$4f$ electrons
are localized and their spectra is mostly contained in Hubbard bands,
which are more than $2\,$eV away from the Fermi level. These electrons
behave as local magnetic moments.
As the temperature is reduced, the moments combine with the conduction
electrons to form a fluid of very heavy quasiparticles, with masses
that are two or three orders of magnitude larger then the mass of the
electrons.

The low temperature physics of 115 materials is very puzzling. The
heavy fermion physics comes primarily from the
Ce-In layer. Indeed, the related material CeIn$_3$ has only the Ce-In
layers (no $X$-In layer), and also displays a similar heavy fermion
properties with superconductivity at very low temperature.
However, 115 materials are very sensitive to the substitution of the
transition metal ion in the $X$-In layer although Co, Rh and Ir ions
have the same valence (they are isovalent). Indeed the three
115 materials have dramatically different low energy properties:
CeCoIn$_5$ is a superconductor with $T_c\sim 2.3\,$K, CeRhIn$_5$ is
antiferromagnet with $T_N \sim 3.5\,$K, while CeIrIn$_5$ is
superconductor with $T_c$ of only $0.4\,$K. A fundamental question
arises: Why are the low energy properties of 115 materials so
different?

A hint to the resolution of this problem was given in
Ref.~\onlinecite{Science}, where the DFT+DMFT calculation for
CeIrIn$_5$ indicated that the Ce $4f$ electrons hybridize stronger to
the out of plane In-$p$ electrons, than the in-plane In $p$ electrons.
Here we carried out the DFT+DMFT calculation for all three 115
materials and we show the difference in electronic structure between
the three materials.  We used the code based on LDA-LMTO code of
Ref.~\onlinecite{Savrasov96} as well as the new LAPW code based on
Wien2K~\cite{Wien2K} code. The results obtained by our DFT+DMFT method
in the two codes are almost indistinguishable. For the impurity
solver, we used both OCA (described above) and CTQMC \cite{CTQMC}. The
analytic continuation of CTQMC results was performed with the method
described in chapter \ref{ancont}.

\begin{figure}[!bt]
\centering{
  \includegraphics[width=0.99\linewidth]{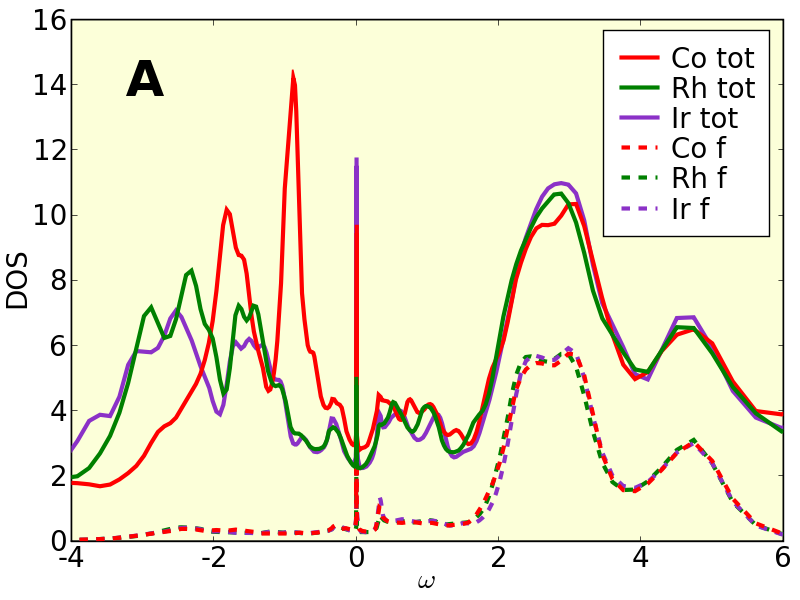}
  \includegraphics[width=0.99\linewidth]{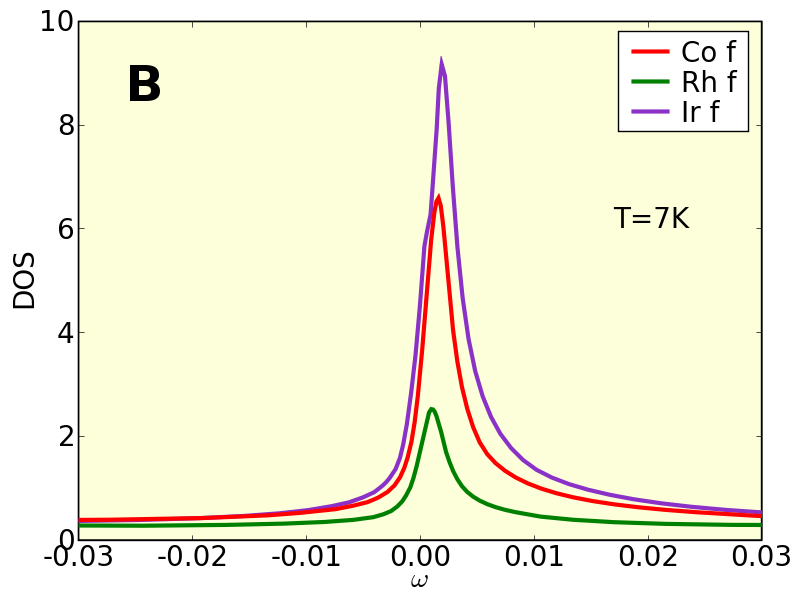}
  }
\caption{Total density of states (full lines) and partial Ce-$4f$
  density of states (dashed lines) for CeCoIn$_5$, CeRhIn$_5$ and CeIrIn$_5$
  materials. The lower pannel show the low energy part of the Ce-$4f$
  density of states for all three compounds. We used OCA solver.
}
\label{DOS}
\end{figure}
Fig.~\ref{DOS}A shows the total density of states (DOS) and the
partial Ce-$4f$ DOS for all three materials at low temperature of
7$\,$K. The transition metal ion DOS is peaked around binding energy
2eV, where the difference of DOS is large. The partial Ce-$4f$ DOS of
the three compound is very similar, except at the very low
energy. Fig.~\ref{DOS}B zooms-in the low energy part of the
spectra. We see that CeIrIn$_5$ compound has the largest quasiparticle
peak, the CeCoIn$_5$ follows, while the CeRhIn$_5$ has substantially
smaller quasiparticle peak at the same temperature of $7\,$K.

\begin{figure}[!bt]
\centering{
  \includegraphics[width=0.99\linewidth]{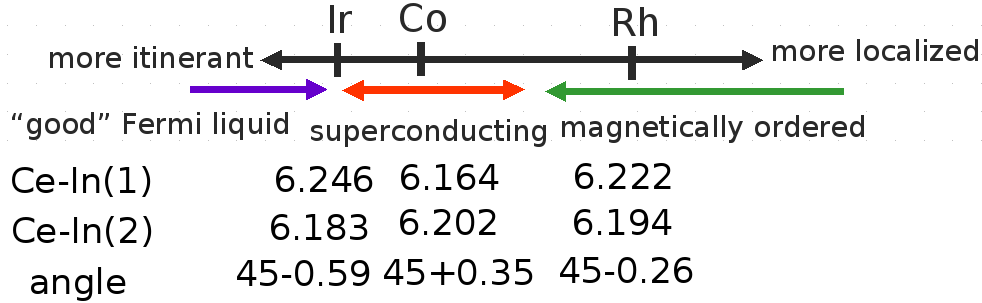}
  }
\caption{
  The sketch of the itinerancy/localization of the three 115
  compounds. In our view, the Ir compound is most itinerant, while the
  Rh compound is most localized.
  The Co compound is not localized
  enough to develop magnetic order at low temperature, while it is nor
  a good metal. It is thus conceivable that it would show tendency towards
  superconductivity. This phenomena is however beyond our current
  theoretical method - the single site DMFT calculation.
  We also show the bond distances between Ce and In atoms, and the
  angle between Ce and out-of-plane In atom. None of these parameters
  can explain the actual order of the compounds, hence the structure
  itself can not explain the trend of localized to itinerant
  transition in these compounds.
}
\label{sketch}
\end{figure}

Our view on the localization-itinerancy in 115 materials is sketched
in Fig.~\ref{sketch}. Rh compound is most localized, while Ir compound
is most itinerant. Co compound is similar to Ir compound, but slightly less
itinerant than Ir-115.

It is well known from the pressure
experiments~\cite{Pressure1_CeRhIn5,Pressure2_CeRhIn5} that Rh
compound is more localized then Co compound. Namely, under pressure of
1~GPa the Rh compound becomes superconducting, and at pressure of
$\sim 2$\,GPa reaches similar maximum T$_C$ as is the maximum T$_C$ of Co
compound \cite{Tusan}. Hence the pressure of the order of GPa
sufficiently increases the Ce-$4f$ hybridization that it overcome the
difference between localization of the electrons in the two compounds.
Experimentally it is a bit less clear what is the relation between Ir
and Co compound, since both compounds are superconductors at low T.
Ir compound has somewhat smaller specific heat coefficient in normal
state than Co compound ($750$~mJ/$(mol \textrm{K}^2)$ for Ir-115
versus 1000~mJ/$(mol \textrm{K}^2)$ for
Co-115)~\cite{Ir115-gamma,cvgamma}.  Ir compound has also somewhat
lower resistivity in the normal state \cite{resistivity}. Moreover,
nuclear quadrupol resonance (NQR) measurements of $1/(T_1 T)$
\cite{Ir-itinerant} suggest that Ir-compound might be more itinerant
than other Ce-compound.
Indeed pressurizing the CeIrIn$_5$ \cite{pressureCeIrIn5} along the
crystallographic $c$-direction, which increases itinerancy
\cite{thermalExp}, decreases T$_C$. Furthermore, it was shown that
Cd-doping acts as reverse pressure in 115's \cite{Cd-doping}. Since
higher Cd-doping is necessary for appearance of antiferromagnetic
phase in CeIrIn$_5$ than in CeCoIn$_5$, this is also suggestive of
more itinerant nature of Ir-compound.

Our results are thus consistent with the resistivity experiments
\cite{resistivity}, NQR experiments \cite{Ir-itinerant} and recent
pressure experiments \cite{pressureCeIrIn5}, and indicate that Ir
compound is on the itinerant side of the phase diagram. Hence the low
superconducting transition temperature might be connected with too
itinerant nature of carriers.

\begin{figure}[!bt]
\centering{ \includegraphics[width=0.99\linewidth]{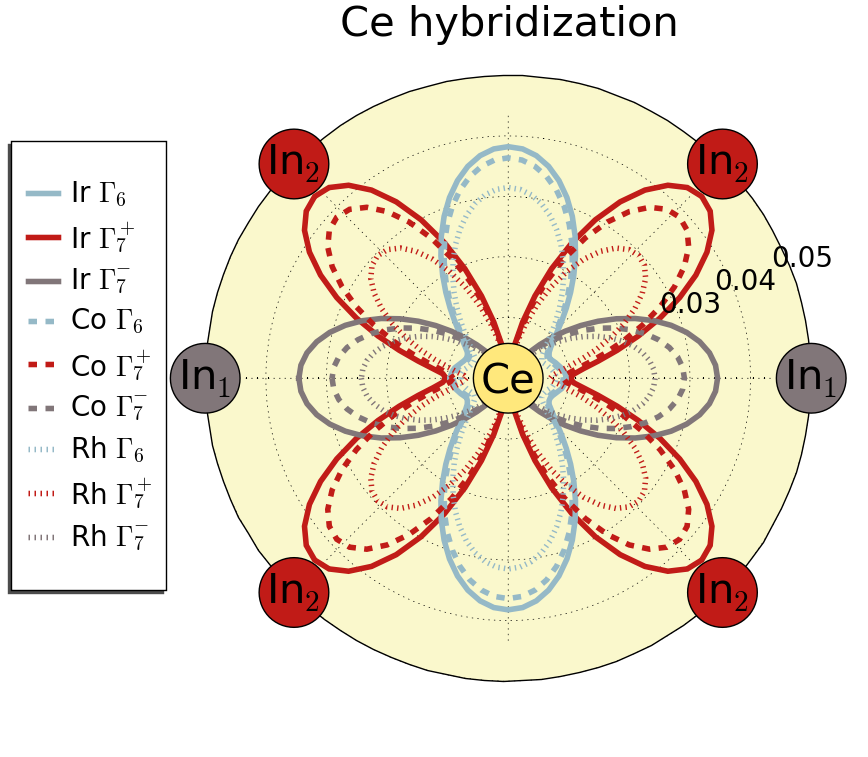}
  \includegraphics[width=0.32\linewidth]{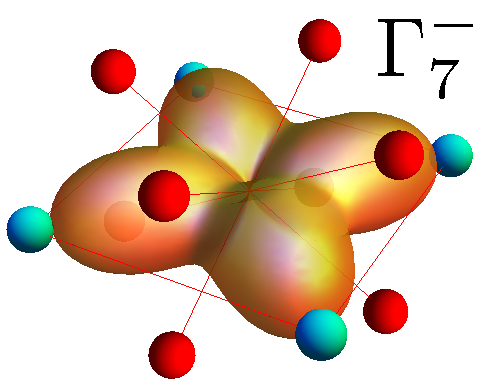}
  \includegraphics[width=0.32\linewidth]{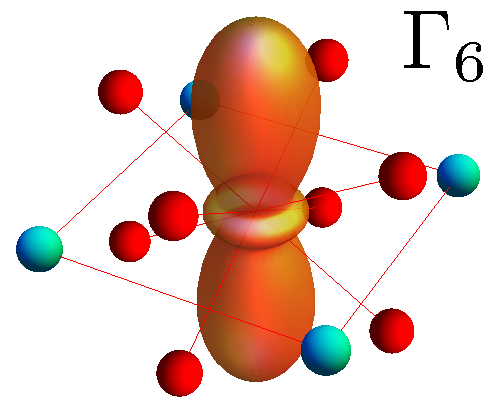}
  \includegraphics[width=0.32\linewidth]{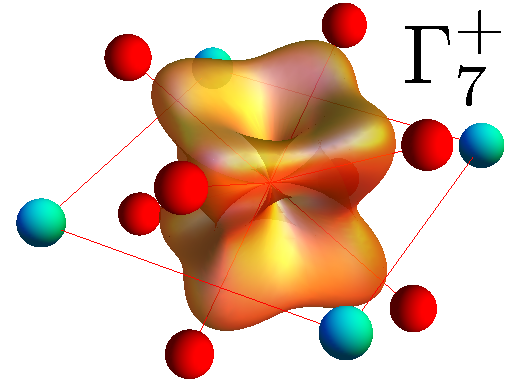} }
\caption{ The Ce-$4f$ Weiss field hybridization function
  $|\Delta^{''}|/\pi$ decomposed into crystal field components of
  tetragonal field. All quantities are in units of eV.  The upper plot
  shows the 2D projection of the three relevant orbitals, while the
  lower pannel shows their 3D shapes (blue dots mark the position of
  the in-plane In atoms, while the red dots the position of the
  out-of-plane In atoms). The radial extend of the orbitals in the
  polar plot of the upper panel is proportional to the value
  $|\Delta^{''}|/\pi$ at zero frequency.  The full/dashed/dotted lines
  correspond to CeIrIn$_5$/CeCoIn$_5$/CeRhIn$_5$. While all three
  components of hybridization $\Gamma_7^-$, $\Gamma_7^+$, and
  $\Gamma_6$, are largest (smallest) in CeIrIn$_5$ (CeRhIn$_5$)
  compound, $\Gamma_7^+$ takes the largest value and also changes
  more than the other two components.  }
\label{115-hybrid}
\end{figure}
We further analize the difference in itinerancy by plotting the
hybridization function at zero frequency $\Delta(\omega=0)_{LL}$
resolved in crystal field basis. The $14$ dimensional matrix of
hybridizations has a 6 dimensional $j=5/2$ component and a 9
dimensional $j=7/2$ component.  The itinerancy (the quasiparticle
peak) is almost entirely from the $j=5/2$ component, hence we will not
analize $j=7/2$ part. The degeneracy of the $5/2$ shell is lifted in
tetragonal crystal environment and hybridization splits into
$\Gamma_7^-$, $\Gamma_7^+$ and $\Gamma_6$ components. The $\Gamma_6$
corresponds to $j_z=\pm 1/2$, while $\Gamma_7^+$ and $\Gamma_7^-$
correspond approximately to $j_z=3/2$ and $j_z=5/2$, respectively.  In
Fig.~\ref{115-hybrid} we plot the hybridization ($-\Im[\Delta(0)]/\pi$)
in polar coordinates with Ce atom in the center and In atoms
around. The plot is the cut in $xz$ direction. The three dimensional
orbitals that correspond to the three crystal fields are plotted in
the lower pannel of Fig.~\ref{115-hybrid} together with the real space
positions of In and Ce atoms. In this decomposition,
hybridizations $\Gamma_7^+$ and $\Gamma_7^-$ are pointing towards out
of plane In (In$_2$) and in-plane In (In$_1$), respectively. The
third component, $\Gamma_6$ is pointing towards transition metal ion.

When comparing hybridization of the three 115 compounds, the Ir
compound has all three components of the hybridization larger than the
other two compounds. In Co-115 all three hybridizations are slightly
smaller, while in Rh-115 all three hybridizations are substantially
smaller.

Furthermore, comparing the strength of the three components of the
hybridization, one can notice that in Ir compound the $\Gamma_7^+$
component, pointing towards out-of plane In, is largest. This is
consistent with the experimental finding of Oeschler~\textit{et
  al.}~\cite{thermalExp} that the Gr\"uneisen parameters in
$c$-direction is 2.5 times bigger that in $a$ direction, resulting in
larger effective coherence scale in $c$-direction.

In Co-compund the $\Gamma_7^+$ and $\Gamma_6$ components have similar
strength, while $\Gamma_7^-$ is smaller, hence the
hybridization in $c$-direction is still more important than
in $ab$-plane, consistent with Gr\"uneisen parameter
measurements~\cite{thermalExp}.
It was shown in Ref.~\onlinecite{Science} that the double peak
structure of the optical conductivity is directly related to the
strength of the two hybridizations. The hybridization gap in one part
of the momentum space is larger, and is primarily due to out of plane
In, and the hybridization gap in some other part of momentum space,
controled mainly by the in-plane In, is smaller, resulting in double
peak structure of the mid-infrared optics peak. Optical measurements
on CeCoIn$_5$ of Singley~\textit{et al}\cite{Singley} demonstrated
very clearly that the mid-infrared peak is split into two peaks, one
at $250\,$cm$^{-1}$ and one at $630\,$cm$^{-1}$, which can hint
towards substantial difference in the two types of hybridization.

Finally, in contrast to Ir and Co compound, Rh compound has largest
$\Gamma_6$ hybridization, followed by $\Gamma_7^+$ and $\Gamma_7^-$.

\begin{figure}[hbt]
\centering{
  \includegraphics[width=0.99\linewidth]{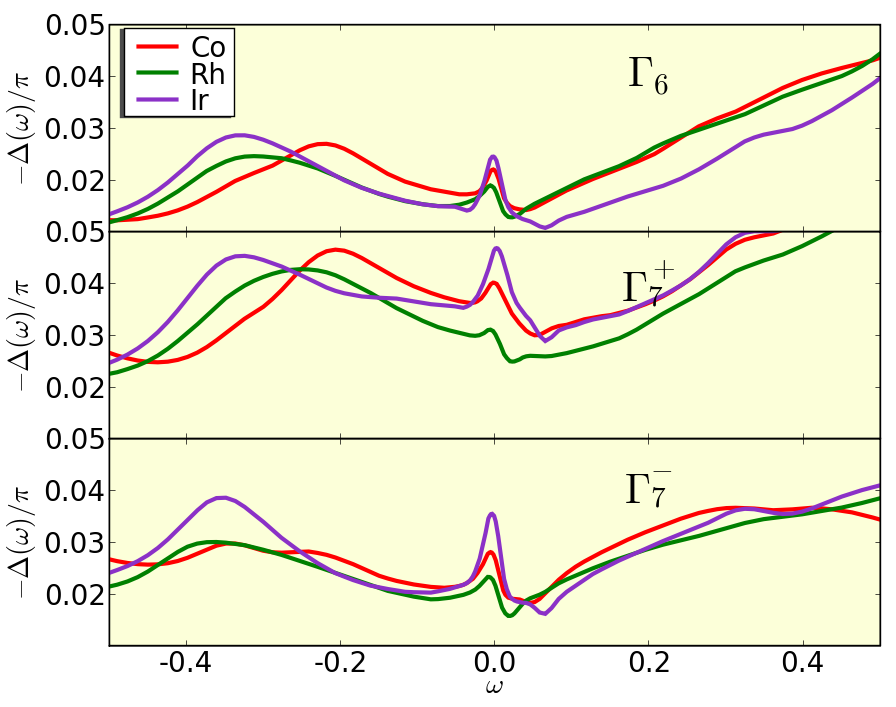}
  }
\caption{
  The frequency dependence of the three most important hybridization
  functions $\Delta^{''}(\omega)/\pi$ in all three 115 compounds:
  CeCoIn$_5$, CeRhIn$_5$ and, CeIrIn$_5$. The frequency is in units of eV.
}
\label{qstm}
\end{figure}
%
In Fig.~\ref{qstm} we show the frequency dependent hybridization
function $-\Im\Delta(\omega)/\pi$ to demonstrate that the retardation
effects in heavy fermion materials are very nontrivial and that the
buildup of the quasiparticle peak in spectral function usually results
in a sharp peak in hybridization, on the background of the depleted
region of hybridization. The peak is sometimes called the collective
hybridization, because it arises from the lattice effects. Namely, the
Ce-$4f$ electrons on neighboring atoms also become delocalized,
enhancing the hybridization at low energy. However, the spd-electrons
need to screen many Ce-$4f$ moments, and therefore the effective $spd$
hybridization is actually slightly reduced, resulting in depletion
away from the Fermi level, sometimes called Kondo hole.

Our results demonstrate that the degree of itinerancy is controlled by
the \textit{collective hybridization}, encoded into the Weiss mean
field hybridization $\Delta(\omega)$ within DMFT. But what is the
origin of the difference between the three compounds? In
Fig.~\ref{sketch} we show the parameters of the lattice structure,
namely the Ce-In(1) distance, the Ce-In(2) distance and the angle
between the CeIn$_3$ plane and out of plane In (In$_2$). From these
numbers, it is clear that none of the three quantities follows the
trend of itinerancy. Hence the difference in the lattice structure is
likely not the key element.

To demonstrate that the difference in the lattice structure is not the
driving force, we performed the DMFT calculation for the three
compounds using the same lattice structure of CeIrIn$_5$. The results
were very similar to the results plotted in Fig.~\ref{DOS}, with only
slight increase in itinerancy of Rh compound. This demonstrates that
the chemistry of the transition metal ion (difference between $3d$,
$4d$ and $5d$ orbitals) is the driving force of the itinerancy, and
not the diffrence in the crystal structure. The latter are the
secondary effects.

\begin{figure}[hbt]
\centering{ \includegraphics[width=0.99\linewidth]{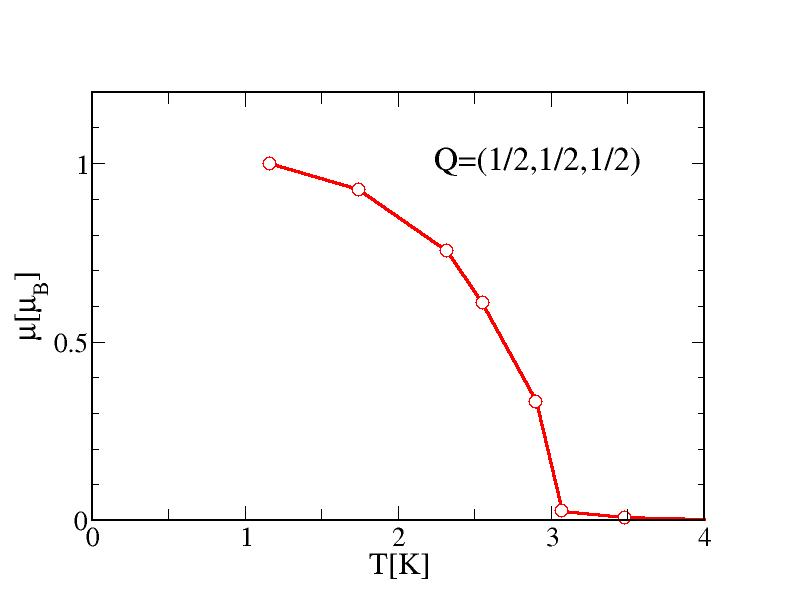}}
\caption{
  Temperature dependence of the magnetic moment of the comensurate AFM
  Neel state in CeRhIn$_5$.
}
\label{Rh-moment}
\end{figure}
Since CeRhIn$_5$ remains in local moment regime down to very low
tempeture of the order of the RKKY interaction, it is worth trying to
stabilize a magnetic solution within DMFT. To this end, we doubled the
unit cell and allowed the comensurate antiferromagnetic ordering with
the wave vector $(1/2,1/2,1/2)$. Experimentally, the order is a
helical spiral with wave vector $(1/2,1/2,0.298)$ and $T_C$ of
$3.8\,$K. The broken symmetry solution can be stabilized below
$T\sim 3\,$K as shown in Fig.~\ref{Rh-moment}. The magnetization has a
typical mean field form, as expected for a theory with spatial
mean-field character like DMFT.

\begin{figure}[hbt]
\centering{
  \includegraphics[width=0.99\linewidth]{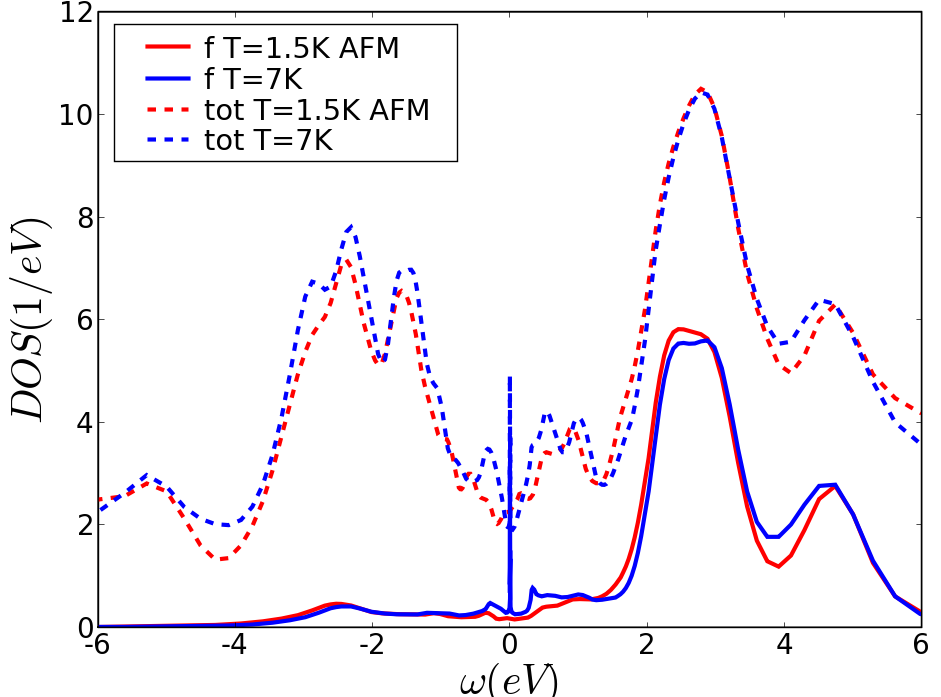}
  \includegraphics[width=0.99\linewidth]{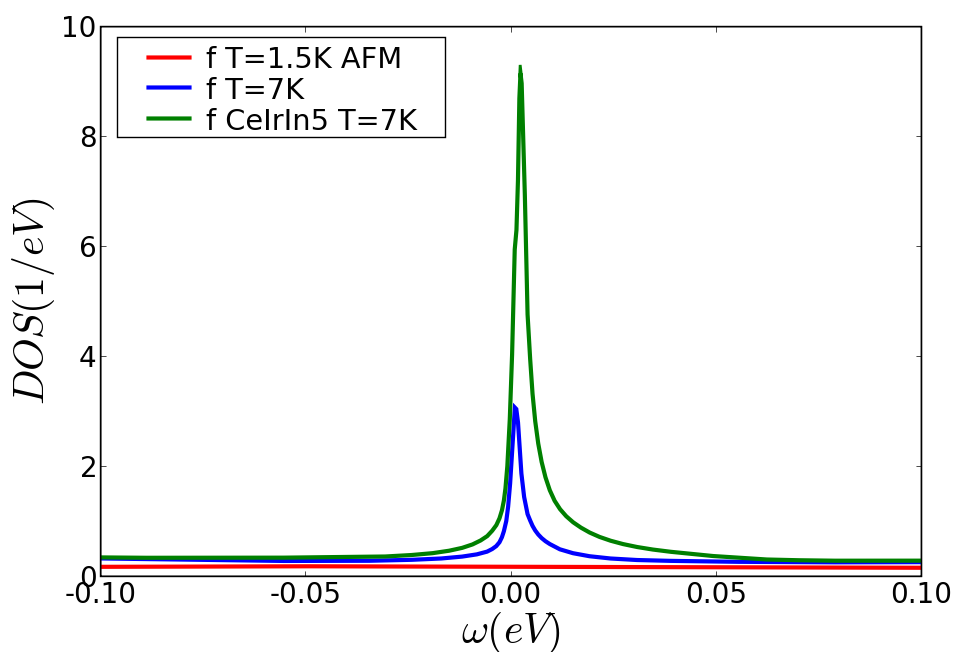}
  }
\caption{ Total and partial Ce-$4f$ density of states for Rh-115 below
  and above the AFM transition. Above the Neel temperature, there is a
  signature of Kondo effect, wich partially screenes magnetic moment
  at elevated temperatures, even though the system develops the long
  range order below 3K. The quasiparticle peak is however much smaller than
  the same peak in CeIrIn$_5$ material. Once in the ordered state, the
  quasiparticle peak dissapears.
}
\label{Rh-dos}
\end{figure}
An interesting question is how does the large moment antiferromagnetic
solution change the emerging quasiparticle peak. We have shown in
Fig.~\ref{DOS} that even in more localized CeRhIn$_5$ a peak starts to
develop at the Fermi level by decreasing temperature, hence coherence starts to
develop at quite high temperature similar to the other two
compounds. However, the height of the quasiparticle peak is smaller
and the scattering rate of Ce-$4f$ orbital (imaginary part of the
self-energy) is higher in CeRhIn$_5$. The long range order state
develops from a state with a partially screened moment. In
Fig.~\ref{Rh-dos} we show the density of states of the two phases, the
paramagnetic state and the Neel state. The latter has no quasiparticle
peak left and only a very broad background of the $f$ spectral weigh
remains at the Fermi level. The lower panel of Fig.~\ref{Rh-dos} compares a very
coherent quasiparticle peak of CeIrIn$_5$ with the partially screened
state of CeRhIn$_5$ above $T_{Neel}$ and in the ordered state below
$T_{Neel}$, to emphasize the dramatic difference in the density of
state at low energy. Because the full coherence of quasiparticles is
not reached to very low temperature in CeRhIn$_5$, and the non-local
RKKY interaction is strong enough, it interrupts the formation of
coherent quasiparticles. Within DMFT, this is reflected in two stable
solutions of DMFT equations, the paramagnetic and the magnetic
solution.
We note that we did not prove the stability of the magnetic solution
compared to the paramagnetic solution, because this would require a
comparison between free energies, a task beyond our current
capabilities. However, our experience from model calculations suggests
that when the magnetic DMFT solution can be stabilized, it usually has
lower free energy than the nonmagnetic solution.

\section{Conclusion}

In the first part of the article, we discussed in detail the
implementation of DFT+DMFT in full potential methods.  We defined the
central object of the DMFT, the local Green's function using a
projection operator. We showed that the projector used in LDA+U
implementations leads to non-causal DMFT equations and that the
straightforward projection to the solution of the Schr\"odinger
equation within the Muffin Tin sphere leads to spectral weight
loss. We suggested an alternative projection that resolves these
shortcomings.

We sketched the algorithmic steps within an implemention of DFT+DMFT
in the full potential methods, using a formulation which avoids the
ambiguities of downfolding or Wannier orbital construction. Hence, the
kinetic energy operator and electron density are \textbf{not
approximated} by a tight-binding parameterization, which allowed us to
carry out a charge density self-consistent calculation.

In the second part of the article, we concentrated on impurity solvers
based on the hybridization expansion. We derived the equations for the
\textbf{bold} continuous time quantum Monte Carlo (CTQMC) method,
which samples the \textit{dressed} propagators, as opposed the bare
propagators sampled in current CTQMC methods. We showed a few test
results for simplified implementation of the method.
In this part of the article we also gave detailed formulas for the
impurity solver called the One-crossing approximation, which can be
viewed as the four kink approximation within the bold CTQMC.

Finally we give details on a new analytic continuation method, which
can continue the self-energy from the imaginary to the real
axis. This step is crucial when computing the response functions
within DMFT, as done in section~\ref{transport} for transport
coefficients.

In the third part of the article, we presented the test results of our
DFT+DMFT implementation on a classical problem of strong correlations, the
isostructural transition of elemental cerium from its $\gamma$ phase
at high temperature to its $\alpha$ phase at low temperature.

In the last part of the article, we applied the DFT+DMFT method to a
group of heavy fermion compounds, namely CeIrIn$_5$, CeCoIn$_5$
and CeRhIn$_5$, collectively dubbed the Ce-115s.  Although the
isovalent substitution of a transition metal ion does not
substantially alter the Ce-In planes, which are believed to be
responsible for the heavy mass in these compounds, the ground state
properties of these materials are very different. 

We analyzed the electronic structure of the three Ce-115 materials and
showed that the Ce-$4f$ electrons in CeRhIn$_5$ are more localized
that those in the other two 115 compounds, in agreement with
experiments. Below $3\,$K, an antiferromagnetic DFT+DMFT solution in
CeRhIn$_5$ is stable, while CeCoIn$_5$ and CeIrIn$_5$ remain
paramagnetic (the AFM solution is not stable) down to the lowest
temperature $T = 1.5\,$K explored in our calculation.

The hybridization in CeIrIn$_5$ is very anisotropic with the largest
component pointing towards the out-of-plane In. The hybridization is
slightly smaller in CeCoIn$_5$, hence we believe CeIrIn$_5$ to be more
itinerant than the other two compounds.

We speculate that the reason CeCoIn$_5$ exhibits the highest
superconducting T$_C$ is due to the fact that it is at the
border between itinerancy and localization, while CeIrIn$_5$ is on the
itinerant side of the phase diagram and CeRhIn$_5$ is on the localized
side. The position of CeRhIn$_5$ in the phase diagram is clear from
the pressure experiments, while the position of CeIrIn$_5$ is less
obvious. We believe that recent uniaxial pressure
experiments~\cite{pressureCeIrIn5} confirm our view, since the 
$c$-axis compression, which makes CeIrIn$_5$ more itinerant, decreases
the superconducting $T_c$.

\section{Acknowledgement}

We thank Gabriel Kotliar for careful reading of the manuscript and
numerous usefull suggestions from the early stage of the project till
its completion. We are grateful to Jim Allen and David Pines for
fruitful discussion.  K.H was supported by Grant NSF NFS DMR-0746395
and DMR-0806937, and Alfred P. Sloan fellowship. C.H.Y was funded by
NSF DMR-0806937 and K.K. by Petroleium Research Fund 48802-DNI10.

\textbf{Note added:} As the writing of this work was being completed,
we became aware of a related work of M.~Aichhorn \textit{et al.}
(arXiv: 0906.3735) also reporting on an implementation of LDA+DMFT in
a LAPW code. However, in contrast to our implementation, the authors
used downfolding method to obtain a tight-binding model Hamiltonian,
and hence could not compute electronic charge self-consistently.
In the process of downfolding, the authors used projection $P^1$
defined in section~\ref{basis_set}.  Furthermore, the impurity solver
did not take into account the full Coulomb interaction
Eq.~(\ref{CoulombUU}). Only the density-density part of the
interaction was considered by M.~Aichhorn \textit{et al.} (only the
$z$ component of the Hund's coupling) which allows substantial
simplification of the impurity solver, but leads to improper
description of the multiplet structure of the correlated atoms.

\appendix

\section{Complex Tetrahdron Method}
\label{complexTetrahedra}

The formulas for tetrahedron integral in case of complex eigenvalues
are very similar to the case of real eigenvalues. However, a special
attention needs to be payed to choose the right branch-cut in
logarithms, such that all terms in the sum are causal.

First step in tetrahedron method consists of dividing the first
Brillouin zone into tetrahedra which fill up whole space. Each
thrahedra has four corners. The energy is thus interpolated
$\varepsilon =
\varepsilon_1+a(\varepsilon_2-\varepsilon_1)+b(\varepsilon_3-\varepsilon_1)+c(\varepsilon_4-\varepsilon_1)$,
where $a$, $b$ and $c$ run between 0 and 1 when visiting corners of
tetrahedra.

For the Green's function we need integral of the form
\begin{equation}
\sum_\vk \frac{C_{\vk}}{\omega-\varepsilon_\vk}
\rightarrow \sum_\vk w(\vk,\omega) C_\vk
\end{equation}
and for the electron density and the chemical potential we need
\begin{equation}
\sum_\vk \int_{\omega_1}^{\omega_2} d\omega
\frac{C_{\vk}}{\omega-\varepsilon_\vk}
\rightarrow \sum_\vk wi(\vk,\omega) C_\vk
\end{equation}

The integral is first written as the sum over all tetrahedra and the
integral in the interior of tetrahedra:
\begin{equation}
  \sum_\vk \frac{C_{\vk}}{\omega-\varepsilon_\vk}=
  \sum_{t}\int_{t}d^3\vk \frac{C_{\vk}}{\omega-\varepsilon_\vk}=
  \sum_t\sum_{k_i=1}^4 w(k_i,\omega) C_{k_i}
\end{equation}
The latter is evaluated analytically using linear interpolation inside
the volume of the tetrahedra for both the nominator and denominator
\begin{eqnarray}
  && w(k_i,\omega) = 6\int_0^1 dc\int_0^{1-c}db\int_0^{1-b-c}da \times \\
  &&\qquad\frac{
    (1-a-b-c)\delta_{k_i,1}+a\delta_{k_i,2}+b\delta_{k_i,3}+c\delta_{k_i,4}}
  {\omega - \varepsilon_1-a(\varepsilon_2-\varepsilon_1)-b(\varepsilon_3-\varepsilon_1)-c(\varepsilon_4-\varepsilon_1)}\nonumber
\end{eqnarray}
Here we used a short notation $\varepsilon_{k_i}\equiv \varepsilon_i$

The integrals are analytic and a closed expression for computing the
green's function is
\begin{eqnarray}
  w(k_i,\omega) = \sum_{j\ne i}\frac{\omega-\varepsilon_j}
  {\prod_{l\ne i,j}(\varepsilon_l-\varepsilon_j)}\mathrm{lv}\left({\omega-\varepsilon_j},{\varepsilon_j-\varepsilon_i}\right)\nonumber
\end{eqnarray}
where
\begin{equation}
 \textrm{lv}(x,y) = \frac{x}{y} \left\{ 1 - \frac{x}{y} \left[\log(x+y) - \log(x)\right]\right\} 
\end{equation}
and $l\ne i,j$ means $l\ne i$ and $l\ne j$. Notice that only
$\log(x+y)$ and $\log(x)$ can appear in $\textrm{lv}(x,y)$ (not $\log(y)$) to ensure
causality. Namely, imaginary part of all $\varepsilon_i$ is strictly
negative, hence the expression
$\mathrm{lv}\left(\omega-\varepsilon_j,\varepsilon_j-\varepsilon_i\right)$
contains $\log(\omega-\varepsilon_j)$ and
$\log(\omega-\varepsilon_i)$, which both have imaginary part in the
interval $[0,\pi]$.
  
Similarly, the formulas for the integral over frequency
$\int_{\omega_1}^{\omega_2}w(k_i,\omega)d\omega$ are
\begin{eqnarray}
  wi(k_i,\omega_2,\omega_1) &=&
  \sum_{j\ne i}\frac{\mathrm{ilv}(\omega_2-\varepsilon_j,\varepsilon_j-\varepsilon_i)}{\prod_{l\ne i,j}(\varepsilon_l-\varepsilon_j)}\nonumber\\
  &-&\sum_{j\ne i}\frac{\mathrm{ilv}(\omega_1-\varepsilon_j,\varepsilon_j-\varepsilon_i)}{\prod_{l\ne i,j}(\varepsilon_l-\varepsilon_j)}
\end{eqnarray}
where
\begin{eqnarray}
\textrm{ilv}(x,y) = \frac{1}{4} y^2 &\{&
u^4[\log(x)-\log(x+y)] + \log(x+y)
\nonumber\\
&+& u^3+\frac{1}{2}u^2-u
\}
\end{eqnarray}
and $u=x/y$

\section{Transport integrals}
\label{AppB}

To compute the transport coefficients, we need to evaluate to high
precision the following integrals
\begin{eqnarray}
P_1(z) &=& \int  dx\left(- \frac{df}{dx}\right)\frac{1}{x-z}
\label{P1}\\
P_2(z,\gamma) &=& \int  dx\left(- \frac{df}{dx}\right)\frac{1}{|x-z+i x^2 \gamma|^2}
\label{P2}\\
Q_2(z,\gamma) &=& \int  dx\left(- \frac{df}{dx}\right)\frac{1}{(x-z+i x^2 \gamma)^2}
\label{Q2}
\end{eqnarray}

The integrals need to be carefully implemented and special care needs
to be taken for the two case: a) $|z|\gg 1$ and b) $|z^{''}|\ll 1$ and
$|\gamma|\ll 1$.

The first integral of Eq.~(\ref{P1}) is computed numerically, except in
the following cases
\begin{eqnarray}
P_1(z) =\left\{
\begin{tabular}{ll}
$-({1}/{z} + {c_0}/{z^3} + {c_1}/{z^5}  +{c_2}/{z^7} + {c_3}/{z^9})$ & $|z|>10$\\
$w_0(z') + i\pi \frac{df}{dx}(z')$& $|z^{''}|\ll 1$  
\end{tabular}
\right.
\nonumber
\end{eqnarray}
where
$$w_0(x)= P\int \frac{\frac{df}{dt}(t)dt}{t-x}$$
is precomputed on a fine mesh and interpolated using cubic spline
interpolation.
The constants $c_i$ are
\begin{eqnarray}
  c_0 = \frac{\pi^2}{3},\quad
  c_1 = \frac{7\pi^4}{15},\quad
  c_2 = \frac{31\pi^6}{21},\quad
  c_3 = \frac{127\pi^8}{15}
\end{eqnarray}

The second integral of Eq.~(\ref{P2}) is computed numerically, except
in the following cases i) $|z^{''}|\ll 1, \gamma|\ll 1 $: In this
limit it becomes  $P_2(z,\gamma)\sim \frac{\pi}{|z^{''}|}\frac{df}{dx}(z^{'})$, ii)
$|z|>14$: In this case, the power expansion in terms of $|z|^2$ is performed
and all terms are analytically evaluated.

Similarly we treat integral Eq.~(\ref{Q2}). For $|z^{''}|\ll 1$, 
$|\gamma|\ll 1 $ we approximate 
$Q_2(z,\gamma)\sim \frac{d^2 f}{dx^2}(a)-\frac{i\pi}{4}\frac{\sinh(a/2)}{\cosh^3(a/2)}$
and for $|z|>8$ we perform the power expansion in terms of $z^2$ and
analytically evaluated the resulting integrals.

\end{document}